\documentclass[final,times]{elsarticle}

\usepackage{graphicx}
\usepackage[table,xcdraw]{xcolor}
\usepackage{caption}
\usepackage{subcaption}
\usepackage{multirow}
\usepackage{hyperref}
\usepackage[export]{adjustbox}%
\usepackage{amsmath}
\usepackage{tabularx}
\usepackage{array}
\usepackage{booktabs}
\usepackage{siunitx}
\usepackage{makecell}
\usepackage[a4paper, lmargin=0.1666\paperwidth, rmargin=0.1666\paperwidth, tmargin=0.1111\paperheight, bmargin=0.1111\paperheight]{geometry} %

\usepackage{amssymb}
\usepackage{lineno}
\usepackage{babel}

\journal{Computers, Environment and Urban Systems}

\usepackage{url}
\usepackage{boxedminipage}
\usepackage{textpos}

\begin{document}
\begin{sloppypar}
\begin{frontmatter}

\title{VoxCity: A Seamless Framework for Open Geospatial Data Integration, Grid-Based Semantic 3D City Model Generation, and Urban Environment Simulation}

\author[doa,tak]{Kunihiko Fujiwara}
\author[nsri]{Ryuta Tsurumi}
\author[ynu]{Tomoki Kiyono}
\author[doa]{Zicheng Fan}
\author[doa]{Xiucheng Liang}
\author[doa]{Binyu Lei}
\author[doa]{Winston Yap}
\author[doa]{Koichi Ito}
\author[doa,dre]{Filip Biljecki\corref{cor1}}

\cortext[cor1]{Corresponding author. Email: filip@nus.edu.sg}

\affiliation[doa]{organization={Department of Architecture, National University of Singapore}, addressline={4 Architecture Drive}, postcode={117566}, country={Singapore}}
\affiliation[tak]{organization={Research \& Development Institute, Takenaka Corporation},addressline={1-5-1 Otsuka, Inzai, Chiba}, postcode={270-1352}, country={Japan}}
\affiliation[nsri]{organization={Nikken Sekkei Research Institute}, addressline={3-7-1 Kanda Ogawamachi, Chiyoda-ku, Tokyo}, postcode={101-0052}, country={Japan}}
\affiliation[ynu]{organization={Institute of Urban Innovation, Yokohama National University}, addressline={79-5 Tokiwadai, Hodogaya, Yokohama, Kanagawa}, postcode={240-8501}, country={Japan}}
\affiliation[dre]{organization={Department of Real Estate, National University of Singapore}, country={Singapore}}

\begin{abstract}%
\begin{textblock*}{\textwidth}(0cm,-11.2cm)
\begin{center}
\begin{footnotesize}
\begin{boxedminipage}{1\textwidth}
This is the Accepted Manuscript version of an article published by Elsevier in the journal \emph{Computers, Environment and Urban Systems} in 2026, which is available at: \url{https://doi.org/10.1016/j.compenvurbsys.2025.102366}\\ Cite as:
Fujiwara K, Tsurumi R, Kiyono T, Fan Z, Liang X, Lei B, Yap W, Ito K, Biljecki F (2026): VoxCity: A seamless framework for open geospatial data integration, grid-based semantic 3D city model generation, and urban environment simulation. \textit{Computers, Environment and Urban Systems}, 123: 102366.
\end{boxedminipage}
\end{footnotesize}
\end{center}
\end{textblock*}

\begin{textblock*}{1.5\textwidth}(-.78cm,17.9cm)
{\tiny{\copyright{ }2026, Elsevier. Licensed under the Creative Commons Attribution-NonCommercial-NoDerivatives 4.0 International (\url{http://creativecommons.org/licenses/by-nc-nd/4.0/})}}
\end{textblock*}
Three-dimensional urban environment simulation is a powerful tool for informed urban planning. However, the intensive manual effort required to prepare input 3D city models has hindered its widespread adoption. To address this challenge, we present VoxCity, an open-source Python package that provides a one-stop solution for grid-based 3D city model generation and urban environment simulation for cities worldwide. VoxCity's `generator' subpackage automatically downloads building heights, tree canopy heights, land cover, and terrain elevation within a specified target area, and voxelizes buildings, trees, land cover, and terrain to generate an integrated voxel city model. The `simulator' subpackage enables users to conduct environmental simulations, including solar radiation and view index analyses. Users can export the generated models using several file formats compatible with external software, such as ENVI-met (INX), Blender, and Rhino (OBJ). We generated 3D city models for eight global cities, and demonstrated the calculation of solar irradiance, sky view index, and green view index. We also showcased microclimate simulation and 3D rendering visualization through ENVI-met and Rhino, respectively, through the file export function. Additionally, we reviewed openly available geospatial data to create guidelines to help users choose appropriate data sources depending on their target areas and purposes. VoxCity can significantly reduce the effort and time required for 3D city model preparation and promote the utilization of urban environment simulations. This contributes to more informed urban and architectural design that considers environmental impacts, and in turn, fosters sustainable and livable cities. VoxCity is released openly at \href{https://github.com/kunifujiwara/VoxCity}{https://github.com/kunifujiwara/VoxCity}.
\end{abstract}


\end{frontmatter}

\section{Introduction}
\label{intro}

Three-dimensional urban environment simulations facilitate assessing various conditions including outdoor heat stress \citep{Lindberg2008-hq, Yuan2022-wf}, wind flow \citep{Paden2022-me}, visual perception \citep{Urech2022-es, Oh1994-ot}, building energy demand \citep{Gros2016-xo, Bouyer2011-mt}, air quality \citep{Katal2022-iq}, and noise propagation \citep{Stoter2008-sa, Zhao2017-cj}, at high spatial and temporal resolution. 
Their results help diverse applications including policy-making, urban planning, city management, architectural and landscape design \citep{Qi2023-os, Qi2011-lw, Gaspari2018-ix}. 
Such urban environment simulations require 3D city models with semantic information for diverse elements. For example, microclimate simulation uses heat-related parameters such as heat conductivity, solar reflectance, transmittance, and evaporation coefficients for each object \citep{Taleghani2016-mf, Robineau2022-qv}. Therefore, models need to define semantic classes of objects, such as buildings, roads, vegetation, and water bodies, that determine these parameters. Simulation of visual perception also requires similar semantic information to evaluate visibility of greenery or buildings \citep{Giannico2022-lh, Virtanen2021-bb, Qiang2019-wg}.  
However, in most cases of urban environment simulations, researchers and engineers must prepare such 3D city models themselves due to a lack of available data sources \citep{Lei2023-cb, Lei2023-dj}.
The intensive manual effort required to create or adapt these models has been a significant bottleneck for urban environment simulations. 
Although some cities have prepared open 3D city models with semantic attributes, the coverage of such data is currently limited. Moreover, existing models may not be simulation-ready because of issues such as incompatible data formats, unsealed solid geometries, and insufficient semantic information \citep{Paden2022-me, Paden2024-fb, Lei2023-cb}. 

Meanwhile, an increasing amount of global geospatial data relevant to 3D city models has been openly released in the last few years, largely due to rapid advancements in deep learning techniques \citep{Kamath2024-at, Tolan2024-an, Zanaga2022-eb, Hawker2022-by}. For instance, \citet{Sirko2023-hb} released building height data with a 4 m spatial resolution for Africa, South Asia, Southeast Asia, Latin America and the Caribbean; \citet{Tolan2024-an} published global canopy height data at a 1 m resolution; \citet{Zanaga2022-eb} released land cover data with worldwide coverage at a 10 m resolution; and \citet{Hawker2022-by} created global terrain elevation data with a 1 Arc-second (30.9 m at the equator) resolution, based on Copernicus DEM, canopy height and building footprint data. 
We can use these open geospatial datasets to create 3D city models with semantic information regarding buildings, trees, land cover, and terrain. 
However, while these datasets can reduce the efforts required to collect necessary information and increase the coverage, they still require intensive manual efforts to integrate them and reconstruct 3D city models of sufficient quality for advanced uses such as simulations. Each dataset has different data types, such as raster and vector, and varying spatial resolutions, making the data integration process complex. 
To scale applications of urban environment simulation globally, it is desirable to automate these manual efforts for data integration.
Moreover, even with simulation-ready models, working in simulation software remains time-consuming, particularly for tasks such as data import, boundary condition settings, and result data export.

Several open-source packages have been developed to address these challenges.
3dfier \citep{Ledoux2021-iz} and City4CFD \citep{Paden2024-fb, Paden2022-me} automate the integration of land use and point cloud data to generate semantic 3D city models. They enable users to prepare input 3D city models for urban environmental simulations; however, point cloud data is not available for most cities, and they do not incorporate tree canopies.
UMEP \citep{Lindberg2018-jg} offers functionality to simulate urban thermal and wind environments using 3D city models; however, users need to prepare input datasets including building digital surface models (DSMs) and vegetation DSMs, which require manual efforts, and are, moreover, not available for most cities. 
Additionally, to the best of our knowledge, there are no open packages that provide seamless functionality encompassing open geospatial data integration, 3D city model generation, and urban environment simulation.

To bridge the gap between 3D model generation and urban environment simulation, a meshing process fundamentally needs to be incorporated.
Many urban environment simulation methods include the process where 3D city models are split into meshes with certain scales. 
For example, Computational Fluid Dynamics (CFD) for wind environment simulation requires volume meshes for fluid (air) domains and surface meshes for solid objects, such as buildings and ground \citep{Katz2011-oj}. Ray-tracing for solar irradiance simulation requires surface meshes of solid objects, where solar radiation is reflected or absorbed \citep{Willenborg2018-kh}. 
There are two principal methods for the meshing process: structured meshing and unstructured meshing. Structured meshing is a grid-based approach to create cuboid volume meshes and rectangular surface meshes \citep{Zhang2023-at, Chen2024-ch}. Unstructured meshing does not rely on a grid and generates polyhedral volume meshes and triangular surface meshes, allowing for more flexibility in shape and proportion \citep{Gargallo-Peiro2016-zk, Katz2011-oj}. Numerous software packages employ structured meshing due to its lower computational load for both calculation and memory, as well as its stability in numerical simulations. Unstructured meshing has the advantage of incorporating curved surfaces; however, it can generate low-quality meshes with overly sharp or blunt vertex angles, leading to divergence or low accuracy in numerical simulations. For city-scale applications, structured meshing should be more desirable than unstructured meshing.
More specifically, the voxel-based approach, which applies the same mesh size for the entire domain in structured meshing \citep{Fisher-Gewirtzman2013-ed, Aleksandrov2019-yv, Gorte2024-bl}, offers advantages in simplicity, as a single 3D array and a specified voxel size can represent the entire mesh.
Moreover, voxels have better compatibility with raster formats that are commonly used in geospatial data, such as land cover, tree canopy height, and terrain elevation. 

\begin{figure}
    \centering    \includegraphics[width=\linewidth]{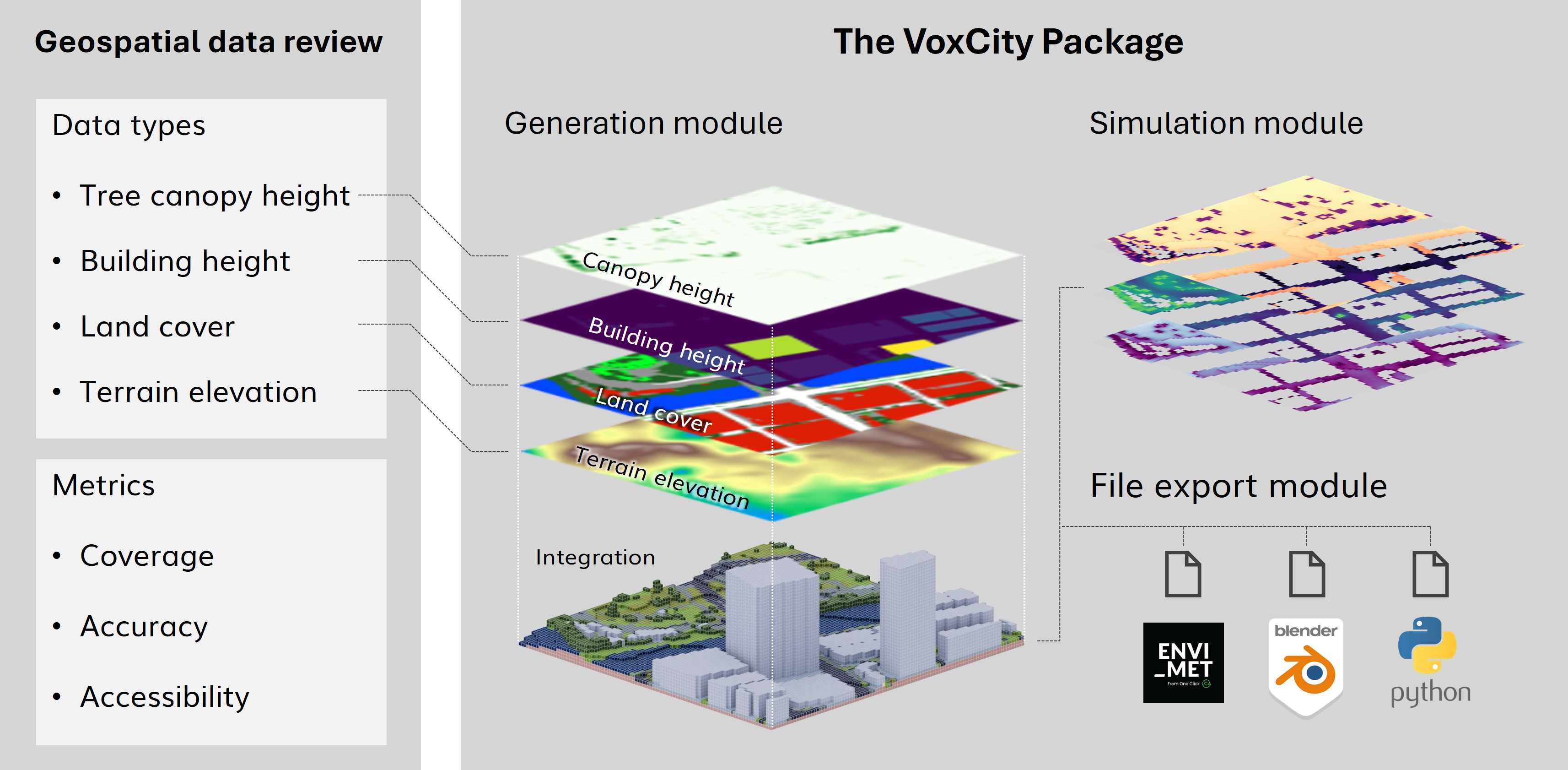}
    \caption{Framework of the development of VoxCity.}
    \label{fig:framework}
\end{figure}

Therefore, this paper introduces `VoxCity', an open-source Python package that provides a seamless solution for voxel-based geospatial data integration, 3D city model generation, and urban environment simulation. 
The framework is illustrated in Figure~\ref{fig:framework}. 
We review open geospatial data, including building height, tree canopy height, land cover, and terrain elevation data types, and select appropriate data sources for our 3D city model generator based on metrics such as coverage, accuracy, format and accessibility.
VoxCity automates the process of downloading, voxelizing, and integrating geospatial data from the selected sources to generate voxel-based 3D city models with semantic information.
The package includes a subpackage for urban environment simulation, providing users an end-to-end workflow from data acquisition and integration through model generation and simulation. 
Furthermore, VoxCity supports multiple export file formats, ensuring compatibility with various external 3D modeling and simulation software packages.

\section{Background and Related Work}
\label{sec:backgr}
\subsection{Urban Environment Simulation}
\label{sec:b_sim}

Climate change \citep{Xiao2022-fe} has intensified heat-related challenges like frequent heat waves \citep{Li2023-fu, Taleghani2019-fk} and urban heat island effects \citep{Kousis2022-dl}, spurring increased assessment of outdoor heat stress and thermal comfort through CFD \citep{Chen2021-te, Yuan2022-wf}, radiation transfer models \citep{Lindberg2008-hq, Kong2022-zw, Li2021-ye}, and heat conduction and energy balance analyses \citep{Forouzandeh2021-fm}.
The objectives of microclimate simulations have encompassed not only heat-related health issues but also other diverse aspects such as energy consumption for heating, ventilation, air conditioning (HVAC), and lighting \citep{Gros2016-xo, Bouyer2011-mt, Malys2015-pi, Sezer2023-yj}, air quality \citep{Kadaverugu2019-sj, Kwak2015-nu}, wind comfort \citep{Blocken2012-kj, Blocken2016-wk}, wind pressure on building surfaces \citep{Mou2017-iq, Stathopoulos1996-nk}, and sunlight and ultraviolet exposure focusing on human skin health \citep{Grant2002-vp}. Additionally, numerous studies have reported impacts of buildings \citep{Galal2020-uc, Zhang2022-av, Hong2015-ik, Wu2019-hp}, trees \citep{Morakinyo2016-pm, Li2023-uq, Wang2015-nb, Oshio2021-vr, Li2024-zi}, low vegetation \citep{He2022-fs, Yang2013, Wang2021-ua}, water bodies \citep{Robitu2006-lp, Du2019-ab, Imam-Syafii2017-to}, and terrain morphology \citep{Sun2012-fw} on the microclimate, indicating the importance of including these urban elements in 3D city models for simulations.

Meanwhile, many studies have conducted urban environment simulations to assess the visual comfort of streetscapes \citep{Oh1994-ot} and window views \citep{Abd-Alhamid2023-on, Yu2016-gb}. 
Specifically, sky view factor (SVF) or sky view index (SVI) \citep{Yi2017-vf} and green view index (GVI) \citep{Fujiwara2022-yb, Yu2016-gb, Labib2021-xj} have mainly been evaluated as quantitative indicators using 3D city models with buildings, trees, and low vegetation land covers. 
Numerous studies have employed street view imagery and computer vision techniques instead of simulations with 3D city models to evaluate view indices \citep{Biljecki2023-rx, Ki2021-qs, Zhang2024-fj, Liang2023-cy, Ito2025-mf}. 
This enables view analyses without preparing detailed 3D city models; however, such analyses cover only viewpoints along street networks, and furthermore, not all streets have sufficient imagery.  
These view indicators have been assessed focusing not only on visual comfort but also on walkability \citep{Lu2018-bu, Ki2021-qs} and bikeability \citep{Ito2021-mk} of streets, the well-being of city dwellers \citep{Wang2020-kx}, and the preferences of residents and real-estate prices \citep{Wu2022-ve, Yang2021-xu}. 
The visibility of water features, including the ocean, lakes, and rivers, has also been discussed in the same context \citep{Luo2022-la}.

Table~\ref{tab:urban_elements} in Appendix illustrates the relationships between simulation categories and the urban elements incorporated into 3D city models used in previous studies. Our literature review revealed that buildings, trees, land cover, water bodies, and terrain elevation have been included as essential information in 3D city models across different simulation objectives. 

\subsection{Creation of 3D city models} 
\label{sec:b_3d}

Numerous studies have prepared their 3D city models by processing point clouds acquired through LiDAR measurements \citep{Kong2022-zw, Wang2024-gg, Peters2022-qq, Ledoux2021-iz}, or by processing polygon data acquired through photogrammetry methods \citep{Rong2020-wk}. While these methods can reduce the manual effort required for 3D city model preparation, their input data---point cloud data from LiDAR or multi-view photography---is not readily available in most cases.

Meanwhile, an increasing number of studies have employed open 3D city models that have been individually released by countries or cities \citep{Matsuoka2024-ij, Zhang2025-bz, Nouvel2015-ph, Harter2023-ar, Malhotra2022-yv}. 
However, in most cases, we still need to create 3D city models ourselves because cities with such available models are limited \citep{Lei2023-cb}. 
Furthermore, the data, even when openly available, is not always directly compatible with urban environment simulations.
For instance, while 3D city models fundamentally include building data, only a portion of them have other semantic information, including trees and land cover \citep{Lei2023-dj}. 
Many software solutions for urban environment simulation require their own input file formats for 3D city models, such as the ENVI-met\footnote{\url{https://envi-met.com/}} geometry file (INX) and the ANSYS\footnote{\url{https://www.ansys.com/}} mesh file (MSH). 
Additionally, some simulations require geometrically complete models for their meshing processes. Software for CFD simulations tends to require precise watertight geometries for solid objects and does not accept polygons with gaps and intersections \citep{Paden2022-me, Paden2024-fb}.  
Therefore, even for studies targeting areas with open 3D city models, many of them need to add missing information from other data sources, convert file formats, or modify them to meet quality standards.  

Our review of the preparation of 3D city models highlights the significant lack of available 3D city models with sufficient quality for urban environment simulations, while there is a strong demand for methods to generate 3D city models from openly available data sources for many cities.

\subsection{Related work} %
\label{sec:b_rel}

There are several open-source tools to generate 3D city models. For instance, 3dfier \citep{Ledoux2021-iz} is an open-source software to generate 3D building models from point cloud data and building footprints. BlenderGIS\footnote{\url{https://blender-addons.org/blendergis-addon/}} is a plug-in tool for Blender\footnote{\url{https://www.blender.org/}}, a free and open-source 3D modeling software, which generates 3D city models by combining building data from OpenStreetMap\footnote{\url{https://www.openstreetmap.org}} and terrain elevation from Shuttle Radar Topography Mission (SRTM) \citep{Farr2000-dl}. However, we were unable to find any open-source tools that could generate 3D city models containing all essential elements and attributes, including buildings, vegetation, water bodies, and terrain.

Several previous studies suggest a potential solution to the demand for a semantic 3D city model generator. \citet{Ding2024-kj} and \citet{Li2021-ye} integrated multiple geospatial data to create 3D city models with semantic information and conducted microclimate simulations using the generated models. Specifically, \citet{Ding2024-kj} combined data on building height, land cover, and terrain elevation, while \citet{Li2021-ye} combined building height and tree canopy height. Although \citet{Ding2024-kj} and \citet{Li2021-ye} created land cover data using aerial imagery and building data using aerial LiDAR data respectively, they directly used openly available data sources for other data types.
If these methods can be extended to encompass all essential data types---building height, land cover, canopy height, and terrain elevation---and if all these data can be collected from openly and globally available data sources, such a tool would be highly beneficial for many researchers and engineers working on urban environment simulations.

Existing open datasets vary considerably in their spatial coverage, accuracy, and resolutions. Despite the importance of these factors for application in 3D city modeling and urban environment simulation, we were unable to find any studies offering a systematic comparison for such datasets. 
This gap underscores the importance of cataloging available open geospatial data, and evaluating their quality and applicability in generating semantic 3D city models. However, no studies have reviewed open geospatial data from this perspective.

Meanwhile, an increasing number of studies have proposed open-source tools as Python packages \citep{Ledoux2021-iz, Yap2023-my, Gholami2024-xs, Balakrishnan2022-pz, Boeing2017-es, Martin2023-mi}. For example, \citet{Yap2023-my} introduced the Python package `Urbanity' to automate the construction of feature-rich urban networks at any geographical scale and any location. \citet{Gholami2024-xs} developed a Python-based approach to assess microscale human thermal stress in urban environments. Python is one of the most dominant programming languages for data science, especially in machine learning. 
Many Python users openly share their scripts and packages through platforms such as GitHub\footnote{\url{https://github.com/}} and PyPI\footnote{\url{https://pypi.org/}}, facilitating easier and more scalable distribution of developed tools.

Therefore, we propose a tool that offers two key contributions: reducing the manual effort required to prepare 3D city models for urban environment simulations, and enabling broader applications of these simulations in environmental research and urban-architectural design projects. To achieve these contributions, we present three main developments:
\begin{enumerate}
    \item Review of open geospatial data. We review publicly available geospatial datasets---including building height, tree canopy height, land cover, and terrain elevation---focusing on crucial metrics for 3D city modeling, such as coverage, spatial resolutions, and accuracy. We identify suitable data sources for the 3D city model generation process and compile them into a comprehensive catalog. This catalog will assist researchers and practitioners in selecting appropriate data sources for 3D city model preparation and urban environment simulation, depending on their purposes and target areas.  
    \item Integration of open geospatial data. We propose a tool that automates the integration of publicly available geospatial data to generate semantic 3D city models. This approach enables users to prepare 3D city models and conduct urban environment simulations for cities worldwide with minimal manual effort.   
    \item Open-source Python package. We release our tool as an open-source Python package, allowing users to easily adopt, modify, and combine it with other Python libraries. Through this release, we aim to streamline 3D city model preparation and encourage broader collaboration and innovation within the community.
\end{enumerate}

\section{Review of open geospatial data}
\label{sec:review}

In this section, we review open geospatial data related to building height, tree canopy height, land cover, and terrain elevation, comparing spatial coverage, resolution, platform, and file format. 
Sections~\ref{sec:r_bh} to \ref{sec:r_te} discuss each data type in detail. 
To select data sources to review, we set two criteria: (1) global or multinational coverage and (2) horizontal resolution finer than 10 m.
We set this 10 m threshold, following previous studies that employ 10 m as the maximum mesh size for wind and microclimate simulation \citep{huttner2012further, Liu2023-vx, Maronga2020-bn}.
Additionally, our data review has not incorporated existing 3D city models such as Project PLATEAU for Japan\footnote{\url{https://www.mlit.go.jp/plateau/open-data/}} because this work specifically focuses on contributing to cities without such available models.
This limitation is discussed in Section~\ref{sec:d_3d}.
Based on this review, we present a comprehensive data catalog in Section~\ref{sec:r_ct} to guide readers in selecting suitable data sources for specific target areas and research purposes.

\subsection{Building height}
\label{sec:r_bh}

Following the aforementioned criteria for spatial coverage and resolution, we selected OpenStreetMap\footnote{\url{https://www.openstreetmap.org}} (OSM), Overture\footnote{\url{https://overturemaps.org}}, EUBUCCO v0.1\footnote{\url{https://eubucco.com/}} \citep{Milojevic-Dupont2023-bw}, UT-GLOBUS \citep{Kamath2024-at}, Open Buildings 2.5D Temporal dataset (OB2.5DT) \citep{Sirko2023-hb}, and Microsoft Building Footprints (MSBF) \citep{microsoft2024} for review. 
Table~\ref{tab:building_height} summarizes their characteristics. 
None of these sources provides complete, worldwide coverage, underscoring the importance of selecting a dataset that best suits the target city. 

\begin{table}[htbp]
\centering
\caption{Comparison of building height data sources. RMSE and MAE denote Root Mean Squared Error and Mean Absolute Error, respectively.}
\resizebox{\textwidth}{!}{%
\begin{tabular}{p{0.17\textwidth} p{0.22\textwidth} p{0.13\textwidth} p{0.34\textwidth} p{0.28\textwidth}}
\toprule[1.1pt]
\textbf{Dataset} & \textbf{Spatial Coverage} & \textbf{Resolution /Accuracy} & \textbf{Platform\newline/File format} & \textbf{Source\newline/Data Acquisition} \\
\midrule[1.1pt]
OpenStreetMap & Worldwide (24\% completeness in city centers \citep{Herfort2023-gj}) & - / Not provided & API / JSON (vector) & Volunteered / updated at irregular intervals\\ %
\midrule
Overture & Worldwide & - / Not provided & API / JSON (vector) & OpenStreetMap, Esri Community Maps Program, Google Open Buildings, etc. / updated at irregular intervals\\ %
\midrule
EUBUCCO v0.1 \citep{Milojevic-Dupont2023-bw} & 27 EU countries and Switzerland (378 regions and 40,829 cities) & - / Not provided & Files on the official website (\url{https://eubucco.com/}), Zenodo / GPKG (vector) & OpenStreetMap, government datasets / 2003-2021 (majority is after 2019) \\ %
\midrule 
UT-GLOBUS \citep{Kamath2024-at} & Worldwide (more than 1200 cities or locales) & - / 7.8 m (RMSE, height) & Files on Zenodo / GPKG (vector) & Prediction from building footprints, population, spaceborne nDSM / not provided \\ %
\midrule 
Microsoft Building Footprints \citep{microsoft2024} & North America, Europe, Australia & - / Not provided & List of download links with QuadKey / GeoJSON (vector) & Prediction from satellite or aerial imagery / 2018-2019 for the majority of the input imagery \\ %
\midrule 
Open Buildings 2.5D Temporal dataset \citep{Sirko2023-hb} & Africa, Latin America, and South and Southeast Asia & 4 m / 1.5 m (MAE, height) & Google Earth Engine, Google Cloud Storage / GeoTIFF (Raster) & Prediction from satellite imagery / 2016-2023 \\ %
\bottomrule[1.1pt] 
\end{tabular}%
}
\label{tab:building_height}
\end{table}

\begin{figure}
    \centering    \includegraphics[width=0.9\linewidth]{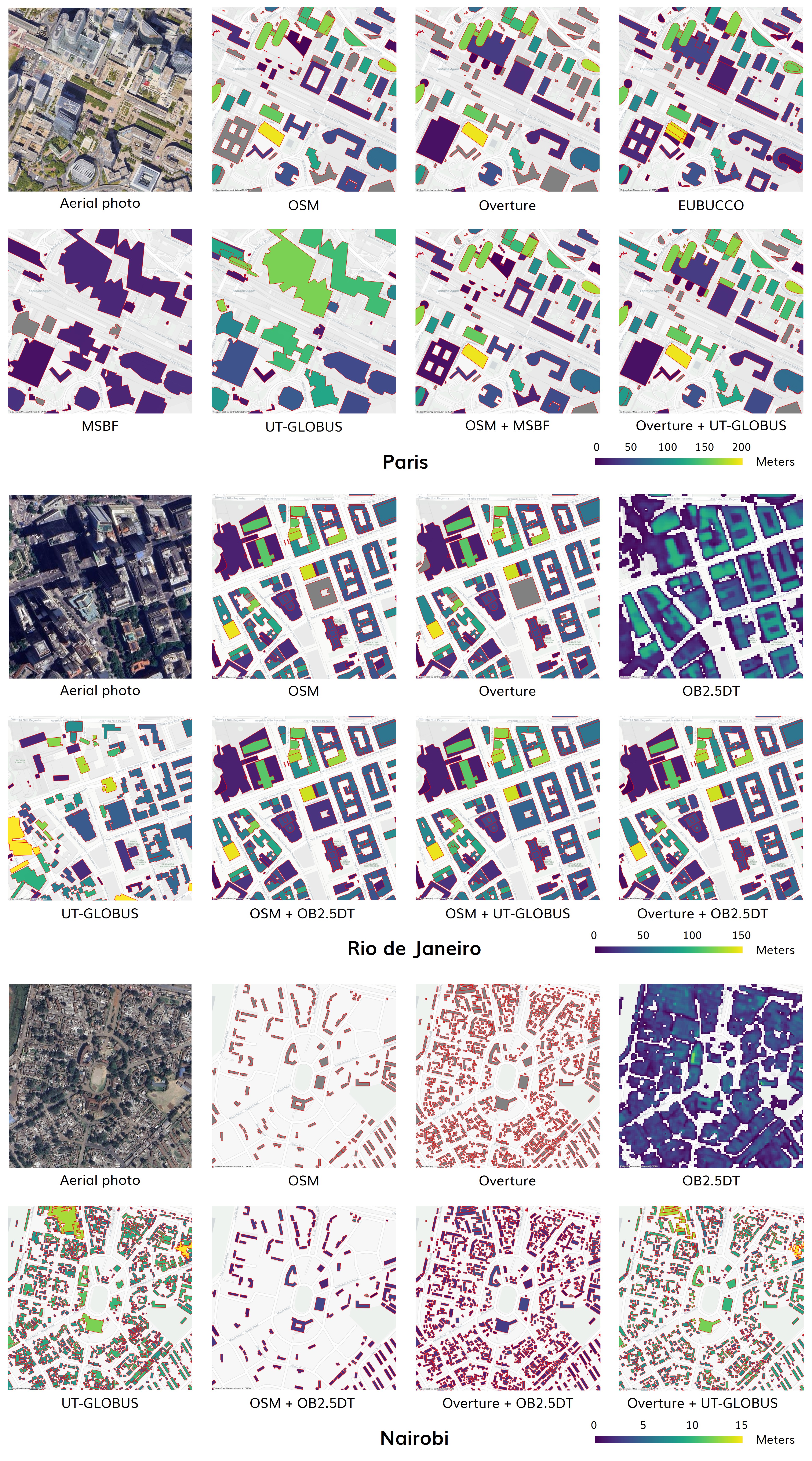}
    \caption{Examples of building height maps for Paris, Rio de Janeiro, and Nairobi. Gray indicates buildings without height data. Basemap: © OpenStreetMap contributors, © CARTO. Imagery: Google satellite tiles.}
    \label{fig:building_example}
\end{figure}

Figure~\ref{fig:building_example} illustrates building height maps from the reviewed sources for Paris, Rio de Janeiro, and Nairobi. 
In OSM, some buildings lack footprints, and some footprints lack height attributes, reflecting its nature as Volunteered Geographic Information (VGI)~\citep{2023_bae_osm_qa}. 
Moreover, while OSM has higher coverage for footprints and heights in central areas of major cities (e.g., Paris and Rio de Janeiro), coverage in rural areas (e.g., Nairobi) is often poor. 
Overture displays coverage similar to OSM in Paris and Rio de Janeiro, but differs in footprint geometries and height values. For Nairobi, it offers substantially better footprint coverage than OSM. We attribute such differences to Overture's data acquisition strategy: it uses OSM as a baseline and augments coverage with other sources, including MSBF, Google Open Buildings, and the Esri Community Maps Program (as detailed in their documentation: \url{https://docs.overturemaps.org/guides/buildings}).
EUBUCCO exhibits higher completeness than both OSM and Overture in Europe, often providing more detailed footprints. This is partly due to the inclusion of governmental datasets in addition to those from OSM.
OB2.5DT covers most buildings and suggests reasonably good height accuracy for low- and mid-rise buildings; however, it underestimates tall buildings exceeding 100 m.
In contrast, MSBF and UT-GLOBUS tend to have lower footprint accuracy than OSM, Overture, and EUBUCCO; some footprints deviate significantly from actual building outlines, and multiple buildings can appear as a single merged footprint. 

Overall, OSM, Overture, and EUBUCCO demonstrate sufficient quality for many applications, whereas MSBF, UT-GLOBUS, and OB2.5DT are less suited as sole sources for 3D city model generation. They can, however, serve as complementary data for missing building heights in EUBUCCO, OSM, and Overture. In Figure~\ref{fig:building_example}, the panels labeled ``OSM + [dataset]'' and ``Overture + [dataset]'' illustrate the integration of OSM or Overture footprints with either MSBF, UT-GLOBUS, or OB2.5DT. Where OSM or Overture footprints lack a height value, the missing attribute is retrieved from the intersecting footprints in a complementary data source. If a building intersects multiple footprints, the final height is determined by a weighted average based on intersection area. The resulting combined data exhibit more complete height coverage than OSM or Overture alone.

Based on this review, we decided to use EUBUCCO, OSM, and Overture as base building height data. OB2.5DT, UT-GLOBUS, and MSBF are used as complementary sources to fill in missing values. 

\subsection{Land cover}
\label{sec:r_lc}

To select data sources for review, we set criteria requiring land cover classes suitable for general urban environment simulations, in addition to the aforementioned criteria for spatial coverage and resolution. Therefore, we excluded datasets that are restricted to specific land cover types (e.g., ice, crop, or forest) \citep{Van-Tricht2023-jd, Shimada2014-xw, GLIMS2005, Raup2007}. 

As a result, we selected ESA World Cover 10 m 2021 V200 (ESA) \citep{Zanaga2022-eb}, Esri 10m Annual Land Cover (2017-2023) (Esri) \citep{Karra2021-tc}, Dynamic World V1 (DW) \citep{Brown2022-ua}, OpenStreetMap\footnote{\url{https://www.openstreetmap.org}} (OSM), OpenEarthMap Japan (OEMJ) \citep{Yokoya2024-me}, and UrbanWatch (UW) \citep{Zhang2022-oa} for our further review. 
Although OEMJ and UW each cover only one nation, we included them because of their exceptionally high (1 m) resolution. 
Table~\ref{tab:land-cover-datasets} compares key metrics for these sources.

\begin{table}[htbp]
\centering
\caption{Comparison of land cover data sources.}
\resizebox{\textwidth}{!}{%
\begin{tabular}{p{0.18\textwidth} p{0.33\textwidth} p{0.12\textwidth} p{0.15\textwidth} p{0.2\textwidth} p{0.23\textwidth}}
\toprule[1.2pt]
\textbf{Dataset} & \textbf{Classes} & \textbf{Spatial Coverage} & \textbf{Resolution\newline/Accuracy} & \textbf{Platform\newline/File format} & \textbf{Source\newline/Data Acquisition} \\
\midrule[1.2pt]
ESA World Cover 10 m 2021 V200 \citep{Zanaga2022-eb} & Tree cover, Shrubland, Grassland, Cropland, Built-up, Bare/Sparse vegetation, Snow and ice, Permanent water bodies, Herbaceous wetland, Mangroves, Moss and lichen & Worldwide & 10 m / 76.7\% & Google Earth Engine, Zenodo / GeoTIFF (Raster) & Prediction from satellite imagery / 2021 \\ %
\midrule
Esri 10m Annual Land Cover (2017-2023) \citep{Karra2021-tc} & Water, Trees, Flooded Vegetation, Crops, Built Area, Bare Ground, Snow/Ice, Clouds, Rangeland & Worldwide & 10 m / 85\% & Google Earth Engine / GeoTIFF (Raster) & Prediction from satellite imagery / 2017-2023 \\ %
\midrule
Dynamic World V1 \citep{Brown2022-ua} & Water, Trees, Grass, Flooded vegetation, Crops,
Shrub and scrub, Built, Bare, Snow and ice & Worldwide & 10 m / 73.8\% & Google Earth Engine, Zenodo / GeoTIFF (Raster) & Prediction from satellite imagery / updated daily \\ %
\midrule
OpenStreetMap & Bare rock, Rock, Sand, Desert, Grass, Park, Industrial, Construction, Railway, Parking, Highway, Wood, Forest, Tree, Water, Waterway, Bay, Ocean, Farmland, Building, etc. & Worldwide & - / Not provided & API / JSON (vector) & Volunteered / updated at irregular intervals\\ %
\midrule
OpenEarthMap Japan \citep{Yokoya2024-me} & Bareland, Rangeland, Developed space, Road, Tree, Water, Agriculture land, Building & Japan & $\sim$1 m / 80\% & Webmap (downloadable as tiles) / PNG (Raster) & Prediction from aerial imagery / 1974-2022 (mostly after 2018 in major cities)\\ %
\midrule
UrbanWatch \citep{Zhang2022-oa} & Building, Road, Parking Lot, Tree Canopy, Grass/Shrub, Agriculture, Water, Barren, Other & 22 major cities in the US & 1 m / 92\% & Google Earth Engine, Google Drive / GeoTIFF (Raster) & Prediction from aerial imagery / 2014–2017\\ %
\bottomrule[1.2pt]
\end{tabular}%
}
\label{tab:land-cover-datasets}
\end{table}

\begin{figure}
    \centering    \includegraphics[width=0.95\linewidth]{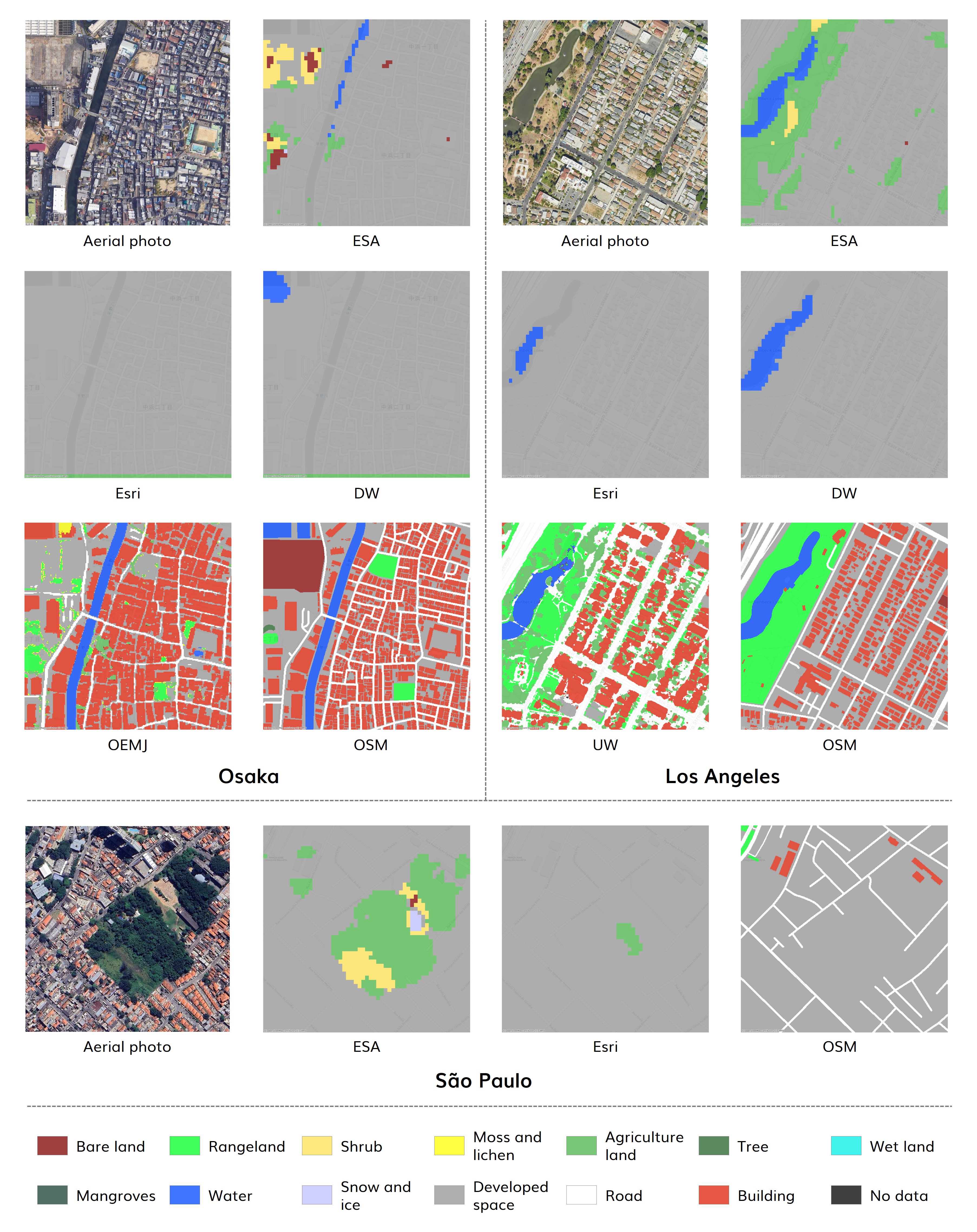}
    \caption{Examples of land cover data for Osaka and Los Angeles. Basemap: © OpenStreetMap contributors, © CARTO. Imagery: Google satellite tiles.}
    \label{fig:lc_exam}
\end{figure}

Figure~\ref{fig:lc_exam} shows example land cover maps from the reviewed datasets for Osaka, Los Angeles, and S\~{a}o Paulo. 
To allow direct comparisons, we harmonized the class definitions across all datasets using the conversion rules presented in Table~\ref{tab:lc_conv}.
Overall, OEMJ (for Osaka) and UW (for Los Angeles) most closely match the actual land cover when compared to aerial imagery. 
OSM also aligns well but tends to miss smaller patches of vegetation (particularly trees). 
ESA captures major water bodies and broad vegetation patterns but is less detailed than OEMJ, UW, or OSM. 
In OSM, some areas (e.g., S\~{a}o Paulo) lack detail in spatial variation, indicating lower completeness than ESA.
Esri and DW indicate low completeness, classifying most pixels as ``Developed space'' despite these areas actually containing diverse classes. We attribute the relatively low performance of Esri and DW to their primary focus not being on urban areas, resulting in training and validation data that lacks detailed annotations for urban regions.
It is important to note that, while this is not applicable to the examples in Figure~\ref{fig:lc_exam}, OSM sometimes misses the water class for oceans because some ocean areas in OSM lack water object patches.

\begin{table}[htbp]
\centering
\caption{Class definition harmonization across different land cover data sources.}
\resizebox{\textwidth}{!}{%
\begin{tabular}{p{0.13\textwidth} p{0.13\textwidth} p{0.13\textwidth} p{0.12\textwidth} p{0.12\textwidth} p{0.13\textwidth} p{0.49\textwidth}}
\toprule
\textbf{Class} & \multicolumn{6}{c}{\textbf{Allocated classes}} \\
\cmidrule(l){2-7}
& ESA & Esri & DW & OEMJ & UW & OSM \\
\midrule
Bareland & Barren/sparse vegetation & Bare Ground & Bare & Bareland & Barren & quarry, brownfield, bare\_rock, scree, shingle, rock, sand, desert, landfill, beach \\
\midrule
Rangeland & Grassland & Grass & Grass & Rangeland & Grass/Shrub & grass, meadow, grassland, heath, garden, park \\
\midrule
Shrub & Shrubland & Scrub/Shrub & Shrub and Scrub & Shrub & -- & scrub, shrubland, bush, thicket \\
\midrule
Agriculture land & Cropland & Crops & Crops & Agriculture & Agriculture & farmland, orchard, vineyard, plant\_nursery, greenhouse\_horticulture, flowerbed, allotments \\
\midrule
Tree & Trees & Trees & Trees & Tree & Tree Canopy & wood, forest, tree, tree\_row \\
\midrule
Moss and lichen & Moss and lichen & -- & -- & -- & -- & moss, lichen, tundra\_vegetation \\
\midrule
Wet land & Herbaceous wetland & Flooded Vegetation & Flooded Vegetation & Wetland & -- & wetland, marsh, swamp, bog, fen \\
\midrule
Mangrove & Mangroves & -- & -- & Mangrove & -- & mangrove, mangrove\_forest, mangrove\_swamp \\
\midrule
Water & Open water & Water & Water & Water & Water, Sea & water, waterway, reservoir, basin, bay, ocean, sea, river, lake \\
\midrule
Snow and ice & Snow and ice & Snow/Ice & Snow and Ice & Snow & -- & glacier, snow, ice, snowfield, ice\_shelf \\
\midrule
Developed space & Built-up & Built Area & Built & Developed & Parking Lot & industrial, retail, commercial, residential, construction, railway, parking, islet, island \\
\midrule
Road & -- & -- & -- & Road & Road & highway, road, path, track, street \\
\midrule
Building & -- & -- & -- & Building & Building & building, house, apartment, commercial\_building, industrial\_building \\
\midrule
No Data & -- & No Data, Clouds & -- & -- & Unknown & unknown, no\_data, clouds, undefined \\
\bottomrule
\end{tabular}%
}
\label{tab:lc_conv}
\end{table}

Based on this review, we excluded Esri and DW, and included the remaining datasets as data source options in our package.

\subsection{Tree canopy height}
\label{sec:r_ch}

To screen datasets, we set criteria: (1) coverage of urban areas and (2) no restriction to specific tree species, in addition to the aforementioned criteria for spatial coverage and resolution. Therefore, we excluded datasets that focus on forested regions or specific tree species \citep{Potapov2021-ff, Simard2019-kw}. 
We also considered open tree inventories, which typically include a database of individual trees (e.g., location, size, age, species) \citep{Ma2021-ky, Nielsen2014-iw, Ossola2020-vd}. 
However, each city or country tends to maintain its own tree inventory format, and coverage rarely extends beyond one locality \citep{paris2024trees, barcelona2015trees, newyork2024trees}. Hence, no standardized globally comprehensive tree inventory dataset was identified. 
Ultimately, we selected the High Resolution 1 m Global Canopy Height Maps (META) \citep{Tolan2024-an} and ETH Global Sentinel-2 10 m Canopy Height (2020) (ETH) \citep{Lang2023-cy} for further review. Table~\ref{tab:ch} provides details on these data sources. 

\begin{table}[htbp]
\centering
\caption{Comparison of tree canopy height data sources}
\resizebox{\textwidth}{!}{%
\begin{tabular}{p{0.3\textwidth} p{0.15\textwidth} p{0.15\textwidth} p{0.25\textwidth} p{0.3\textwidth}}
\toprule
\textbf{Dataset} & \textbf{Coverage} & \textbf{Resolution\newline/Accuracy} & \textbf{Platform\newline/File format} & \textbf{Source\newline/Data Acquisition} \\
\midrule
High Resolution 1 m Global Canopy Height Maps \citep{Tolan2024-an} & Worldwide & 1 m / 2.8 m (MAE) & Google Earth Engine / GeoTIFF (Raster) & Prediction from satellite imagery / 2009 and 2020 (80\% are 2018-2020) \\
\midrule
ETH Global Sentinel-2 10 m Canopy Height (2020) \citep{Lang2023-cy} & Worldwide & 10 m / 6.0 m (RMSE) & Google Earth Engine / GeoTIFF (Raster) & Prediction from satellite imagery / 2020 \\
\bottomrule
\end{tabular}%
}
\label{tab:ch}
\end{table}

\begin{figure}
    \centering    \includegraphics[width=0.95\linewidth]{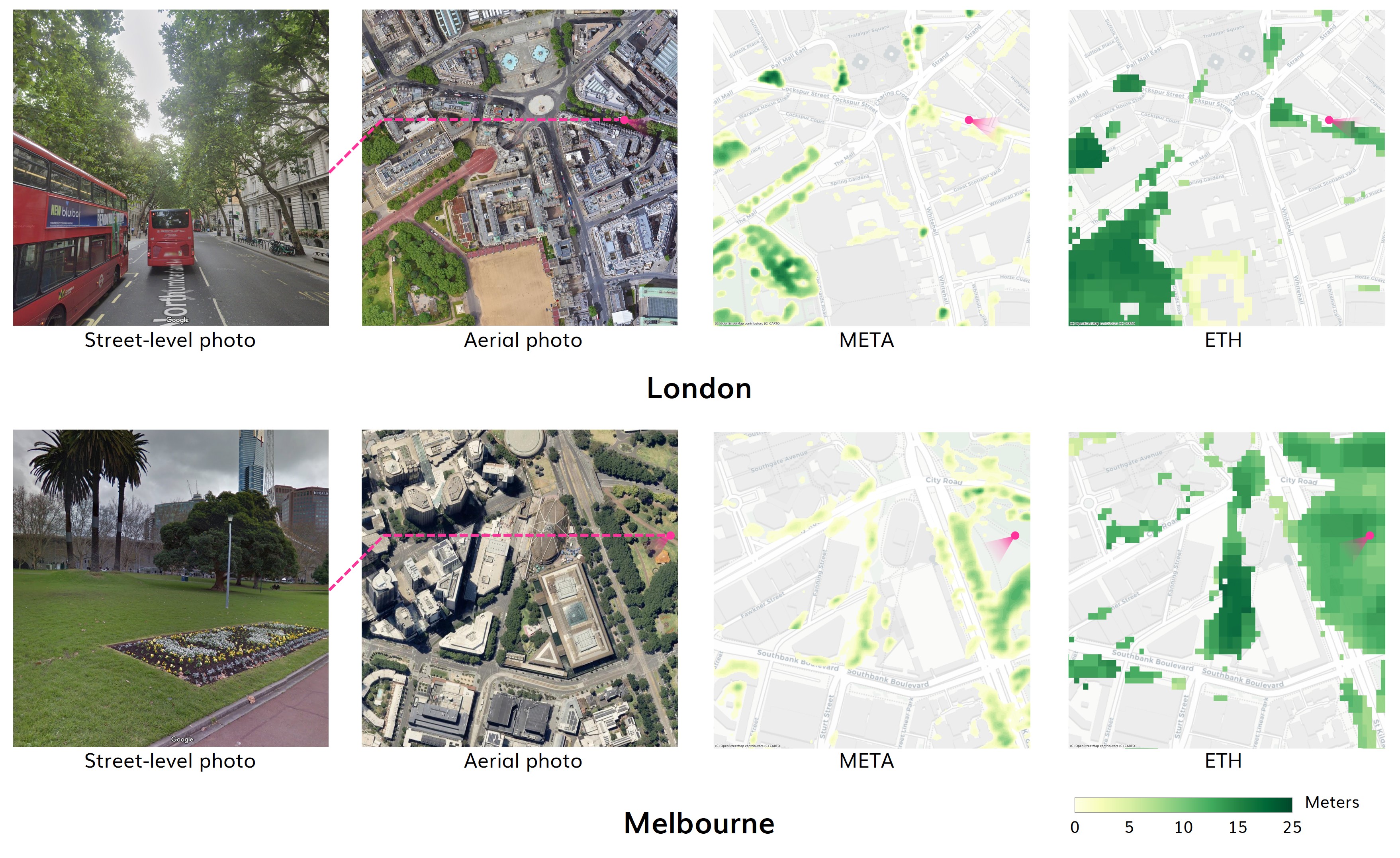}
    \caption{Examples of canopy height data for London and Melbourne. Basemap: © OpenStreetMap contributors, © CARTO. Imagery: Google street view, Google satellite tiles.}
    \label{fig:exa_ch}
\end{figure}

Figure~\ref{fig:exa_ch} compares canopy height maps from META and ETH for areas in London and Melbourne, alongside corresponding street-level and satellite images.
In London, the street-level photo indicates trees over 10 m tall (judging from the height of double-decker buses), while META reports around 3 m and ETH 12 m for the same location, suggesting underestimation by META and more accurate performance by ETH.
Meanwhile in Melbourne, ETH detects a canopy height of around 10 m at a location that is primarily grass in street-level and aerial imagery, whereas META shows almost no canopy. In this case, META appears more accurate, and ETH overestimates.
In short, META and ETH underestimate and overestimate canopy height, respectively, in certain situations.
Neither dataset consistently outperforms the other, so the choice depends on regional characteristics and research requirements.
We therefore include both options in our package, allowing users to select according to their needs.

\subsection{Terrain elevation}
\label{sec:r_te}

To select terrain elevation datasets, we set a resolution threshold of 1 Arc second (30.9 m at the equator) instead of 10 m for other data types. 
This exception was made because, for terrain elevation data with global coverage, only datasets with resolutions coarser than 1 Arc second are available. We allowed this coarser resolution because terrain elevation in urban areas rarely changes drastically at spatial scales less than 30 m. 
Additionally, we established a criterion requiring a bare-earth model to exclude DEMs or DSMs that retain building and vegetation heights \citep{nasadem2020, copernicus2024dem, Tadono2014-xo, O-Loughlin2016-ju}. 
Consequently, we selected Forest And Buildings removed Copernicus 30m DEM (FABDEM)~\citep{Hawker2022-by}, DeltaDTM~\citep{Pronk2024-sp}, USGS 3DEP 1m DEM (USDEM)~\citep{usgs2023}, England 1m Composite DTM (ENGDTM)~\citep{envagency2024}, Australian 5M DEM (AUSDEM)~\citep{geoscienceaus2015dem}, and RGE Alti (FRADEM)~\citep{rgealti2024} for further review. 
While USDEM, ENGDTM, AUSDEM, and FRADEM have only national coverage, they were incorporated because of their exceptionally fine spatial resolution, ranging from 1 to 5 m.

Table~\ref{tab:dem} summarizes the characteristics of the selected datasets. 
FABDEM provides the most extensive worldwide coverage, followed by DeltaDTM, which covers global coastal areas. 
While the other datasets are limited to country-scale coverage, they offer superior horizontal resolutions ranging from 1 to 5 m---significantly finer than those of FABDEM and DeltaDTM.

\begin{table}[htbp]
\centering
\caption{Comparison of terrain elevation data sources}
\resizebox{\textwidth}{!}{%
\begin{tabular}{p{0.18\textwidth} p{0.17\textwidth} p{0.18\textwidth} p{0.23\textwidth} p{0.3\textwidth}}
\toprule
\textbf{Dataset} & \textbf{Coverage} & \textbf{Resolution\newline/Accuracy} & \textbf{Platform\newline/File format} & \textbf{Source\newline/Data Acquisition} \\
\midrule
FABDEM \citep{Hawker2022-by} & Worldwide & 30 m / Built-up areas: 1.12 m, forests: 2.88 m (MAE) & Google Earth Engine / GeoTIFF (Raster) & Correction of Copernicus DEM using canopy height and building footprints data / 2011-2015 (Copernicus DEM) \\
\midrule
DeltaDTM \citep{Pronk2024-sp} & Worldwide (Only for coastal areas below 10m + mean sea level) & 30 m / 0.45 m (MAE) & Google Earth Engine / GeoTIFF (Raster) & Copernicus DEM, spaceborne LiDAR / 2011-2015 (Copernicus DEM) \\
\midrule
USGS 3DEP 1m DEM \citep{usgs2023} & United States & 1 m / Not provided & Google Earth Engine, government website / GeoTIFF (Raster) & Aerial LiDAR / 2004-2024 (mostly after 2015) \\ %
\midrule
England 1m Composite DTM \citep{envagency2024} & England & 1 m / 0.15 m (RMSE, a vertical accuracy of LiDAR used) & Google Earth Engine, government website / GeoTIFF (Raster) & Aerial LiDAR / 2000-2022 \\
\midrule %
Australian 5M DEM \citep{geoscienceaus2015dem} & Australia & 5 m / 0.30 m (error metric not specified) & Google Earth Engine, government website / GeoTIFF (Raster) & Aerial LiDAR / 2001-2015 \\ %
\midrule
RGE Alti \citep{rgealti2024} & France & 1 m / 0.2 m (RMSE, coastal areas), 0.5 m (RMSE, large forest areas) & Google Earth Engine / GeoTIFF (Raster) & Aerial LiDAR / Not provided \\ %
\bottomrule
\end{tabular}%
}
\label{tab:dem}
\end{table}

\begin{figure}
    \centering    \includegraphics[width=0.95\linewidth]{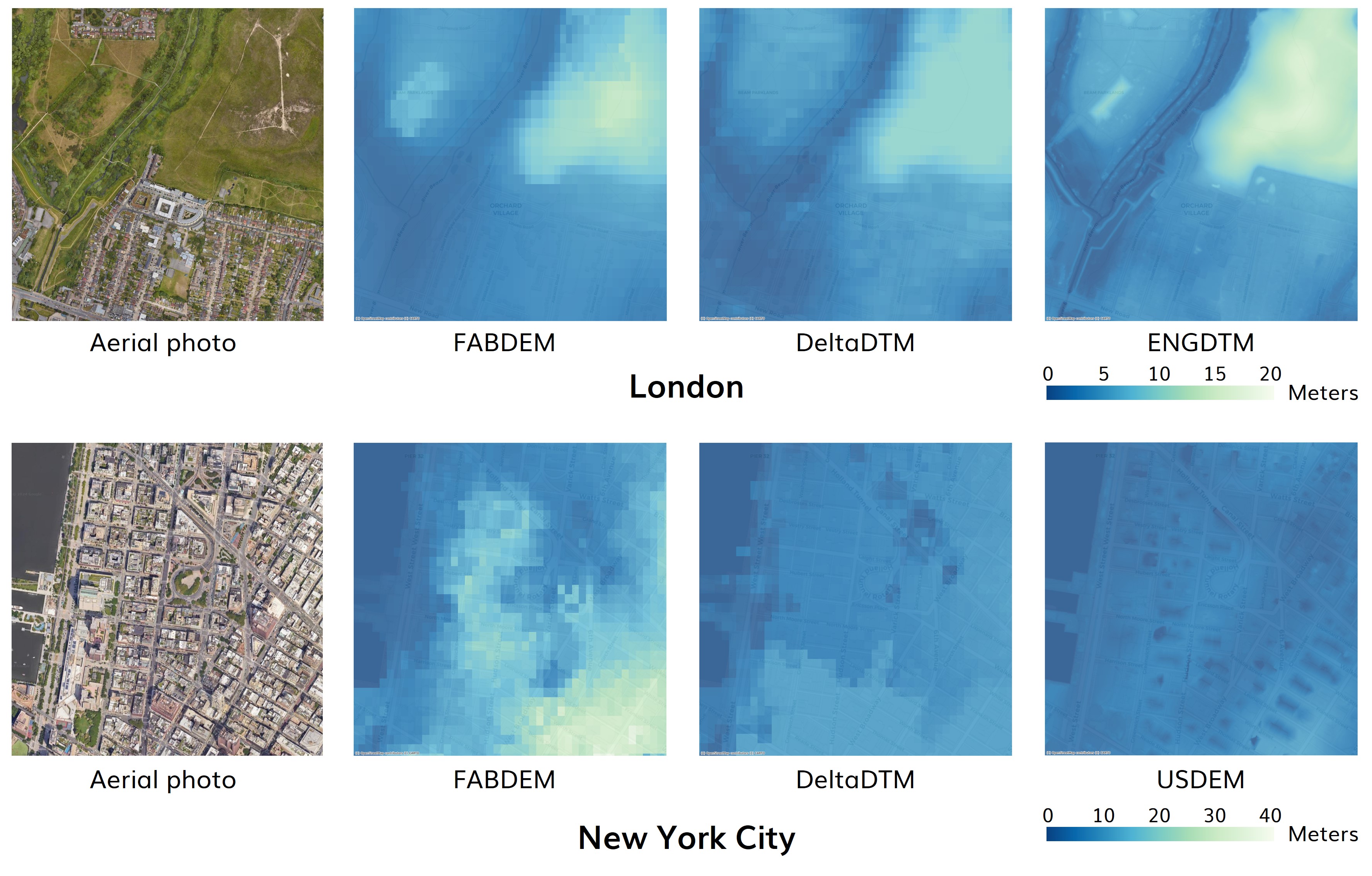}
    \caption{Examples of terrain elevation data for London and New York City. Basemap: © OpenStreetMap contributors, © CARTO. Imagery: Google satellite tiles.}
    \label{fig:exa_dem}
\end{figure}

Figure~\ref{fig:exa_dem} illustrates example terrain elevation maps for London and New York City derived from the reviewed sources.
ENGDTM and USDEM show more detailed spatial variation than the other datasets, consistent with their finer resolutions. 
The London maps display similar elevation values across all sources, whereas the New York City maps show significant discrepancies, particularly between FABDEM and USDEM.
We attribute these differences to the varying methods and resulting accuracy in removing building heights in areas with dense high-rise constructions. 

Based on the observed advantages and disadvantages of each dataset in terms of coverage and resolution, we decided to include all reviewed sources as options in our package. 

\subsection{Data catalog with guidelines for data selection}
\label{sec:r_ct}

After reviewing data sources for building height, land cover, tree canopy height, and terrain elevation, we compile a comprehensive data catalog for our package (Table~\ref{tab:catalog}). This catalog includes guidelines to select appropriate data sources for each data type. Users can refer to this catalog to identify the most suitable data sources based on their specific target areas and research objectives.  

\begin{table}[htbp]
\caption{Catalog of data types for 3D city model generation.}
\resizebox{\textwidth}{!}{%
\begin{tabular}
{p{0.2\textwidth}p{0.29\textwidth}p{0.81\textwidth}}
\toprule
Type & Data sources & Guidelines for selection \\
\midrule
Building height & Base: EUBUCCO \citep{Milojevic-Dupont2023-bw}, OSM, Overture\newline Complementary: OB2.5DT \citep{Sirko2023-hb}, UT-GLOBUS \citep{Kamath2024-at}, MSBF \citep{microsoft2024} & 
(1) Use EUBUCCO for EU countries; for other regions, employ OSM or Overture as the base source. \newline(2) If the target area is covered by any complementary dataset (MSBF for the USA, Europe, and Australia; OB2.5DT for Africa, South Asia, South-East Asia, Latin America and the Caribbean; UT-GLOBUS for 1,200 global cities), use it as the complementary source. \\
\midrule
Land cover & UW \citep{Zhang2022-oa}, OEMJ \citep{Yokoya2024-me}, OSM, ESA \citep{Zanaga2022-eb} & 
(1) Use UW for U.S.\ cities and OEMJ for Japanese cities, where available. \newline(2) For all other regions, rely on OSM by default. \newline(3) If OSM coverage is insufficient, switch to ESA. \\
\midrule
Canopy height & META \citep{Tolan2024-an}, ETH \citep{Lang2023-cy} & 
Select either META or ETH according to the target areas and research objectives, keeping in mind that META typically underestimates canopy height while ETH tends to overestimate. \\
\midrule
Terrain elevation & USDEM \citep{usgs2023}, ENGDTM \citep{envagency2024}, AUSDEM \citep{geoscienceaus2015dem}, FRADEM \citep{rgealti2024}, DeltaDTM \citep{Pronk2024-sp}, FABDEM \citep{Hawker2022-by} & 
(1) For cities covered by high-resolution datasets (ENGDTM, USDEM, AUSDEM, and FRADEM), use those data sources. \newline(2) For other regions, use FABDEM or DeltaDTM. Note that, in areas with dense high-rise buildings, these data may include significant errors. \\
\bottomrule
\end{tabular}
}
\label{tab:catalog}
\end{table}

\section{The VoxCity package}
\label{sec:voxcity}

\begin{figure}[tpb]
    \centering    \includegraphics[width=0.85\linewidth]{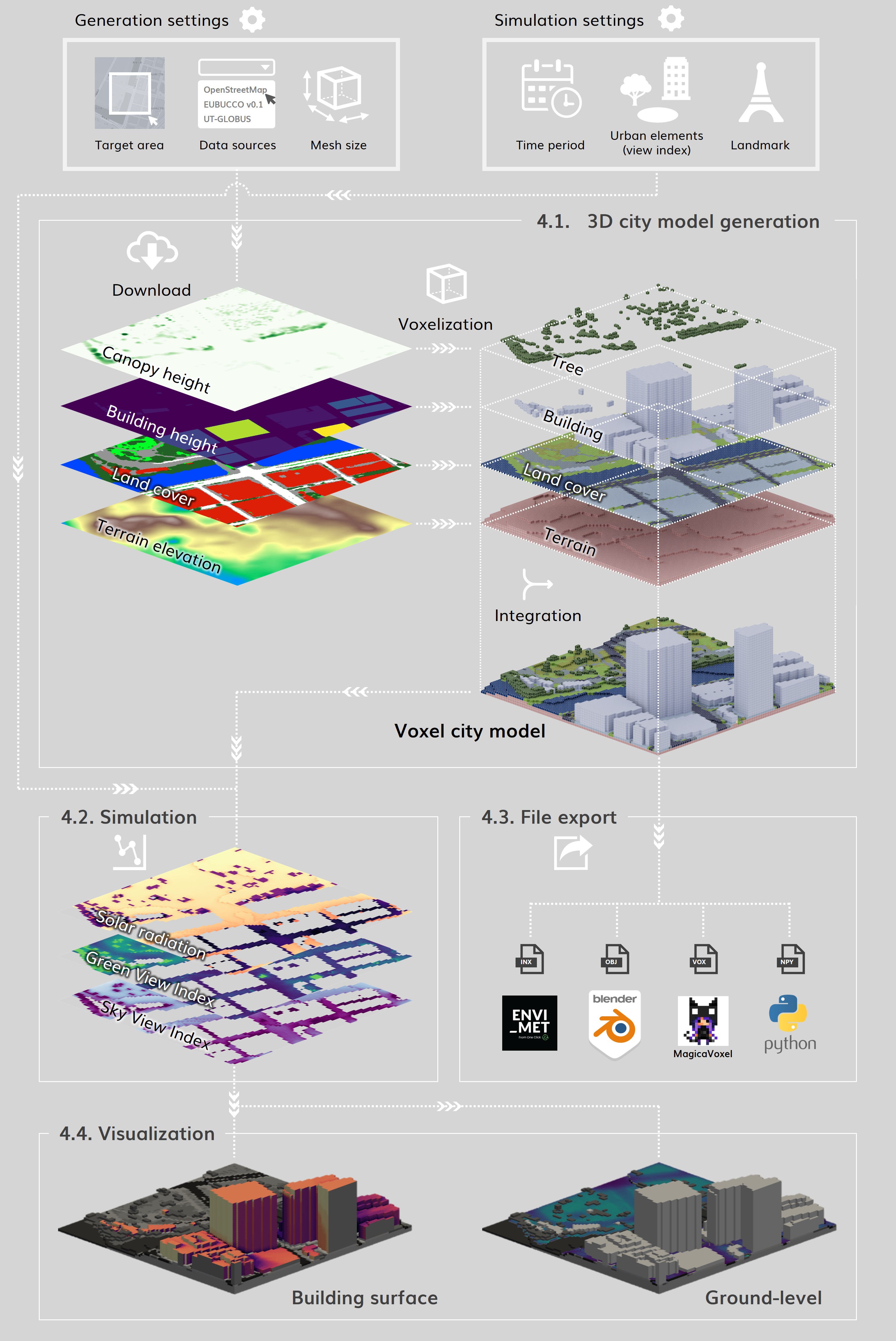}
    \caption{Workflow of data processing in `VoxCity' -- from bringing together disparate urban data sources to conducting complex multi-modal simulations -- all in a single software package.}
    \label{fig:voxcity}
\end{figure}

In this section, we introduce `VoxCity', an open-source Python package for open geospatial data integration, grid-based 3D city model generation, and urban environment simulation. 
Figure~\ref{fig:voxcity} illustrates the framework of VoxCity. Users start the process by specifying the target area and voxel size and selecting data sources from our catalog. 
The workflow consists of four main sub-processes: (1) 3D city model generation, (2) simulation, (3) file export, and (4) visualization.
In (1) 3D city model generation, VoxCity downloads building height, land cover, tree canopy height, and terrain elevation data from the selected sources within the specified target area. It then voxelizes all the downloaded data and integrates them into a semantic 3D city model. 
In (2) simulation, the output models can be used directly to conduct urban environment simulations through VoxCity's built-in simulation functions.
In (3) file export, VoxCity can export the output model in multiple file formats that are compatible with external software. 
In (4) visualization, VoxCity offers 3D visualization functionality for not only generated city models but also simulation results.

\subsection{3D city model generation}
\subsubsection{Download}
\label{sec:voxcity_dl}

VoxCity automatically downloads the required data within a target area through either file links or APIs depending on selected data sources. 
Users can define the target area as a rectangular region.
Downloaded data are saved in users' environments and utilized in subsequent processes. 
Vector and raster data are saved as GeoJSON and GeoTIFF files, respectively.
Users who need these intermediate files can also use VoxCity solely as a downloader for various geospatial datasets, similar to OSMnx \citep{Boeing2017-es}, which often serves as a downloader for road networks from OSM.

\subsubsection{Voxelization and integration}
\label{sec:voxcity_vi}

Downloaded data are then voxelized using voxel units through the following processes.

VoxCity first aggregates values using a two-dimensional horizontal grid defined by the voxel size (in meters).
For building height, tree canopy height, and land cover, each cell's value is determined by the dominant value within that cell. For instance, focusing on land cover, when a cell includes multiple classes, the class that covers the largest area is assigned as the representative value of the cell. 
In this process, the land cover class harmonization shown in Table~\ref{tab:lc_conv} is applied, enabling users to easily compare the generated 3D models across different data sources and select the desirable sources.
For terrain elevation, the representative value of a cell is calculated as the average within the cell.

A voxel model is then generated for each data type by extruding the aggregated cell values. 
For building height, tree canopy height, and terrain elevation, two-dimensional grids are extruded corresponding to the value of each cell and its voxel size. 
For example, if a cell representing building height has a value of 50 m and the voxel size is 5 m, it is extruded into 10 (=50/5) voxels vertically. Tree and terrain voxels are generated using the same method. For trees, the canopy height value includes a gap between the terrain surface and the bottom of the canopy. Tree voxels occupy only the space between the top and bottom of the canopy, while void voxels fill the space below the canopy. The height of this gap is determined by multiplying the canopy height by a trunk height ratio, which can be adjusted to reflect different regions or tree species. For land cover, the topmost (surface) terrain voxel in each cell of the 2D grid is replaced with a land cover voxel. 

Finally, voxel models of buildings, trees, and terrain with land cover are integrated into a single 3D city model. In this integration process, building and tree models are placed on top of the land cover voxels (the surface voxels of the terrain model). 

In OSM, some buildings consist of multiple footprint polygons with detailed height information for both the top and the bottom. Such footprints represent more complex building shapes than simple extrusions based on terrain surfaces.
For these buildings, our method uses both top and bottom height values to fill building voxels only between the two heights, leaving voxels between the terrain surface and the building's bottom as void (see Figure~\ref{fig:vox_eo}-Singapore as an example).
Additionally, VoxCity generally does not support civil engineering structures such as bridges and elevated highways, except for some rare cases where such objects are registered with height values in OSM.   

VoxCity's 3D city models are structured as three-dimensional arrays, whose dimensions correspond to the geospatial x, y, and z axes as shown in Figure~\ref{fig:dstructure} in Appendix. Each cell corresponds to a voxel, and its value represents an element type such as a building, tree, or water.
While some previous studies implemented octrees for voxel model structures to enhance efficiency in data processing \citep{Gorte2024-bl, Liang2017-yu}, VoxCity does not currently use this approach.
Incorporating octrees could reduce memory usage and improve ray-tracing-based simulation performance; therefore, we plan to implement this structure in future work.

\subsubsection{Example outputs}
\label{sec:voxcity_eo}

\begin{figure}
    \centering    \includegraphics[width=0.95\linewidth]{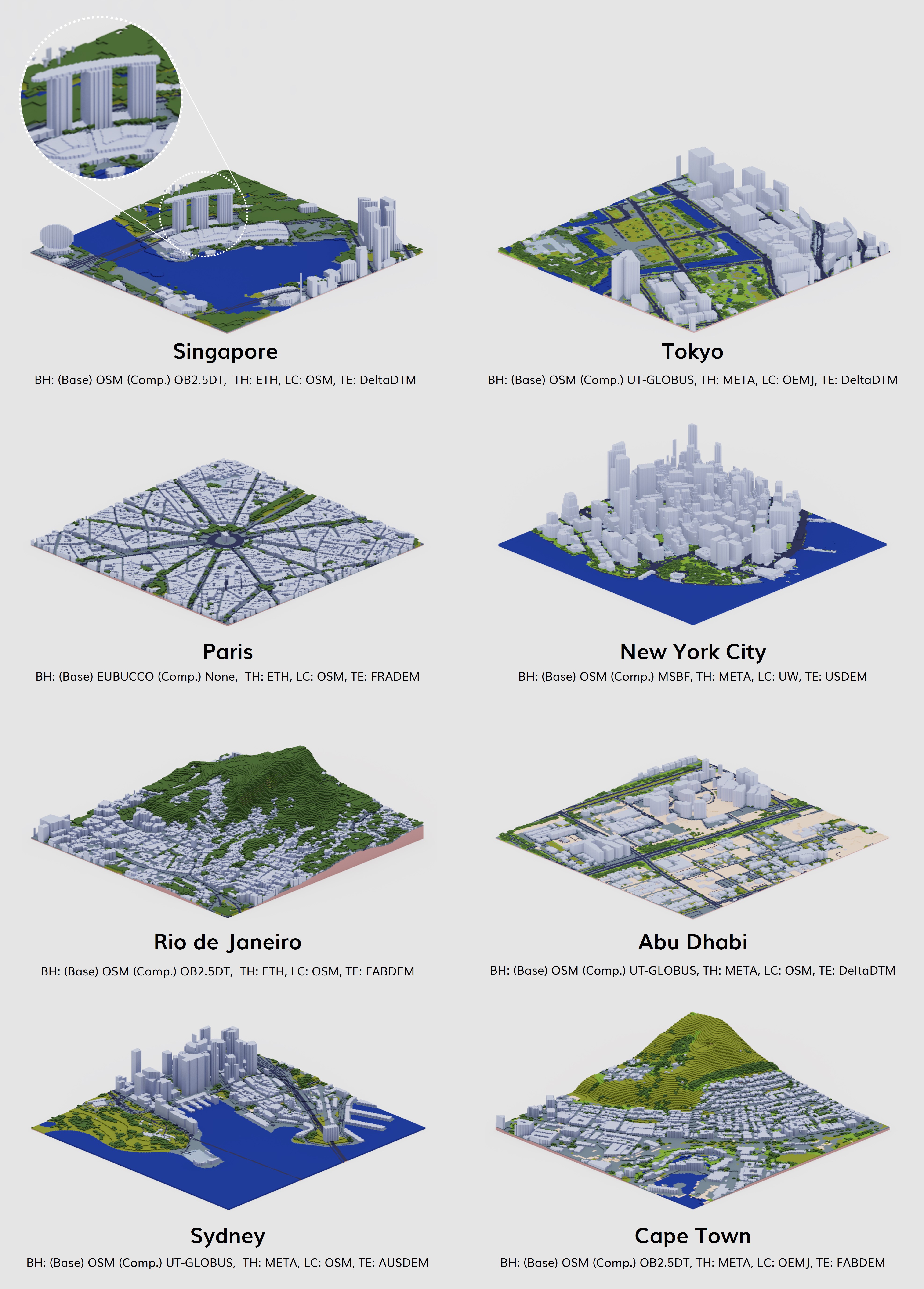}
    \caption{Examples of output 3D city models from VoxCity. 3D rendering performed in MagicaVoxel. BH, TH, LC, and TE represent building height, tree canopy height, land cover, and terrain elevation, respectively.}
    \label{fig:vox_eo}
\end{figure}

Figure~\ref{fig:vox_eo} shows example 3D city models generated by VoxCity. These include models of Singapore, Tokyo, Paris, New York City, Rio de Janeiro, Abu Dhabi, Sydney, and Cape Town, demonstrating VoxCity's capability to generate 3D city models worldwide and in varied urban morphologies.
These outputs also illustrate VoxCity's ability to represent building configurations more complex than simple extrusions of building outlines. This is particularly evident in the Singapore model, which features a rooftop park supported by three towers.

VoxCity's outputs capture the morphological characteristics of each city by combining building shapes and arrangements, land cover, tree canopies, and terrain elevations. Notable examples include the radiating block arrangements and street networks around Arc de Triomphe in Paris; dense skyscrapers surrounded by water in Lower Manhattan, New York City; tightly packed houses on a mountainside in Rio de Janeiro; and sparse high-rise buildings set amid sandy terrain in Abu Dhabi. Additionally, to further demonstrate VoxCity's capability for diverse geographies worldwide, we generated models for eight cities: Vancouver, San Francisco, Medellín, La Paz, Barcelona, Nairobi, Mumbai, and Bangkok (Figure~\ref{fig:gsouth} in Appendix).

To demonstrate the efficiency of our approach, we evaluated the time requirements for manual and computational processing to generate 3D city models, as shown in Table~\ref{tab:process_time} in Appendix. 
Manual processing includes specifying the target area and setting data sources, while computational processing includes downloading the required data and generating the voxel city models.
In our test environment---equipped with an Intel Core i9-13900 processor and an internet connection with a measured download speed of 180 Mbps---VoxCity required 90 seconds for manual and 40 seconds for computational processing for the New York City model shown in Figure~\ref{fig:vox_eo}. 
Computational processing times can vary substantially depending on the urban morphology and selected data sources.
For instance, generation of the Singapore model shown in Figure~\ref{fig:vox_eo}, which exhibits lower building density than that of New York City, required a computational processing time of 25 seconds. 

\subsection{Simulation}
\label{sec:voxcity_sim}

VoxCity includes a built-in subpackage `simulator' that calculates solar irradiance, view indices, and landmark visibility in 3D city models. 
Combined with the `generator' subpackage, VoxCity provides a one-stop solution for everything from 3D city model preparations to urban environment simulations, reducing the manual effort typically required to transfer data between 3D modeling software and specialized simulation tools.
The following subsections detail methodologies used for each simulation.

\subsubsection{Solar radiation}
\label{sec:voxcity_sim_solar}
A module, `solar', provides functionality for calculating global irradiance at ground-level and on building surfaces.
Solar radiation is a critical factor influencing outdoor heat stress~\citep{Park2021-dw}, thermal comfort~\citep{Huang2018-ak, Coccolo2016-jd}, urban farming~\citep{Palliwal2021-ft}, photovoltaic power generation~\citep{Hofierka2009-eb, Mondol2008-hh}, and building energy consumption by HVAC and lighting~\citep{Lam1999-im, Causone2010-ug}. Consequently, it has profound impacts on urban sustainability and the well-being of city dwellers---underscoring its importance in urban and architectural planning.
We employ a ray-tracing based calculation method for solar irradiance introduced by \citet{Pruzinec2022-uo, Monsi2004-eb}.
The method is summarized in \ref{sec:appendix_solar}.
The `solar' module offers two calculation options: (1) instantaneous solar irradiance (Wm$^{-2}$) at a specific timestamp and (2) cumulative solar irradiance (Whm$^{-2}$) over a specific period, such as a day, a month, or a year.

\begin{figure}
    \centering    \includegraphics[width=0.95\linewidth]{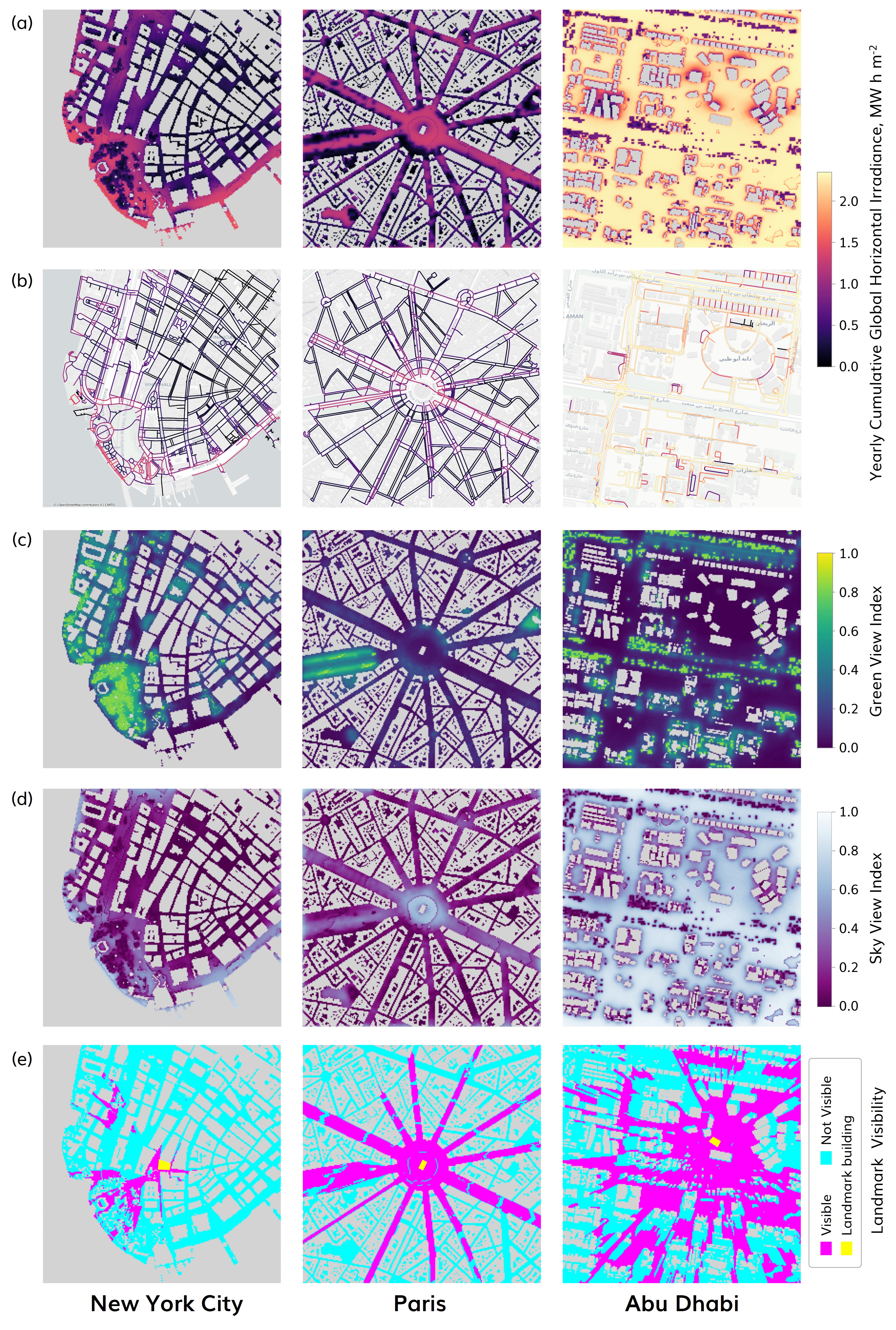}
    \caption{Example urban environment simulations conducted using 3D city models from VoxCity. (a, b) Solar irradiance. (c) Green View Index, GVI. (d) Sky View Index, SVI. (e) Landmark Visibility. Panel (b) employed walking path networks downloaded from OpenStreetMap. Basemap: © OpenStreetMap contributors, © CARTO.}
    \label{fig:simulation}
\end{figure}

Figure~\ref{fig:simulation}a illustrates example results of ground-level cumulative solar irradiance for three target areas: New York City, Paris, and Abu Dhabi. These simulations used the same 3D city models shown in Figure~\ref{fig:vox_eo} and employed the nearest EPW files downloaded from Climate.OneBuilding.Org. Each target area exhibits distinct spatial variations in solar irradiance, reflecting differences in urban morphology, climate, and weather conditions. These results demonstrate the potential of VoxCity's simulation subpackage for evidence-based urban and architectural design that accounts for local conditions.  

Additionally, VoxCity provides a feature to aggregate grid-based simulation results along the edges of road networks, as shown in Figure~\ref{fig:simulation}b. The functionality uses OSMnx to download road networks from OSM, enabling various network analyses within simulated urban environments---including not only solar irradiance but also view indices and landmark visibility, which are detailed in Section~\ref{sec:voxcity_sim_view}.  

\subsubsection{View index}
\label{sec:voxcity_sim_view}
A module, `view', provides functionality for conducting view index analyses, where the ratio of specific object types---such as vegetation, sky, and buildings---visible from a given location is quantitatively evaluated. 
The module calculates the view index by applying a ray-tracing technique to 3D city models.
Specifically, we implemented a line of sight evaluation method introduced by \citet{Mor2021-di} for voxel city models. 
Users can specify the target object type, the vertical angle of view (i.e., the ranges in which rays are cast), and the total number of rays. 
The horizontal angle of view is set to 360 degrees, assuming uniform visibility in all directions.

We demonstrated the calculations of the Green View Index (GVI) and Sky View Index (SVI) using the module on 3D city models for Paris, New York City, and Abu Dhabi, as shown in Figure~\ref{fig:simulation}c and d.
For GVI, our method casts 600 rays per location, covering 60 degrees vertically. Specifically, the vertical angles range from -30 to +30 degrees, with the horizontal direction defined as 0 degrees. Rays are cast at six-degree intervals both horizontally and vertically, resulting in 60 × 10 = 600 rays per location.
In contrast, for SVI, areas below the horizon do not influence the results; therefore, the vertical angle of view is set from 0 to +30 degrees. This configuration produces 300 rays per location.
We employed the same 3D city models shown in Figure~\ref{fig:vox_eo}.
The results reflect the distinct urban morphological features of the target areas: New York City exhibits a notably low SVI due to its dense high-rise buildings, whereas Abu Dhabi shows a relatively low GVI and high SVI, aligning with its limited greenery and more sparse buildings.

The view module also provides functionality for evaluating the visibility of specified landmark buildings from ground-level locations using a ray-tracing technique.
Specifically, we implemented the ground-level viewshed calculation introduced by \citet{Wrozynski2024-cd} for voxel city models.
This enables users to understand from which locations within a target area the landmark buildings can be seen.
Figure~\ref{fig:simulation}e shows example visibility maps.
Landmarks play important roles in pedestrians' perception and wayfinding \citep{filomena_modelling_2021,yesiltepe_landmarks_2021}, influencing visual comfort and walkability \citep{Sabesan2024-vy,Yuan2024-ls}. The access to the view of landmarks has been proved to add property value \citep{turan_development_2021}. The visibility simulations could support informed urban planning considering such effects of landmarks.

\subsection{File export}
\label{sec:voxcity_fe}

VoxCity's `exporter' subpackage can save the generated 3D city model data in several file formats to support downstream applications and data exchange. The available file formats include INX for ENVI-met \citep{Liu2021, Ouyang2022-tv}, the Wavefront OBJ for 3D modeling software, the MagicaVoxel\footnote{\url{https://ephtracy.github.io/index.html}} voxel model format (VOX), and the Pickle (PKL) for Python.
ENVI-met is a widely used microclimate simulation software.
MagicaVoxel offers GPU-accelerated 3D rendering for voxel models.
PKL files enable model data transfer between different Python environments without any format conversion.
When using the PKL format, the data is directly saved as a three-dimensional NumPy \citep{harris2020array} array. For other formats, the data is first converted to each format's specific data structure.
In the case of VOX, the 3D array from VoxCity is reformatted into the VOX structure, including color palette information corresponding to voxel classes.
The subsequent subsections describe the export functionality for the INX and OBJ formats in more detail.

\subsubsection{INX for ENVI-met}
\label{sec:uc_mc}

Building height and terrain elevation from VoxCity are directly converted to the INX format, while canopy height and land cover data are translated into corresponding vegetation ID and material ID in ENVI-met before being integrated into the INX format. Table~\ref{tab:envimet_mapping} in Appendix summarizes mapping of voxel classes in VoxCity to ENVI-met material and vegetation IDs in this study. 
Additionally, users can specify the trunk height ratio relative to total canopy height and the leaf area density (LAD) of trees. These tree-specific settings are then exported separately as a project database (EDB) file.    

\begin{figure}
    \centering    \includegraphics[width=0.95\linewidth]{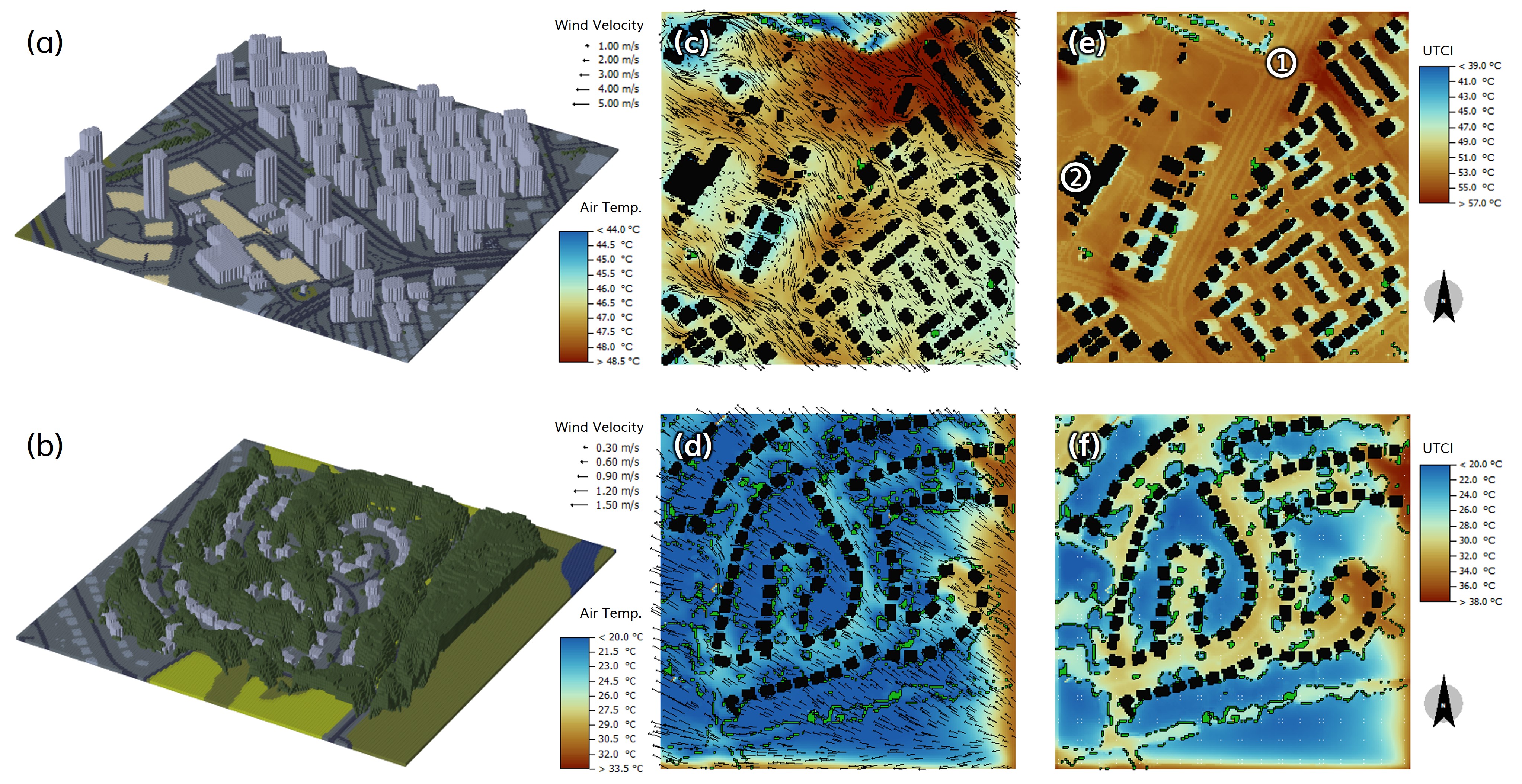}
    \caption{Examples of ENVI-met microclimate simulation in Abu Dhabi and Greenbelt. (a, b) 3D models and land-use land-cover maps obtained from VoxCity. (c, d) Horizontal wind vectors and air temperature, and (e, f) universal thermal climate index (UTCI) distributions at the 1.5-m height on typical summer afternoons.}
    \label{fig:envimet}
\end{figure}

Figure~\ref{fig:envimet} shows microclimate simulation examples using the widely-used ENVI-met code (V5.7.1, ENVI-met Gmbh, Essen) in two characteristic sites, high-rise Al Zahiya district in Abu Dhabi and low-rise residential in Greenbelt, MD (USA). Since ENVI-met is based on the structured mesh (i.e., Cartesian coordinates), VoxCity can provide basic 3D geometries (Figures \ref{fig:envimet}a and \ref{fig:envimet}b) for the simulation. The calculation domains were 800 m × 800 m × 310 m for Abu Dhabi, and 500 m × 500 m × 50 m for Greenbelt. They comprised inner areas with buildings and trees with horizontally homogeneous grid spacing (4 m for Abu Dhabi and 2 m for Greenbelt here), which is off-the-shelf from VoxCity. Because perimeter margins are necessary for ENVI-met simulation to avoid unphysical results (as detailed in their documentation: \url{https://envi-met.com/tutorials-plugins-faqs-helpful-info/}), VoxCity provides a function to remove perimeter buildings and trees. For vertical grid spacing, VoxCity provides options that use the same size for the horizontal grid or set an arbitrary number of layers. Here, we set 19 layers for the Greenbelt case and 50 for the Abu Dhabi case. One should note that before running the simulation, ENVI-met automatically resizes the meshes within the five layers near the ground to a finer resolution. We also note that ENVI-met V5.7.1 often predicts unphysical results for complex terrains (as detailed on their website: \url{http://www.envi-hq.com/}), so we strongly recommend VoxCity users not to use DEM when conducting ENVI-met simulation.

VoxCity automatically sets default values of the thermal properties of buildings, ground, and trees by providing the INX file. Although overwriting the INX file, e.g., setting detailed building surface materials, can give more accurate predictions, we show the results with the default values here to verify their adequacy. For meteorological forcings, we used the 2004–2018 typical meteorological year data of the EnergyPlus Weather file (\url{https://www.ladybug.tools/epwmap/}) and selected the nearest weather stations for each site.	
    
Figures~\ref{fig:envimet}c–\ref{fig:envimet}f show thermal environments at 1.5 m height on typical summer afternoons (2:00 p.m.), namely the horizontal wind vectors and the universal thermal climate index (UTCI) (derived from the air temperature, humidity, radiation, and wind speed) predicted by ENVI-met. In the Abu Dhabi case, the inlet wind was the northwesterly sea breeze from the Persian Gulf. The speed was relatively high above the roads and the open areas and reached 5 m/s at the maximum (Figure~\ref{fig:envimet}c), probably because the city block pattern is along the direction of the inlet wind. However, the inlet air temperature was 44°C, hotter than the human body, and therefore, as seen around point \textcircled{1} in Figure~\ref{fig:envimet}e for example, the increase in wind speed rather increased UTCI by convective heat transfer (see Fig. 7 of \citet{Brode2012-yj}). Shading effects of buildings and trees on UTCI are evident, and the difference between sunlit and shaded areas reached $>$ 20 °C (Figure~\ref{fig:envimet}e). Weak and complex diverged flows with lower air temperature and UTCI values are found in the leeward of the buildings (e.g., around building \textcircled{2} in Figures~\ref{fig:envimet}c and ~\ref{fig:envimet}e and the southeast city block), indicating cooler above-roof air moved to the ground level by the downdrafts. In the Greenbelt case, the wind speed was generally lower than the inlet value (1.5 m/s) due to the momentum sink by trees (Figure~\ref{fig:envimet}d). The UTCI distribution seems to correspond simply to the shading pattern, and the contrast between roads and under-tree areas is evident (Figure~\ref{fig:envimet}f). We verified the validity of various parametric scenarios in Abu Dhabi, Greenbelt, and a few other sites. The calculations were successfully completed in all cases, while physically unrealistic results were obtained when, as mentioned above, OSM's LULC data included water surfaces or when using DEM data. 
Note that the validation results and limitations of the recent ENVI-met code (version 4 or above) are explained in the literature \citep{Yang2013,Tsoka2018,Liu2021}.

\subsubsection{OBJ for 3D modeling and rendering}
\label{sec:uc_3dr}

The 3D array from VoxCity is converted into surface polygons of the voxel model and then formatted according to the OBJ structure. Colors corresponding to voxel classes are assigned to polygons, and the color palettes are saved as Material Template Library (MTL) files. This allows 3D modeling software to automatically apply the assigned surface colors when loading OBJ files and helps preserve semantic information that is not native to OBJ~\citep{2015_udmv_citygml_obj}. Simulation results from VoxCity's internal functions can also be exported as OBJ and MTL files.

\begin{figure}
    \centering    \includegraphics[width=0.95\linewidth]{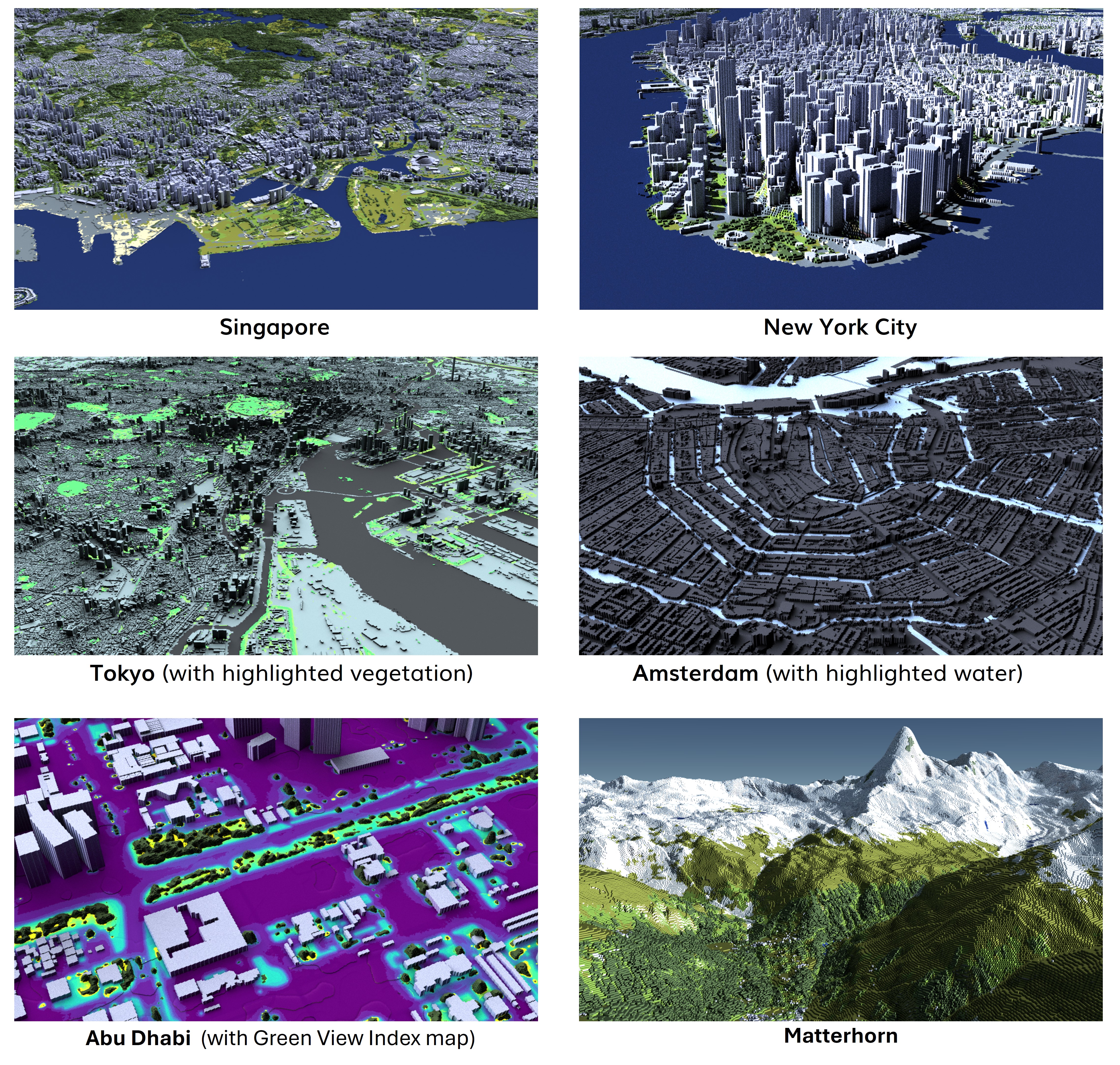}
    \caption{Examples of 3D renderings created in Rhino using output 3D city models from VoxCity.}
    \label{fig:3dr}
\end{figure}

Figure~\ref{fig:3dr} shows examples of OBJ outputs rendered in Rhino 7, illustrating VoxCity's capabilities to produce diverse visualization styles for different purposes. 
The renderings for Singapore and New York City aim to provide photo-realistic visualization of large-scale urban morphologies.
Those for Tokyo and Amsterdam employ vibrant colors for specific urban objects, vegetation, and water, highlighting distinct urban morphological features.
The example of Abu Dhabi demonstrates how urban environment simulation results can be integrated with 3D city models by exporting simulation data as OBJ files. This makes it easier for viewers to understand the relationship between simulated environments and the underlying urban morphology.
Although VoxCity primarily focuses on urban areas, it can also be used to visualize natural environments. In the Matterhorn example, VoxCity accurately captures the mountain's sharp peak and ice-covered terrain.

\subsection{Visualization}
\label{sec:voxcity_vis}

\begin{figure}
    \centering    \includegraphics[width=0.95\linewidth]{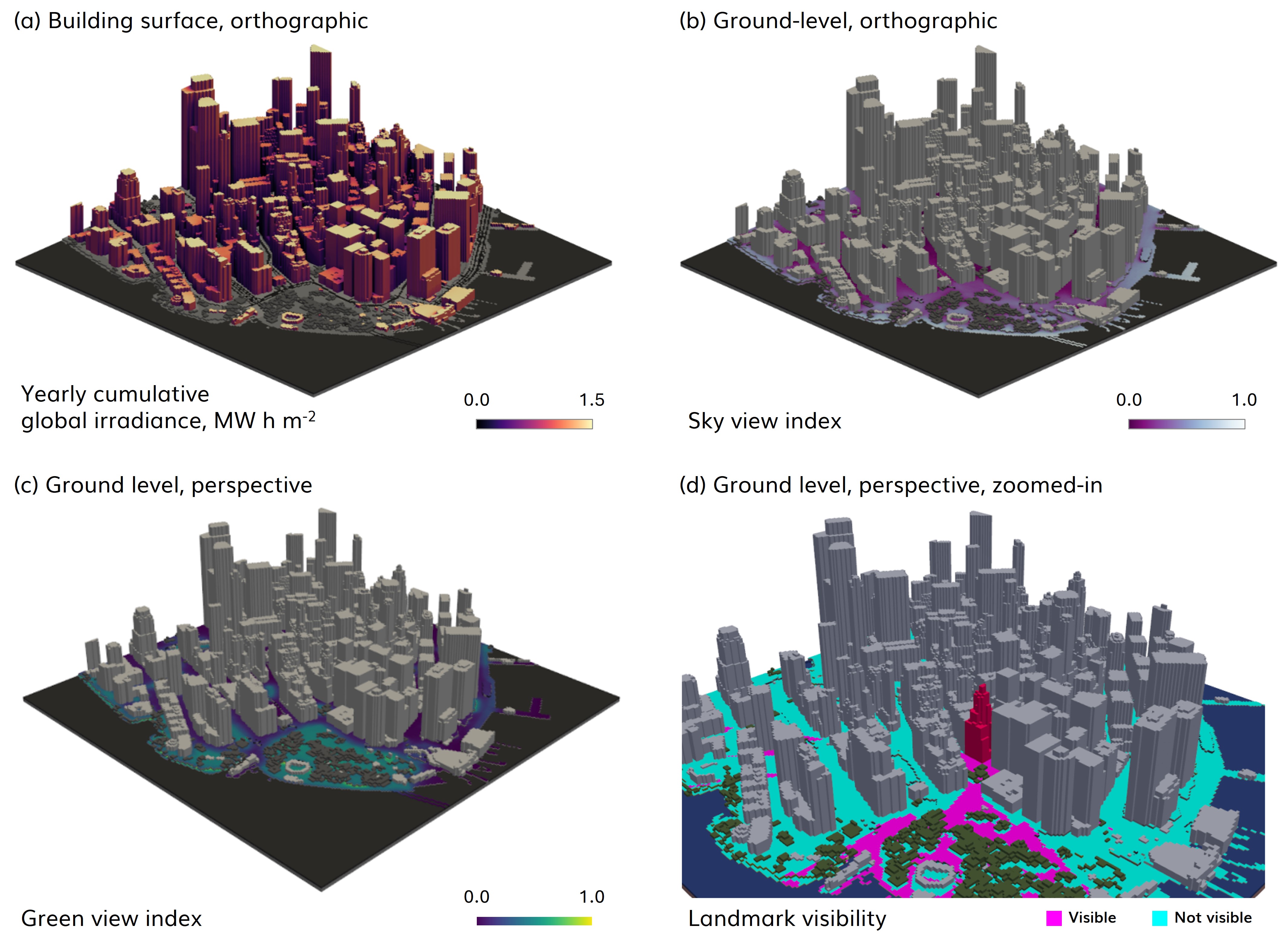}
    \caption{Example outputs for New York City produced by VoxCity's built-in 3D visualization function.}
    \label{fig:vis}
\end{figure}

VoxCity provides a built-in function for 3D visualization. Figure~\ref{fig:vis} shows example outputs produced by this functionality.
Users can visualize generated 3D city models as well as simulation results.
For instance, Figure~\ref{fig:vis}a, b, c, and d indicate the yearly cumulative global irradiance on building surfaces, the ground-level sky view index, the ground-level green view index, and the ground-level landmark visibility, respectively.
Users can specify projection types from two options: `orthographic' and `perspective', adjust the zoom factor, and change the camera position and angle.

\section{Limitations and future directions}
\label{sec:discuss}

\subsection{Support and integration of existing semantic 3D city models}
\label{sec:d_3d}

Considering the aspiring role of VoxCity in 3D GIS, an integration of advanced 3D data models (e.g., CityGML and CityJSON) could facilitate its broader interdisciplinary adoption. In this section, we will discuss such integration in two aspects --- (1) incorporating semantic 3D models as input data sources; and (2) extending export outputs to standardized 3D data formats.

First, integrating 3D models has the potential to enrich geospatial information embedded within VoxCity, contributing to more comprehensive and tailored research. VoxCity currently generates 3D city models from a series of openly available 2D and 2.5D datasets, which can result in a lack of semantic and geometric details especially for research at block or building scale. Integration of publicly available 3D city models as supplementary input may be beneficial for such cases. For example, 3D datasets, such as 3DBAG in the Netherlands~\citep{ledoux2019cityjson,vitalis2020cityjson+,Peters2022-qq} offer higher levels of detail achieved through advanced acquisition techniques (e.g., high-resolution airborne laser scanning) and state-of-the-art reconstruction algorithms. Such 3D datasets can address current limitations in semantics and geometry, particularly in the representation of building forms (e.g., roof-related semantics) and vegetation (e.g., classification)~\citep{petrova2024towards}. Nevertheless, the integration of detailed semantic 3D models introduces certain trade-offs in the future development of VoxCity. Large and complex 3D datasets often entail substantial computational demands, which may hinder overall efficiency. Future research needs to evaluate the balance between the benefits of enhanced semantic detail and the associated technical costs when scaling to an extensive city scale.

Second, in the current implementation, VoxCity provides four export formats (i.e., PKL, VOX, INX, and OBJ) to interface with downstream applications. Fostering data standardization and interoperability, in the future, we plan to implement additional export options (e.g., CityGML or CityJSON). For example, exporting 3D city models in standardized formats is of value for use cases that involve data exchange and sharing with practitioners or systems, which require compliance with established schemas. By offering CityGML or CityJSON as export options, VoxCity can enhance compatibility and ensure broader usability across various domains. Further, VoxCity incorporates semantic information during the generation process; therefore, preserving these attributes through standardized formats is important. Formats such as CityJSON are ideal for this purpose, providing a compact, structured, and easy-to-use approach to storing and managing semantic data~\citep{ledoux2019cityjson}. Therefore, such properties can be retained and effectively communicated for downstream practices. For example, a recent study by \citet{lei2024integrating} extended the CityJSON schema to accommodate human perception of architectural appearance in 3D buildings, enabling potential use cases. In this work, VoxCity features its novelty in automatic generation and tangible usability. Enabling standardized export can strengthen the role of VoxCity as a versatile tool, supporting the consistent and interoperable development of 3D city models. However, challenges should be considered, such as the complexity of translating a voxel-based representation into a hierarchical structure, as well as computational and storage costs compared to current formats (e.g., PKL or OBJ).

\subsection{Integration of building surface material information}
\label{sec:d_bmat}

Integrating building surface attributes, such as surface materials and window-to-wall ratio (WWR), into VoxCity is another promising direction for the next iteration of the package.
VoxCity currently does not support the integration of building surface attributes, primarily due to limitations in data sources. 
To the best of our knowledge, no global open datasets currently provide surface attributes for individual buildings. While some buildings in OSM include material information, the proportion is relatively small~\citep{2023_bae_osm_qa}.
However, building materials and WWR can significantly influence the surrounding microclimate and building energy consumption in various ways, including heat transfer, air leakage or ventilation, and offsetting daylighting demand \citep{marino_does_2017,troup_effect_2019, chi_investigation_2020}. 
This highlights the importance of incorporating material and WWR information in 3D city model generation for detailed urban environment simulations.
A potential approach to address this gap is to use computer vision techniques on street view imagery to infer facade features.

\begin{figure}
    \centering    \includegraphics[width=0.65\linewidth]{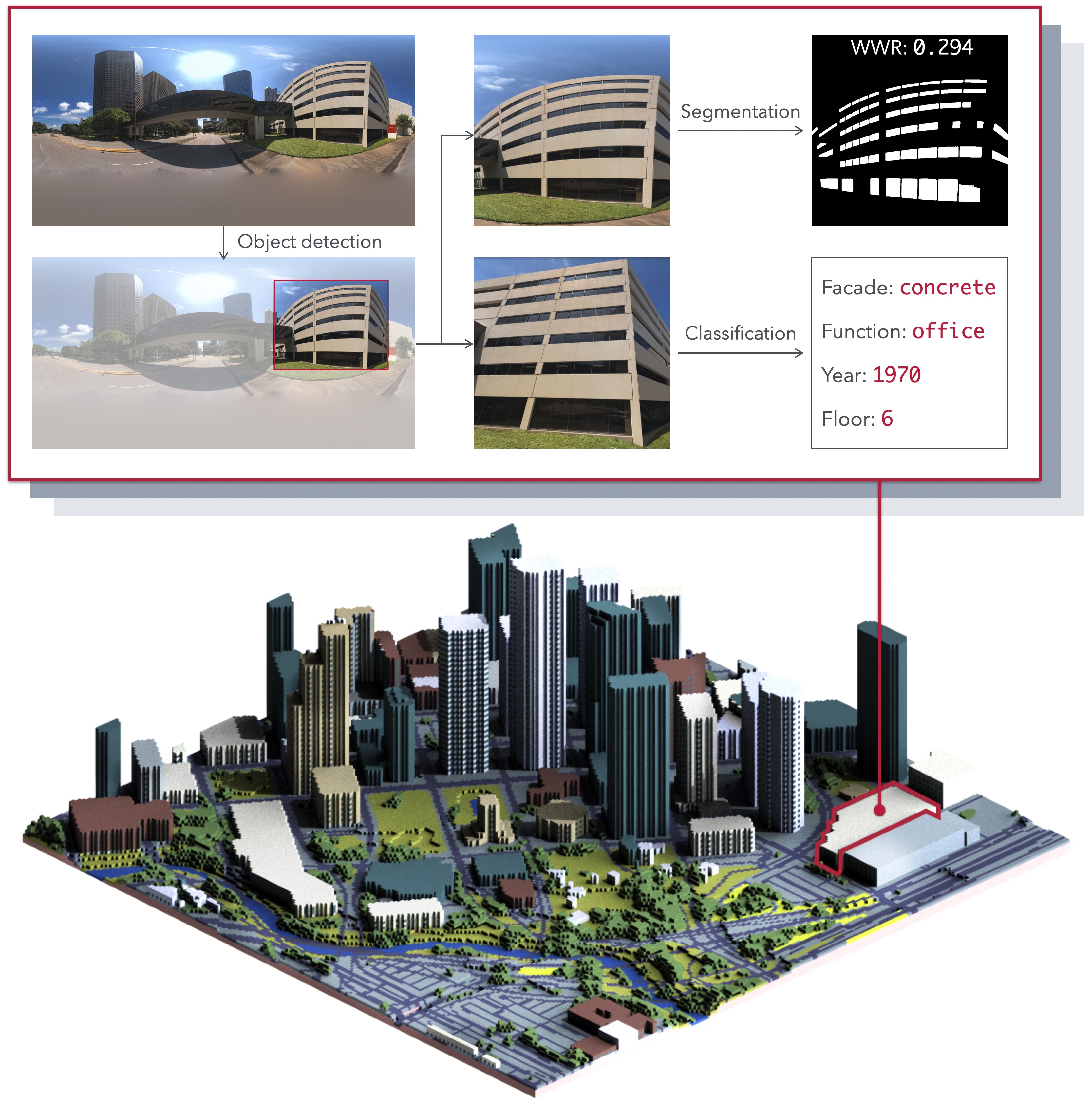}
    \caption{3D city model of a target area in Houston, United States, from a trial incorporating building surface materials and window ratio information extracted from street view imagery. Image source: Mapillary. 3D rendering performed in Rhino.}
    \label{fig:material_window}
\end{figure}

In light of this, we conducted a trial to incorporate (1) building surface materials and (2) WWR information into VoxCity's 3D city model generation.
Figure~\ref{fig:material_window} provides an overview of the process. 
First, street view images were obtained from Mapillary for a target area in Houston, United States. 
OpenFACADES pipeline~\citep{liang2025openfacades} was then employed to geolocate and detect buildings within the street view images, assigning them to their corresponding building footprints.
Building on previous facade classification studies~\citep{kang_building_2018, sun_understanding_2022, raghu_towards_2023}, we applied fine-tuned models to categorize the building materials depicted in these images~\citep{fabbri2020effect, harish2016review}, with each footprint linked to multiple images.
For WWR, window pixels were identified using the Grounded Segment Anything \citep{ren_grounded_2024}, and the percentage of window pixels within the building walls was calculated. 
Based on the inferred materials and WWR, we assigned categorical values to building voxels (e.g., 10 for glass, 11 for concrete, 12 for brick, and 13 for wood).

3D city models with detailed building material attributes can improve the accuracy of these simulations and expand their range of potential applications. Consequently, we plan to develop an additional subpackage that provides this functionality, after refining the methods and conducting a thorough validation. 

\subsection{Data availability and functionality constraints}
\label{sec:d_cover}
In Section~\ref{sec:voxcity}, we have demonstrated VoxCity's capability to generate 3D models of a variety of cities worldwide.
However, we acknowledge that data availability remains a critical challenge, and VoxCity does not incorporate all available data sources. Consequently, users may find it challenging to produce accurate models for regions beyond its current scope. 
Notwithstanding, comprehensive tools such as VoxCity play a crucial role in highlighting where data gaps remain, thereby supporting the development of targeted use cases.
We are further addressing this challenge by continuously updating VoxCity and expanding the range of datasets it supports. 
Meanwhile, to broaden the utilization of building data, efforts to combine individual local datasets into comprehensive global collections---such as EUBUCCO~\citep{Milojevic-Dupont2023-bw}---should be further encouraged. 
Additionally, we have provided only qualitative comparisons of data sources in Section~\ref{sec:review}. To help users select the best data source options for their target areas, future work will conduct comparative experiments that provide quantitative comparisons of data sources in coverage and completeness. 

Our method currently considers only tree canopy height and does not account for variations in shape and density, which can differ by species, season, and health condition. For an accurate assessment of urban environments, it is crucial to include these factors. However, such detailed information is often unavailable in most cities, although some have created tree inventories containing data on individual trees. We plan to add functionality to parse such inventories and incorporate them into model generation. As mentioned in Section~\ref{sec:r_ch}, no globally comprehensive tree inventory datasets or standardized data formats currently exist. Both the integration of local inventories into a global repository and the development of standardized data structures must be pursued in parallel.

VoxCity currently does not support the ability to edit data source files. 
For example, users cannot edit the arrangement of buildings, trees, and land cover from the original datasets. 
Therefore, users cannot conduct simulations to compare different design scenarios. 
To further enhance usability in urban studies and urban and architectural planning practices, future work will address this limitation by developing, for instance, a user interface to interactively edit data source files and voxel city models. Additionally, the current version does not support multi-dimensional data overlays, or integration with GIS platforms.
These limitations represent important areas for future development of VoxCity.

\subsection{Validity and reliability of generated models}
\label{sec:d_validity}

The validity of the 3D city models generated by VoxCity depends on the accuracy of the input datasets. Each incorporated dataset contains its own uncertainty levels, as shown in Tables~\ref{tab:building_height}, \ref{tab:land-cover-datasets}, \ref{tab:ch}, and \ref{tab:dem}. For example, OB2.5DT (building height) has an MAE of approximately 1.5 m \citep{Sirko2023-hb}, META (tree canopy height) exhibits an MAE of 2.8 m \citep{Tolan2024-an}, and FABDEM (terrain elevation) shows MAEs of 1–3 m depending on land cover \citep{Hawker2022-by}. Consequently, the accuracy of the generated 3D city models is constrained by these inherent uncertainties in the source data. Users are therefore encouraged to interpret simulation outputs---such as solar irradiance or view indices---within the expected uncertainty range of the input data. Nonetheless, VoxCity ensures reliability through its deterministic workflow---given the same inputs and parameters, it consistently generates identical results.

\section{Conclusions}
\label{sec:concl}

Urban environment simulations using 3D city models are powerful instruments for informed urban planning and policymaking, particularly for assessing environmental benefits and risks that affect the health and well-being of city dwellers. However, the intensive manual effort required to prepare 3D city models, complicated application-specific data requirements, and fragmented data availability often hinder their broader utilization. To address this, we developed `VoxCity', a one-stop Python package for open geospatial data integration, 3D city model generation, and urban environment simulation. Focusing on four key data types--- building height, tree canopy height, land cover, and terrain elevation---we reviewed existing open datasets and compiled them into a catalog. VoxCity automatically downloads these datasets, voxelizes buildings, trees, land cover, and terrain, and creates an integrated voxel-based city model ready for a variety of environmental simulations. Additionally, VoxCity enables users to perform urban environment simulations through its built-in simulation subpackage and to export the generated 3D models in various file formats compatible with external software. The key contributions of this holistic and integrated work are as follows.

\begin{enumerate}
    \item This paper presents a review of globally available geospatial data relevant to 3D city models---including building height, tree canopy height, land cover, and terrain elevation. This review not only helps VoxCity users select appropriate datasets but also provides readers with an overview of such datasets.
    \item VoxCity provides a streamlined and automatic method to prepare 3D city models. This is particularly advantageous in cities without openly available 3D city models. Additionally, VoxCity's capability to generate ready-to-use models for urban environment simulations benefits even those cities with existing 3D city models.
    \item VoxCity can integrate four geospatial data types --- building height, tree canopy height, land cover, and terrain elevation ---  to generate semantic 3D city models. Users can conduct simulations that account for buildings, vegetation, water bodies, and terrain geometry, all of which significantly affect urban environments. 
    \item VoxCity's built-in simulation and visualization functions provide a comprehensive, end-to-end solution---from 3D city model generation to running urban environment simulations including solar irradiance and visibility analyses and visualizing the simulation results---thereby significantly reducing the time and effort required for such tasks.  
    \item By leveraging data from OSM, which is continuously updated by volunteers, VoxCity ensures its 3D city models remain relatively current. Many open 3D city models are not continuously updated and can quickly become obsolete; however, VoxCity mitigates this limitation by providing a more up-to-date representation of the urban form.
    \item VoxCity can also serve as a comprehensive data downloader, sourcing information from various providers for users who require only intermediate data.
\end{enumerate}

\section*{CRediT authorship contribution statement}
\textbf{Kunihiko Fujiwara}: Conceptualization, Methodology, Software, Validation, Formal analysis, Investigation, Data Curation, Writing – original draft, Writing – review \& editing. \textbf{Ryuta Tsurumi}: Formal analysis, Investigation, Data Curation, Writing – review \& editing. \textbf{Tomoki Kiyono}: Formal analysis, Investigation, Data Curation, Writing – original draft, Writing – review \& editing. \textbf{Zicheng Fan}: Formal analysis, Investigation, Data Curation, Writing – original draft, Writing – review \& editing. \textbf{Xiucheng Liang}: Formal analysis, Investigation, Data Curation, Writing – original draft, Writing – review \& editing. \textbf{Binyu Lei}: Writing – original draft, Writing – review \& editing. \textbf{Winston Yap}: Methodology, Software, Writing – review \& editing. \textbf{Koichi Ito}: Software, Writing – review \& editing. \textbf{Filip Biljecki}: Conceptualization, Methodology, Supervision, Project administration, Funding acquisition, Writing – review \& editing.

\section*{Declaration of Competing Interest}
The authors declare that they have no known competing financial interests or personal relationships that could have appeared to influence the work reported in this paper.

\section*{Declaration of Generative AI and AI-assisted technologies in the writing process}
During the preparation of this work the authors used ChatGPT in order to proofread the text. After using this tool, the authors reviewed and edited the content as needed and take full responsibility for the content of the publication.

\section*{Acknowledgements}
We express our gratitude to the members of the NUS Urban Analytics Lab for the valuable discussions. 
This research has been supported by Takenaka Corporation.
This research is part of the projects (i) Large-scale 3D Geospatial Data for Urban Analytics, which is supported by the National University of Singapore under the Start Up Grant; and (ii) Multi-scale Digital Twins for the Urban Environment: From Heartbeats to Cities, which is supported by the Singapore Ministry of Education Academic Research Fund Tier 1.
We would like to thank the Singapore International Graduate Award (SINGA) scholarship provided by the Agency for Science, Technology, and Research (A*STAR), and NUS Research Scholarship and the President’s Graduate Fellowship (PGF) provided by NUS.

\section*{Data availability}
The geospatial data used in this paper is openly available (see Section~\ref{sec:review}).

\section*{Code availability}
\label{sec:appendix_code}
The source code of VoxCity is openly available on GitHub (\href{https://github.com/kunifujiwara/VoxCity}{https://github.com/kunifujiwara/VoxCity}), with comprehensive documentation (\href{https://voxcity.readthedocs.io/en/latest/}{https://voxcity.readthedocs.io/en/latest/}), a demo Google Colab notebook (\href{https://colab.research.google.com/drive/1Lofd3RawKMr6QuUsamGaF48u2MN0hfrP?usp=sharing}{https://colab.research.google.com/drive/1Lofd3RawKMr6QuUsamGaF48u2MN0hfrP?usp=sharing}), and a tutorial video (\href{https://youtu.be/qHusvKB07qk}{https://youtu.be/qHusvKB07qk}). 
VoxCity is distributed under the MIT License, a permissive open-source license that allows for free use, modification, and distribution, including for commercial purposes, with minimal restrictions. The package is available via PyPI (\texttt{pip install voxcity}). 

Users can report issues, request features, and contribute code through GitHub. We welcome community contributions through pull requests. We plan to provide long-term support for stable releases, ensuring the package remains functional as underlying data sources and dependencies evolve.

\newpage
\appendix

\section{Environmental simulations and urban elements}
\label{sec:appendix_element}

\begin{table}[htbp]
\centering
\caption{Assessment categories for environmental simulation and their corresponding urban elements}
\resizebox{\textwidth}{!}{%
\begin{tabular}{p{0.25\textwidth} p{0.45\textwidth} p{0.30\textwidth}}
\toprule[1.1pt]
\textbf{Assessment categories} & \textbf{References} & \textbf{Urban elements} \\
\midrule[1.1pt]
Thermal comfort / heat stress & \citet{Chen2021-te, Yuan2022-wf, Lindberg2008-hq, Kong2022-zw, Li2021-ye, Galal2020-uc, Zhang2022-av, Hong2015-ik, Wu2019-hp, Morakinyo2016-pm, Li2023-uq, Wang2015-nb, Robitu2006-lp, Du2019-ab, Imam-Syafii2017-to} & Buildings, trees, land cover, terrain elevation \\
\midrule
Sunlight exposure / photovoltaic potential & \citet{Grant2002-vp, Fath2015-jw} & Buildings, trees \\
\midrule
Wind comfort / wind energy potential / wind pressure & \citet{Blocken2012-kj, Mou2017-iq, Hong2015-ik} & Buildings, trees \\
\midrule
Building energy performance & \citet{Forouzandeh2021-fm, Gros2016-xo, Bouyer2011-mt} & Buildings, land cover (low vegetation, etc.) \\
\midrule
Air quality & \citet{Kwak2015-nu} & Buildings, terrain elevation, air pollutant emissions \\
\midrule
Visual comfort / green view index / sky view factor & \citet{Oh1994-ot, Yi2017-vf, Yu2016-gb, Labib2021-xj, Fujiwara2022-yb} & Buildings, trees, land cover, terrain elevation \\
\bottomrule[1.1pt] 
\end{tabular}%
}
\label{tab:urban_elements}
\end{table}

\section{Data structure of voxel city models}
\label{sec:appendix_dstructure}

\begin{figure}[htbp]
    \centering    \includegraphics[width=0.8\linewidth]{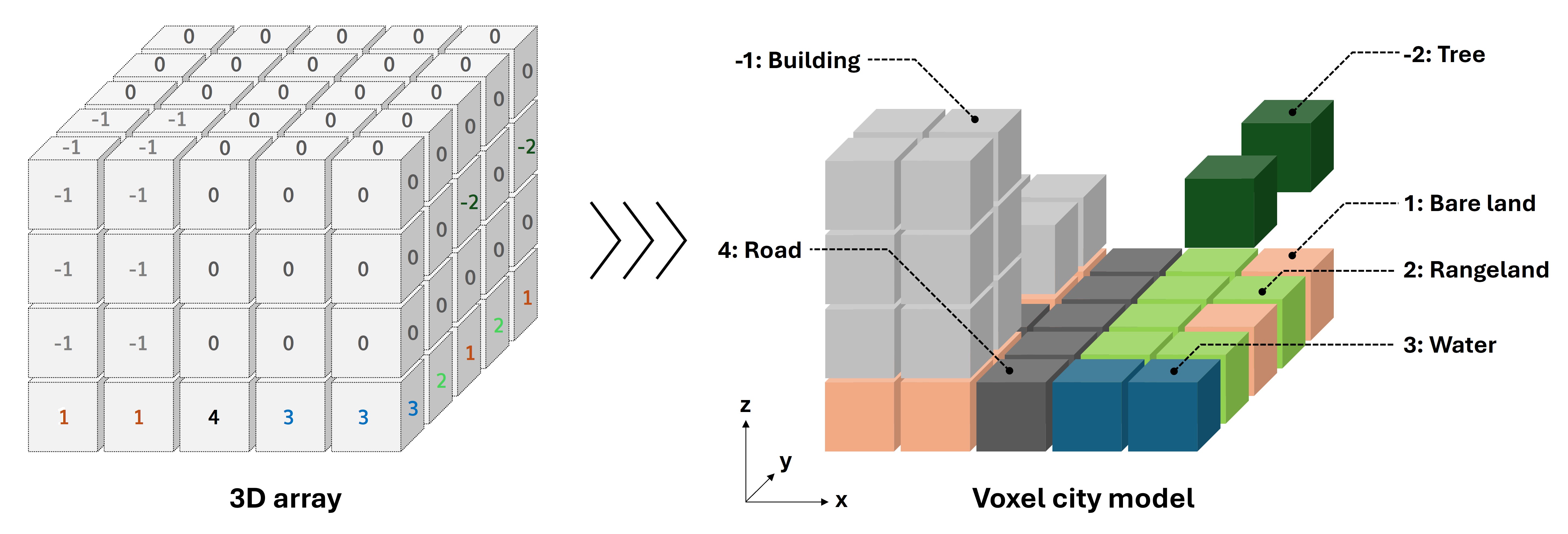}
    \caption{3-dimensional array representing semantic voxel city models.}
    \label{fig:dstructure}
\end{figure}

\newpage

\section{3D models for diverse geographies worldwide}
\label{sec:appendix_gsouth}

\begin{figure}[htbp]
    \centering    \includegraphics[width=0.9\linewidth]{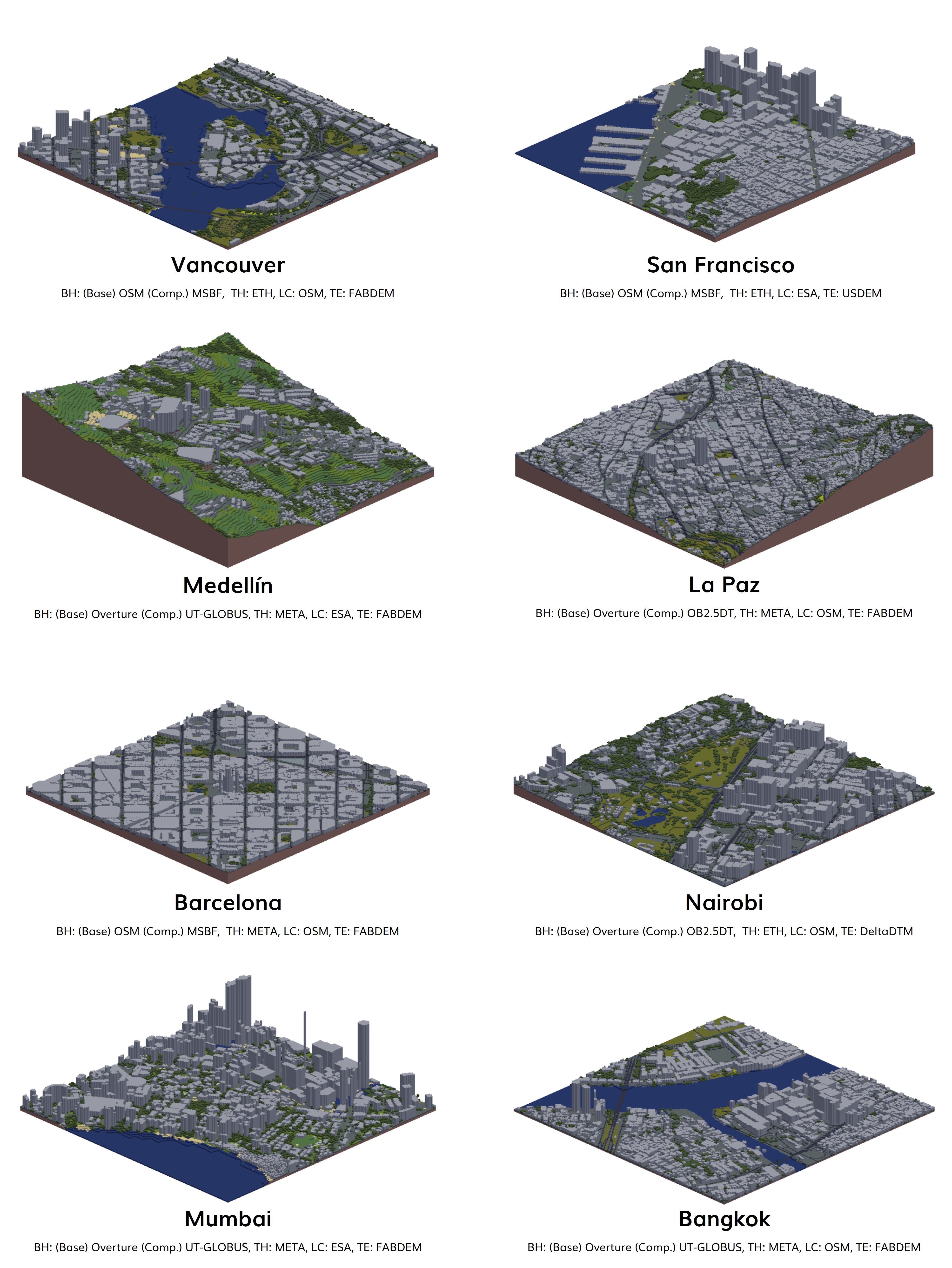}
    \caption{Additional examples of output 3D city models from VoxCity. The 3D rendering was performed in VoxCity's visualization module. BH, TH, LC, and TE represent building height, tree canopy height, land cover, and terrain elevation, respectively.}
    \label{fig:gsouth}
\end{figure}

\newpage
\section{Calculation of solar irradiance}
\label{sec:appendix_solar}

We implemented a calculation combining the methods by \citet{Pruzinec2022-uo} and \citet{Monsi2004-eb}.
\citet{Pruzinec2022-uo} introduced a ray-tracing and voxel-based method, while \citet{Monsi2004-eb} introduced a method to calculate transmittance of trees from leaf area density (LAD).
For instantaneous values, the module first calculates solar azimuth and elevation angles using the location (longitude and latitude) of the target area and the target timestamp, employing a Python package `astral'\footnote{\url{https://github.com/sffjunkie/astral}}.
Direct and diffuse irradiance are then calculated using a ray-tracing technique, incorporating the computed sun position as well as direct normal and diffuse horizontal irradiance that is not affected by shading.
In the ray-tracing process, a transmittance determined by the Beer-Lambert law (Equation~\ref{eq:beerlambert}) is incorporated when rays pass through tree voxels:
\begin{equation}
    \tau = e^{-K \cdot LAD \cdot l} \quad \text{(\citet{Monsi2004-eb})}
    \label{eq:beerlambert}
\end{equation}
where $\tau$ is the transmittance of solar radiation through tree canopies, $e$ is the base of the natural logarithm, $K$ is the extinction coefficient, $LAD$ represents the leaf area density (m$^2$
m$^{-3}$), and $l$ is the path length through tree voxels (m). In this paper, we employed $K$ = 0.5, and $LAD$ = 1.0 m$^2$ m$^{-3}$.

For cumulative values, the module calculates instantaneous irradiance at each timestamp from the start to the end of a specified period and then sums the results. To incorporate locally relevant climate conditions (including direct normal and diffuse horizontal irradiance), the module can use EnergyPlus Weather (EPW) files. Users may supply their own EPW files; otherwise, the module can automatically download the nearest EPW file from Climate.OneBuilding.Org \footnote{\url{https://climate.onebuilding.org/}}.

\section{Time requirements for 3D city model generation and simulations}
\label{sec:appendix_time}

\begin{table}[htbp]
\centering
\caption{Time requirements for 3D city model generation and simulations using VoxCity for an area in New York City. The workflow includes both manual user inputs (specifying parameters, target areas, etc.) and automated computational processes (data downloading, ray-tracing calculations, etc.). The target area, data sources, and simulation settings are the same as those in Figures~\ref{fig:vox_eo} and \ref{fig:simulation}. Processing was performed on our system with an Intel Core i9-13900 processor and an internet connection with a measured download speed of 180 Mbps.}
\resizebox{\textwidth}{!}{%
\begin{tabular}{p{0.20\textwidth} p{0.18\textwidth} p{0.45\textwidth} p{0.17\textwidth}}
\toprule[1.1pt]
\textbf{Process} & \textbf{Type} & \textbf{Breakdown} & \textbf{Time required} \\
\midrule[1.1pt]
\multirow{2}{*}{Model generation} & Manual & Specifying target area, setting data sources and parameters & 90 sec \\
\cmidrule(lr){2-4}
& Computational & Downloading data sources, generating a voxel city model & 40 sec \\
\midrule
\multirow{2}{*}{\parbox{0.18\textwidth}{Solar irradiance \\ simulation}} & Manual & Setting target time range & 30 sec \\
\cmidrule(lr){2-4}
& Computational & Downloading the nearest weather file, ray-tracing calculation (for 8760 time steps) & 50 sec \\
\midrule
\multirow{2}{*}{\parbox{0.18\textwidth}{Green view index \\ calculation}} & Manual & - &  0 sec  \\
\cmidrule(lr){2-4}
& Computational & Ray-tracing calculation & 1 sec \\
\midrule
\multirow{2}{*}{\parbox{0.22\textwidth}{Landmark visibility \\ analysis}} & Manual & Specifying landmark buildings & 30 sec \\
\cmidrule(lr){2-4}
& Computational & Ray-tracing calculation & 1 sec \\
\bottomrule[1.1pt] 
\end{tabular}%
}
\label{tab:process_time}
\end{table}

\newpage

\section{ENVI-met Material Mapping and Parameters}
\label{sec:appendix_envimet}

\begin{table}[htbp]
\centering
\caption{Mapping of voxel classes in VoxCity to ENVI-met material and vegetation IDs with their basic properties. The model uses a layered structure where vegetation and building objects are placed on top of ground/soil profiles (e.g., rangeland = grass on sandy loam; wet land = grass on water; building = wall on sandy loam).}
\resizebox{\textwidth}{!}{%
\begin{tabular}{p{0.22\textwidth} p{0.12\textwidth} p{0.20\textwidth} p{0.40\textwidth}}
\toprule[1.1pt]
\textbf{Voxel classes in\newline VoxCity} & \textbf{Material type} & \textbf{ENVI-met material/\newline vegetation ID} & \textbf{Basic properties} \\
\midrule[1.1pt]
Bareland, Rangeland, Shrub, Moss and lichen, Agriculture land, Tree, Snow and ice, Building, No Data & Ground/soil & 000000 (default sandy loam) & Albedo: 0.20, Emissivity: 0.98, Roughness: 0.015 m, Sandy loam soil profile (19 layers) \\ %
\midrule
Wet land, Mangroves, Water & Ground/soil & 0200WW (deep water) & Albedo: 0.00, Emissivity: 0.96, Roughness: 0.01 m, Water mixing coeff: 0.001, Turbidity: 2.1 \\ %
\midrule
Road & Ground/soil & 0200ST (asphalt road) & Albedo: 0.12, Emissivity: 0.90, Roughness: 0.01 m, 9 layers asphalt over 10 layers sandy loam \\ %
\midrule 
Rangeland, Moss and lichen, Agriculture land, Wet land & Vegetation & 0200XX (grass 25 cm aver. dense) & Height: 0.25 m, Albedo: 0.20, Emissivity: 0.97, Transmittance: 0.30, rs\_min: 200 s/m, LAD: uniform 0.30 m²/m³, Root depth: 0.20 m \\ %
\midrule 
Shrub & Vegetation & 0200H1 (hedge dense, 1 m) & Height: 1.00 m, Albedo: 0.20, Emissivity: 0.97, Transmittance: 0.30, rs\_min: 400 s/m, LAD: uniform 1.00 m²/m³, Root depth: 1.00 m \\ %
\midrule 
Tree & Vegetation & - (user-defined 3D plants by canopy height) & Albedo: 0.18, Transmittance: 0.30, rs\_min: 0.0 s/m, LAD: 1.0 m$^{-1}$, Wood density: 690 kg/m³ \\ %
\midrule 
Building & Building wall & 000000 (default wall -- moderate insulation) & 3-layer (plaster/insulation/concrete), Total thickness: 0.31 m, Roughness: 0.02 m, Can be greened \\ %
\bottomrule[1.1pt] 
\end{tabular}%
}
\label{tab:envimet_mapping}
\end{table}

\end{sloppypar}

\newpage

\begin{thebibliography}{172}
\expandafter\ifx\csname natexlab\endcsname\relax\def\natexlab#1{#1}\fi
\providecommand{\url}[1]{\texttt{#1}}
\providecommand{\href}[2]{#2}
\providecommand{\path}[1]{#1}
\providecommand{\DOIprefix}{doi:}
\providecommand{\ArXivprefix}{arXiv:}
\providecommand{\URLprefix}{URL: }
\providecommand{\Pubmedprefix}{pmid:}
\providecommand{\doi}[1]{\href{http://dx.doi.org/#1}{\path{#1}}}
\providecommand{\Pubmed}[1]{\href{pmid:#1}{\path{#1}}}
\providecommand{\bibinfo}[2]{#2}
\ifx\xfnm\relax \def\xfnm[#1]{\unskip,\space#1}\fi
\bibitem[{Lindberg et~al.(2008)Lindberg, Holmer, and
  Thorsson}]{Lindberg2008-hq}
\bibinfo{author}{F.~Lindberg}, \bibinfo{author}{B.~Holmer},
  \bibinfo{author}{S.~Thorsson},
\newblock \bibinfo{title}{{SOLWEIG} 1.0--modelling spatial variations of {3D}
  radiant fluxes and mean radiant temperature in complex urban settings},
\newblock \bibinfo{journal}{Int. J. Biometeorol.} \bibinfo{volume}{52}
  (\bibinfo{year}{2008}) \bibinfo{pages}{697--713}.
\bibitem[{Yuan et~al.(2022)Yuan, Masuko, Shimazaki, Yamanaka, and
  Kobayashi}]{Yuan2022-wf}
\bibinfo{author}{J.~Yuan}, \bibinfo{author}{S.~Masuko},
  \bibinfo{author}{Y.~Shimazaki}, \bibinfo{author}{T.~Yamanaka},
  \bibinfo{author}{T.~Kobayashi},
\newblock \bibinfo{title}{Evaluation of outdoor thermal comfort under different
  building external-wall-surface with different reflective directional
  properties using {CFD} analysis and model experiment},
\newblock \bibinfo{journal}{Build. Environ.} \bibinfo{volume}{207}
  (\bibinfo{year}{2022}) \bibinfo{pages}{108478}.
\bibitem[{Pađen et~al.(2022)Pađen, García-Sánchez, and
  Ledoux}]{Paden2022-me}
\bibinfo{author}{I.~Pađen}, \bibinfo{author}{C.~García-Sánchez},
  \bibinfo{author}{H.~Ledoux},
\newblock \bibinfo{title}{Towards automatic reconstruction of {3D} city models
  tailored for urban flow simulations},
\newblock \bibinfo{journal}{Front. Built Environ.} \bibinfo{volume}{8}
  (\bibinfo{year}{2022}) \bibinfo{pages}{899332}.
\bibitem[{Urech et~al.(2022)Urech, Mughal, and Bartesaghi-Koc}]{Urech2022-es}
\bibinfo{author}{P.~R.~W. Urech}, \bibinfo{author}{M.~O. Mughal},
  \bibinfo{author}{C.~Bartesaghi-Koc},
\newblock \bibinfo{title}{A simulation-based design framework to iteratively
  analyze and shape urban landscapes using point cloud modeling},
\newblock \bibinfo{journal}{Comput. Environ. Urban Syst.} \bibinfo{volume}{91}
  (\bibinfo{year}{2022}) \bibinfo{pages}{101731}.
\bibitem[{Oh(1994)}]{Oh1994-ot}
\bibinfo{author}{K.~Oh},
\newblock \bibinfo{title}{A perceptual evaluation of computer-based landscape
  simulations},
\newblock \bibinfo{journal}{Landsc. Urban Plan.} \bibinfo{volume}{28}
  (\bibinfo{year}{1994}) \bibinfo{pages}{201--216}.
\bibitem[{Gros et~al.(2016)Gros, Bozonnet, Inard, and Musy}]{Gros2016-xo}
\bibinfo{author}{A.~Gros}, \bibinfo{author}{E.~Bozonnet},
  \bibinfo{author}{C.~Inard}, \bibinfo{author}{M.~Musy},
\newblock \bibinfo{title}{Simulation tools to assess microclimate and building
  energy – a case study on the design of a new district},
\newblock \bibinfo{journal}{Energy Build.} \bibinfo{volume}{114}
  (\bibinfo{year}{2016}) \bibinfo{pages}{112--122}.
\bibitem[{Bouyer et~al.(2011)Bouyer, Inard, and Musy}]{Bouyer2011-mt}
\bibinfo{author}{J.~Bouyer}, \bibinfo{author}{C.~Inard},
  \bibinfo{author}{M.~Musy},
\newblock \bibinfo{title}{Microclimatic coupling as a solution to improve
  building energy simulation in an urban context},
\newblock \bibinfo{journal}{Energy Build.} \bibinfo{volume}{43}
  (\bibinfo{year}{2011}) \bibinfo{pages}{1549--1559}.
\bibitem[{Katal et~al.(2022)Katal, Mortezazadeh, Wang, and Yu}]{Katal2022-iq}
\bibinfo{author}{A.~Katal}, \bibinfo{author}{M.~Mortezazadeh},
  \bibinfo{author}{L.~l. Wang}, \bibinfo{author}{H.~Yu},
\newblock \bibinfo{title}{Urban building energy and microclimate modeling –
  from {3D} city generation to dynamic simulations},
\newblock \bibinfo{journal}{Energy (Oxf.)} \bibinfo{volume}{251}
  (\bibinfo{year}{2022}) \bibinfo{pages}{123817}.
\bibitem[{Stoter et~al.(2008)Stoter, de~Kluijver, and Kurakula}]{Stoter2008-sa}
\bibinfo{author}{J.~Stoter}, \bibinfo{author}{H.~de~Kluijver},
  \bibinfo{author}{V.~Kurakula},
\newblock \bibinfo{title}{{3D} noise mapping in urban areas},
\newblock \bibinfo{journal}{Geogr. Inf. Syst.} \bibinfo{volume}{22}
  (\bibinfo{year}{2008}) \bibinfo{pages}{907--924}.
\bibitem[{Zhao et~al.(2017)Zhao, Liu, Poh, Wang, Gao, Png, Li, and
  Chong}]{Zhao2017-cj}
\bibinfo{author}{W.-J. Zhao}, \bibinfo{author}{E.-X. Liu},
  \bibinfo{author}{H.~J. Poh}, \bibinfo{author}{B.~Wang},
  \bibinfo{author}{S.-P. Gao}, \bibinfo{author}{C.~E. Png},
  \bibinfo{author}{K.~W. Li}, \bibinfo{author}{S.~H. Chong},
\newblock \bibinfo{title}{{3D} traffic noise mapping using unstructured surface
  mesh representation of buildings and roads},
\newblock \bibinfo{journal}{Appl. Acoust.} \bibinfo{volume}{127}
  (\bibinfo{year}{2017}) \bibinfo{pages}{297--304}.
\bibitem[{Qi et~al.(2023)Qi, Ding, and Lim}]{Qi2023-os}
\bibinfo{author}{J.~Qi}, \bibinfo{author}{L.~Ding}, \bibinfo{author}{S.~Lim},
\newblock \bibinfo{title}{Application of a decision-making framework for
  multi-objective optimisation of urban heat mitigation strategies},
\newblock \bibinfo{journal}{Urban Clim.} \bibinfo{volume}{47}
  (\bibinfo{year}{2023}) \bibinfo{pages}{101372}.
\bibitem[{Qi and Altinakar(2011)}]{Qi2011-lw}
\bibinfo{author}{H.~Qi}, \bibinfo{author}{M.~S. Altinakar},
\newblock \bibinfo{title}{A {GIS}-based decision support system for integrated
  flood management under uncertainty with two dimensional numerical
  simulations},
\newblock \bibinfo{journal}{Environ. Model. Softw.} \bibinfo{volume}{26}
  (\bibinfo{year}{2011}) \bibinfo{pages}{817--821}.
\bibitem[{Gaspari et~al.(2018)Gaspari, Fabbri, and Lucchi}]{Gaspari2018-ix}
\bibinfo{author}{J.~Gaspari}, \bibinfo{author}{K.~Fabbri},
  \bibinfo{author}{M.~Lucchi},
\newblock \bibinfo{title}{The use of outdoor microclimate analysis to support
  decision making process: Case study of bufalini square in cesena},
\newblock \bibinfo{journal}{Sustain. Cities Soc.} \bibinfo{volume}{42}
  (\bibinfo{year}{2018}) \bibinfo{pages}{206--215}.
\bibitem[{Taleghani et~al.(2016)Taleghani, Sailor, and
  Ban-Weiss}]{Taleghani2016-mf}
\bibinfo{author}{M.~Taleghani}, \bibinfo{author}{D.~Sailor},
  \bibinfo{author}{G.~A. Ban-Weiss},
\newblock \bibinfo{title}{Micrometeorological simulations to predict the
  impacts of heat mitigation strategies on pedestrian thermal comfort in a los
  angeles neighborhood},
\newblock \bibinfo{journal}{Environ. Res. Lett.} \bibinfo{volume}{11}
  (\bibinfo{year}{2016}) \bibinfo{pages}{024003}.
\bibitem[{Robineau et~al.(2022)Robineau, Rodler, Morille, Ramier, Sage, Musy,
  Graffin, and Berthier}]{Robineau2022-qv}
\bibinfo{author}{T.~Robineau}, \bibinfo{author}{A.~Rodler},
  \bibinfo{author}{B.~Morille}, \bibinfo{author}{D.~Ramier},
  \bibinfo{author}{J.~Sage}, \bibinfo{author}{M.~Musy},
  \bibinfo{author}{V.~Graffin}, \bibinfo{author}{E.~Berthier},
\newblock \bibinfo{title}{Coupling hydrological and microclimate models to
  simulate evapotranspiration from urban green areas and air temperature at the
  district scale},
\newblock \bibinfo{journal}{Urban Clim.} \bibinfo{volume}{44}
  (\bibinfo{year}{2022}) \bibinfo{pages}{101179}.
\bibitem[{Giannico et~al.(2022)Giannico, Stafoggia, Spano, Elia, Dadvand, and
  Sanesi}]{Giannico2022-lh}
\bibinfo{author}{V.~Giannico}, \bibinfo{author}{M.~Stafoggia},
  \bibinfo{author}{G.~Spano}, \bibinfo{author}{M.~Elia},
  \bibinfo{author}{P.~Dadvand}, \bibinfo{author}{G.~Sanesi},
\newblock \bibinfo{title}{Characterizing green and gray space exposure for
  epidemiological studies: Moving from {2D} to {3D} indicators},
\newblock \bibinfo{journal}{Urban For. Urban Greening} \bibinfo{volume}{72}
  (\bibinfo{year}{2022}) \bibinfo{pages}{127567}.
\bibitem[{Virtanen et~al.(2021)Virtanen, Jaalama, Puustinen, Julin, Hyyppä,
  and Hyyppä}]{Virtanen2021-bb}
\bibinfo{author}{J.-P. Virtanen}, \bibinfo{author}{K.~Jaalama},
  \bibinfo{author}{T.~Puustinen}, \bibinfo{author}{A.~Julin},
  \bibinfo{author}{J.~Hyyppä}, \bibinfo{author}{H.~Hyyppä},
\newblock \bibinfo{title}{Near real-time semantic view analysis of {3D} city
  models in web browser},
\newblock \bibinfo{journal}{ISPRS Int. J. Geoinf.} \bibinfo{volume}{10}
  (\bibinfo{year}{2021}) \bibinfo{pages}{138}.
\bibitem[{Qiang et~al.(2019)Qiang, Shen, and Chen}]{Qiang2019-wg}
\bibinfo{author}{Y.~Qiang}, \bibinfo{author}{S.~Shen},
  \bibinfo{author}{Q.~Chen},
\newblock \bibinfo{title}{Visibility analysis of oceanic blue space using
  digital elevation models},
\newblock \bibinfo{journal}{Landsc. Urban Plan.} \bibinfo{volume}{181}
  (\bibinfo{year}{2019}) \bibinfo{pages}{92--102}.
\bibitem[{Lei et~al.(2023{\natexlab{a}})Lei, Stouffs, and
  Biljecki}]{Lei2023-cb}
\bibinfo{author}{B.~Lei}, \bibinfo{author}{R.~Stouffs},
  \bibinfo{author}{F.~Biljecki},
\newblock \bibinfo{title}{Assessing and benchmarking {3D} city models},
\newblock \bibinfo{journal}{Int. J. Geogr. Inf. Sci.} \bibinfo{volume}{37}
  (\bibinfo{year}{2023}{\natexlab{a}}) \bibinfo{pages}{788--809}.
\bibitem[{Lei et~al.(2023{\natexlab{b}})Lei, Janssen, Stoter, and
  Biljecki}]{Lei2023-dj}
\bibinfo{author}{B.~Lei}, \bibinfo{author}{P.~Janssen},
  \bibinfo{author}{J.~Stoter}, \bibinfo{author}{F.~Biljecki},
\newblock \bibinfo{title}{Challenges of urban digital twins: A systematic
  review and a delphi expert survey},
\newblock \bibinfo{journal}{Autom. Constr.} \bibinfo{volume}{147}
  (\bibinfo{year}{2023}{\natexlab{b}}) \bibinfo{pages}{104716}.
\bibitem[{Pađen et~al.(2024)Pađen, Peters, García-Sánchez, and
  Ledoux}]{Paden2024-fb}
\bibinfo{author}{I.~Pađen}, \bibinfo{author}{R.~Peters},
  \bibinfo{author}{C.~García-Sánchez}, \bibinfo{author}{H.~Ledoux},
\newblock \bibinfo{title}{Automatic high-detailed building reconstruction
  workflow for urban microscale simulations},
\newblock \bibinfo{journal}{Build. Environ.} \bibinfo{volume}{265}
  (\bibinfo{year}{2024}) \bibinfo{pages}{111978}.
\bibitem[{Kamath et~al.(2024)Kamath, Singh, Malviya, Martilli, He, Aliaga, He,
  Chen, Magruder, Yang, and Niyogi}]{Kamath2024-at}
\bibinfo{author}{H.~G. Kamath}, \bibinfo{author}{M.~Singh},
  \bibinfo{author}{N.~Malviya}, \bibinfo{author}{A.~Martilli},
  \bibinfo{author}{L.~He}, \bibinfo{author}{D.~Aliaga},
  \bibinfo{author}{C.~He}, \bibinfo{author}{F.~Chen}, \bibinfo{author}{L.~A.
  Magruder}, \bibinfo{author}{Z.-L. Yang}, \bibinfo{author}{D.~Niyogi},
\newblock \bibinfo{title}{{GLObal} building heights for urban studies
  ({UT}-{GLOBUS}) for city- and street- scale urban simulations: Development
  and first applications},
\newblock \bibinfo{journal}{Sci. Data} \bibinfo{volume}{11}
  (\bibinfo{year}{2024}) \bibinfo{pages}{886}.
\bibitem[{Tolan et~al.(2024)Tolan, Yang, Nosarzewski, Couairon, Vo, Brandt,
  Spore, Majumdar, Haziza, Vamaraju, Moutakanni, Bojanowski, Johns, White,
  Tiecke, and Couprie}]{Tolan2024-an}
\bibinfo{author}{J.~Tolan}, \bibinfo{author}{H.-I. Yang},
  \bibinfo{author}{B.~Nosarzewski}, \bibinfo{author}{G.~Couairon},
  \bibinfo{author}{H.~V. Vo}, \bibinfo{author}{J.~Brandt},
  \bibinfo{author}{J.~Spore}, \bibinfo{author}{S.~Majumdar},
  \bibinfo{author}{D.~Haziza}, \bibinfo{author}{J.~Vamaraju},
  \bibinfo{author}{T.~Moutakanni}, \bibinfo{author}{P.~Bojanowski},
  \bibinfo{author}{T.~Johns}, \bibinfo{author}{B.~White},
  \bibinfo{author}{T.~Tiecke}, \bibinfo{author}{C.~Couprie},
\newblock \bibinfo{title}{Very high resolution canopy height maps from {RGB}
  imagery using self-supervised vision transformer and convolutional decoder
  trained on aerial lidar},
\newblock \bibinfo{journal}{Remote Sens. Environ.} \bibinfo{volume}{300}
  (\bibinfo{year}{2024}) \bibinfo{pages}{113888}.
\bibitem[{Zanaga et~al.(2022)Zanaga, Van De~Kerchove, Daems, De~Keersmaecker,
  Brockmann, Kirches, Wevers, Cartus, Santoro, Fritz, Lesiv, Herold,
  Tsendbazar, Xu, Ramoino, and Arino}]{Zanaga2022-eb}
\bibinfo{author}{D.~Zanaga}, \bibinfo{author}{R.~Van De~Kerchove},
  \bibinfo{author}{D.~Daems}, \bibinfo{author}{W.~De~Keersmaecker},
  \bibinfo{author}{C.~Brockmann}, \bibinfo{author}{G.~Kirches},
  \bibinfo{author}{J.~Wevers}, \bibinfo{author}{O.~Cartus},
  \bibinfo{author}{M.~Santoro}, \bibinfo{author}{S.~Fritz},
  \bibinfo{author}{M.~Lesiv}, \bibinfo{author}{M.~Herold},
  \bibinfo{author}{N.-E. Tsendbazar}, \bibinfo{author}{P.~Xu},
  \bibinfo{author}{F.~Ramoino}, \bibinfo{author}{O.~Arino},
  \bibinfo{title}{{ESA} {WorldCover} 10 m 2021 {v200}}, \bibinfo{year}{2022}.
\bibitem[{Hawker et~al.(2022)Hawker, Uhe, Paulo, Sosa, Savage, Sampson, and
  Neal}]{Hawker2022-by}
\bibinfo{author}{L.~Hawker}, \bibinfo{author}{P.~Uhe},
  \bibinfo{author}{L.~Paulo}, \bibinfo{author}{J.~Sosa},
  \bibinfo{author}{J.~Savage}, \bibinfo{author}{C.~Sampson},
  \bibinfo{author}{J.~Neal},
\newblock \bibinfo{title}{A {30m} global map of elevation with forests and
  buildings removed},
\newblock \bibinfo{journal}{Environ. Res. Lett.} \bibinfo{volume}{17}
  (\bibinfo{year}{2022}) \bibinfo{pages}{024016}.
\bibitem[{Sirko et~al.(2023)Sirko, Brempong, Marcos, Annkah, Korme, Hassen,
  Sapkota, Shekel, Diack, Nevo, Hickey, and Quinn}]{Sirko2023-hb}
\bibinfo{author}{W.~Sirko}, \bibinfo{author}{E.~A. Brempong},
  \bibinfo{author}{J.~T.~C. Marcos}, \bibinfo{author}{A.~Annkah},
  \bibinfo{author}{A.~Korme}, \bibinfo{author}{M.~A. Hassen},
  \bibinfo{author}{K.~Sapkota}, \bibinfo{author}{T.~Shekel},
  \bibinfo{author}{A.~Diack}, \bibinfo{author}{S.~Nevo},
  \bibinfo{author}{J.~Hickey}, \bibinfo{author}{J.~Quinn},
\newblock \bibinfo{title}{High-resolution building and road detection from
  sentinel-2},
\newblock \bibinfo{journal}{arXiv [cs.CV]}  (\bibinfo{year}{2023}).
\bibitem[{Ledoux et~al.(2021)Ledoux, Biljecki, Dukai, Kumar, Peters, Stoter,
  and Commandeur}]{Ledoux2021-iz}
\bibinfo{author}{H.~Ledoux}, \bibinfo{author}{F.~Biljecki},
  \bibinfo{author}{B.~Dukai}, \bibinfo{author}{K.~Kumar},
  \bibinfo{author}{R.~Y. Peters}, \bibinfo{author}{J.~E. Stoter},
  \bibinfo{author}{T.~J.~F. Commandeur},
\newblock \bibinfo{title}{{3dfier}: automatic reconstruction of {3D} city
  models},
\newblock \bibinfo{journal}{J. Open Source Softw.} \bibinfo{volume}{6}
  (\bibinfo{year}{2021}) \bibinfo{pages}{2866}.
\bibitem[{Lindberg et~al.(2018)Lindberg, Grimmond, Gabey, Huang, Kent, Sun,
  Theeuwes, Järvi, Ward, Capel-Timms, Chang, Jonsson, Krave, Liu, Meyer,
  Olofson, Tan, Wästberg, Xue, and Zhang}]{Lindberg2018-jg}
\bibinfo{author}{F.~Lindberg}, \bibinfo{author}{C.~S.~B. Grimmond},
  \bibinfo{author}{A.~Gabey}, \bibinfo{author}{B.~Huang},
  \bibinfo{author}{C.~W. Kent}, \bibinfo{author}{T.~Sun},
  \bibinfo{author}{N.~E. Theeuwes}, \bibinfo{author}{L.~Järvi},
  \bibinfo{author}{H.~C. Ward}, \bibinfo{author}{I.~Capel-Timms},
  \bibinfo{author}{Y.~Chang}, \bibinfo{author}{P.~Jonsson},
  \bibinfo{author}{N.~Krave}, \bibinfo{author}{D.~Liu},
  \bibinfo{author}{D.~Meyer}, \bibinfo{author}{K.~F.~G. Olofson},
  \bibinfo{author}{J.~Tan}, \bibinfo{author}{D.~Wästberg},
  \bibinfo{author}{L.~Xue}, \bibinfo{author}{Z.~Zhang},
\newblock \bibinfo{title}{Urban multi-scale environmental predictor ({UMEP}):
  An integrated tool for city-based climate services},
\newblock \bibinfo{journal}{Environ. Model. Softw.} \bibinfo{volume}{99}
  (\bibinfo{year}{2018}) \bibinfo{pages}{70--87}.
\bibitem[{Katz and Sankaran(2011)}]{Katz2011-oj}
\bibinfo{author}{A.~Katz}, \bibinfo{author}{V.~Sankaran},
\newblock \bibinfo{title}{Mesh quality effects on the accuracy of {CFD}
  solutions on unstructured meshes},
\newblock \bibinfo{journal}{J. Comput. Phys.} \bibinfo{volume}{230}
  (\bibinfo{year}{2011}) \bibinfo{pages}{7670--7686}.
\bibitem[{Willenborg et~al.(2018)Willenborg, Pültz, and
  Kolbe}]{Willenborg2018-kh}
\bibinfo{author}{B.~Willenborg}, \bibinfo{author}{M.~Pültz},
  \bibinfo{author}{T.~H. Kolbe},
\newblock \bibinfo{title}{Integration of semantic {3D} city models and {3D}
  mesh models for accuracy improvements of solar potential analyses},
\newblock \bibinfo{journal}{The International Archives of the Photogrammetry,
  Remote Sensing and Spatial Information Sciences} \bibinfo{volume}{42}
  (\bibinfo{year}{2018}) \bibinfo{pages}{223--230}.
\bibitem[{Zhang et~al.(2023)Zhang, Liu, Qin, Li, Yang, and Liu}]{Zhang2023-at}
\bibinfo{author}{J.~Zhang}, \bibinfo{author}{Y.~Liu}, \bibinfo{author}{B.~Qin},
  \bibinfo{author}{D.~Li}, \bibinfo{author}{S.~Yang}, \bibinfo{author}{Q.~Liu},
\newblock \bibinfo{title}{A structured mesh generation tool for {CFD}
  simulations in process metallurgy},
\newblock \bibinfo{journal}{Metall. Mater. Trans. B} \bibinfo{volume}{54}
  (\bibinfo{year}{2023}) \bibinfo{pages}{481--486}.
\bibitem[{Chen et~al.(2024)Chen, Guo, Mou, Xu, and Wang}]{Chen2024-ch}
\bibinfo{author}{J.~Chen}, \bibinfo{author}{J.~Guo}, \bibinfo{author}{C.~Mou},
  \bibinfo{author}{Z.~Xu}, \bibinfo{author}{J.~Wang},
\newblock \bibinfo{title}{A structured mesh generation based on improved
  ray-tracing method for finite difference time domain simulation},
\newblock \bibinfo{journal}{Electronics (Basel)} \bibinfo{volume}{13}
  (\bibinfo{year}{2024}) \bibinfo{pages}{1189}.
\bibitem[{Gargallo-Peiró et~al.(2016)Gargallo-Peiró, Folch, and
  Roca}]{Gargallo-Peiro2016-zk}
\bibinfo{author}{A.~Gargallo-Peiró}, \bibinfo{author}{A.~Folch},
  \bibinfo{author}{X.~Roca},
\newblock \bibinfo{title}{Representing urban geometries for unstructured mesh
  generation},
\newblock \bibinfo{journal}{Procedia Eng.} \bibinfo{volume}{163}
  (\bibinfo{year}{2016}) \bibinfo{pages}{175--185}.
\bibitem[{Fisher-Gewirtzman et~al.(2013)Fisher-Gewirtzman, Shashkov, and
  Doytsher}]{Fisher-Gewirtzman2013-ed}
\bibinfo{author}{D.~Fisher-Gewirtzman}, \bibinfo{author}{A.~Shashkov},
  \bibinfo{author}{Y.~Doytsher},
\newblock \bibinfo{title}{Voxel based volumetric visibility analysis of urban
  environments},
\newblock \bibinfo{journal}{Surv. Rev.} \bibinfo{volume}{45}
  (\bibinfo{year}{2013}) \bibinfo{pages}{451--461}.
\bibitem[{Aleksandrov et~al.(2019)Aleksandrov, Zlatanova, Kimmel, Barton, and
  Gorte}]{Aleksandrov2019-yv}
\bibinfo{author}{M.~Aleksandrov}, \bibinfo{author}{S.~Zlatanova},
  \bibinfo{author}{L.~Kimmel}, \bibinfo{author}{J.~Barton},
  \bibinfo{author}{B.~Gorte},
\newblock \bibinfo{title}{Voxel-based visibility analysis for safety assessment
  of urban environments},
\newblock \bibinfo{journal}{ISPRS Ann. Photogramm. Remote Sens. Spat. Inf.
  Sci.} \bibinfo{volume}{IV-4/W8} (\bibinfo{year}{2019})
  \bibinfo{pages}{11--17}.
\bibitem[{Gorte et~al.(2024)Gorte, Zlatanova, Pilouk, Diakite, and
  Barton}]{Gorte2024-bl}
\bibinfo{author}{B.~Gorte}, \bibinfo{author}{S.~Zlatanova},
  \bibinfo{author}{M.~Pilouk}, \bibinfo{author}{A.~Diakite},
  \bibinfo{author}{J.~Barton},
\newblock \bibinfo{title}{{3D} data integration in the voxel domain},
\newblock \bibinfo{journal}{ISPRS Ann. Photogramm. Remote Sens. Spat. Inf.
  Sci.} \bibinfo{volume}{X-4-2024} (\bibinfo{year}{2024})
  \bibinfo{pages}{133--140}.
\bibitem[{Xiao and Yuizono(2022)}]{Xiao2022-fe}
\bibinfo{author}{J.~Xiao}, \bibinfo{author}{T.~Yuizono},
\newblock \bibinfo{title}{Climate-adaptive landscape design: Microclimate and
  thermal comfort regulation of station square in the hokuriku region, japan},
\newblock \bibinfo{journal}{Build. Environ.} \bibinfo{volume}{212}
  (\bibinfo{year}{2022}) \bibinfo{pages}{108813}.
\bibitem[{Li et~al.(2023)Li, Ouyang, Yin, Tan, and Ren}]{Li2023-fu}
\bibinfo{author}{Y.~Li}, \bibinfo{author}{W.~Ouyang}, \bibinfo{author}{S.~Yin},
  \bibinfo{author}{Z.~Tan}, \bibinfo{author}{C.~Ren},
\newblock \bibinfo{title}{Microclimate and its influencing factors in
  residential public spaces during heat waves: An empirical study in hong
  kong},
\newblock \bibinfo{journal}{Build. Environ.} \bibinfo{volume}{236}
  (\bibinfo{year}{2023}) \bibinfo{pages}{110225}.
\bibitem[{Taleghani et~al.(2019)Taleghani, Marshall, Fitton, and
  Swan}]{Taleghani2019-fk}
\bibinfo{author}{M.~Taleghani}, \bibinfo{author}{A.~Marshall},
  \bibinfo{author}{R.~Fitton}, \bibinfo{author}{W.~Swan},
\newblock \bibinfo{title}{Renaturing a microclimate: The impact of greening a
  neighbourhood on indoor thermal comfort during a heatwave in manchester,
  {UK}},
\newblock \bibinfo{journal}{Solar Energy} \bibinfo{volume}{182}
  (\bibinfo{year}{2019}) \bibinfo{pages}{245--255}.
\bibitem[{Kousis et~al.(2022)Kousis, Manni, and Pisello}]{Kousis2022-dl}
\bibinfo{author}{I.~Kousis}, \bibinfo{author}{M.~Manni}, \bibinfo{author}{A.~L.
  Pisello},
\newblock \bibinfo{title}{Environmental mobile monitoring of urban
  microclimates: A review},
\newblock \bibinfo{journal}{Renewable Sustainable Energy Rev.}
  \bibinfo{volume}{169} (\bibinfo{year}{2022}) \bibinfo{pages}{112847}.
\bibitem[{Chen et~al.(2021)Chen, Rong, and Zhang}]{Chen2021-te}
\bibinfo{author}{G.~Chen}, \bibinfo{author}{L.~Rong},
  \bibinfo{author}{G.~Zhang},
\newblock \bibinfo{title}{Unsteady-state {CFD} simulations on the impacts of
  urban geometry on outdoor thermal comfort within idealized building arrays},
\newblock \bibinfo{journal}{Sustain. Cities Soc.} \bibinfo{volume}{74}
  (\bibinfo{year}{2021}) \bibinfo{pages}{103187}.
\bibitem[{Kong et~al.(2022)Kong, Chen, Middel, Yin, Li, Sun, Zhang, Huang, Liu,
  Zhou, and Ma}]{Kong2022-zw}
\bibinfo{author}{F.~Kong}, \bibinfo{author}{J.~Chen},
  \bibinfo{author}{A.~Middel}, \bibinfo{author}{H.~Yin},
  \bibinfo{author}{M.~Li}, \bibinfo{author}{T.~Sun},
  \bibinfo{author}{N.~Zhang}, \bibinfo{author}{J.~Huang},
  \bibinfo{author}{H.~Liu}, \bibinfo{author}{K.~Zhou}, \bibinfo{author}{J.~Ma},
\newblock \bibinfo{title}{Impact of 3-{D} urban landscape patterns on the
  outdoor thermal environment: A modelling study with {SOLWEIG}},
\newblock \bibinfo{journal}{Comput. Environ. Urban Syst.} \bibinfo{volume}{94}
  (\bibinfo{year}{2022}) \bibinfo{pages}{101773}.
\bibitem[{Li and Wang(2021)}]{Li2021-ye}
\bibinfo{author}{X.~Li}, \bibinfo{author}{G.~Wang},
\newblock \bibinfo{title}{Examining runner's outdoor heat exposure using urban
  microclimate modeling and {GPS} trajectory mining},
\newblock \bibinfo{journal}{Comput. Environ. Urban Syst.} \bibinfo{volume}{89}
  (\bibinfo{year}{2021}) \bibinfo{pages}{101678}.
\bibitem[{Forouzandeh(2021)}]{Forouzandeh2021-fm}
\bibinfo{author}{A.~Forouzandeh},
\newblock \bibinfo{title}{Prediction of surface temperature of building
  surrounding envelopes using holistic microclimate {ENVI}-met model},
\newblock \bibinfo{journal}{Sustain. Cities Soc.} \bibinfo{volume}{70}
  (\bibinfo{year}{2021}) \bibinfo{pages}{102878}.
\bibitem[{Malys et~al.(2015)Malys, Musy, and Inard}]{Malys2015-pi}
\bibinfo{author}{L.~Malys}, \bibinfo{author}{M.~Musy},
  \bibinfo{author}{C.~Inard},
\newblock \bibinfo{title}{Microclimate and building energy consumption: study
  of different coupling methods},
\newblock \bibinfo{journal}{Adv. Build. Energy Res.} \bibinfo{volume}{9}
  (\bibinfo{year}{2015}) \bibinfo{pages}{151--174}.
\bibitem[{Sezer et~al.(2023)Sezer, Yoonus, Zhan, Wang, Hassan, and
  Rahman}]{Sezer2023-yj}
\bibinfo{author}{N.~Sezer}, \bibinfo{author}{H.~Yoonus},
  \bibinfo{author}{D.~Zhan}, \bibinfo{author}{L.~l. Wang},
  \bibinfo{author}{I.~G. Hassan}, \bibinfo{author}{M.~A. Rahman},
\newblock \bibinfo{title}{Urban microclimate and building energy models: A
  review of the latest progress in coupling strategies},
\newblock \bibinfo{journal}{Renew. Sustain. Energy Rev.} \bibinfo{volume}{184}
  (\bibinfo{year}{2023}) \bibinfo{pages}{113577}.
\bibitem[{Kadaverugu et~al.(2019)Kadaverugu, Sharma, Matli, and
  Biniwale}]{Kadaverugu2019-sj}
\bibinfo{author}{R.~Kadaverugu}, \bibinfo{author}{A.~Sharma},
  \bibinfo{author}{C.~Matli}, \bibinfo{author}{R.~Biniwale},
\newblock \bibinfo{title}{High resolution urban air quality modeling by
  coupling {CFD} and mesoscale models: A review},
\newblock \bibinfo{journal}{Asia Pac. J. Atmos. Sci.} \bibinfo{volume}{55}
  (\bibinfo{year}{2019}) \bibinfo{pages}{539--556}.
\bibitem[{Kwak et~al.(2015)Kwak, Baik, Ryu, and Lee}]{Kwak2015-nu}
\bibinfo{author}{K.-H. Kwak}, \bibinfo{author}{J.-J. Baik},
  \bibinfo{author}{Y.-H. Ryu}, \bibinfo{author}{S.-H. Lee},
\newblock \bibinfo{title}{Urban air quality simulation in a high-rise building
  area using a {CFD} model coupled with mesoscale meteorological and
  chemistry-transport models},
\newblock \bibinfo{journal}{Atmos. Environ. (1994)} \bibinfo{volume}{100}
  (\bibinfo{year}{2015}) \bibinfo{pages}{167--177}.
\bibitem[{Blocken et~al.(2012)Blocken, Janssen, and van Hooff}]{Blocken2012-kj}
\bibinfo{author}{B.~Blocken}, \bibinfo{author}{W.~D. Janssen},
  \bibinfo{author}{T.~van Hooff},
\newblock \bibinfo{title}{{CFD} simulation for pedestrian wind comfort and wind
  safety in urban areas: General decision framework and case study for the
  eindhoven university campus},
\newblock \bibinfo{journal}{Environ. Model. Softw.} \bibinfo{volume}{30}
  (\bibinfo{year}{2012}) \bibinfo{pages}{15--34}.
\bibitem[{Blocken et~al.(2016)Blocken, Stathopoulos, and van
  Beeck}]{Blocken2016-wk}
\bibinfo{author}{B.~Blocken}, \bibinfo{author}{T.~Stathopoulos},
  \bibinfo{author}{J.~P. A.~J. van Beeck},
\newblock \bibinfo{title}{Pedestrian-level wind conditions around buildings:
  Review of wind-tunnel and {CFD} techniques and their accuracy for wind
  comfort assessment},
\newblock \bibinfo{journal}{Build. Environ.} \bibinfo{volume}{100}
  (\bibinfo{year}{2016}) \bibinfo{pages}{50--81}.
\bibitem[{Mou et~al.(2017)Mou, He, Zhao, and Chau}]{Mou2017-iq}
\bibinfo{author}{B.~Mou}, \bibinfo{author}{B.-J. He}, \bibinfo{author}{D.-X.
  Zhao}, \bibinfo{author}{K.-W. Chau},
\newblock \bibinfo{title}{Numerical simulation of the effects of building
  dimensional variation on wind pressure distribution},
\newblock \bibinfo{journal}{Eng. Appl. Comput. Fluid Mech.}
  \bibinfo{volume}{11} (\bibinfo{year}{2017}) \bibinfo{pages}{293--309}.
\bibitem[{Stathopoulos and Baskaran(1996)}]{Stathopoulos1996-nk}
\bibinfo{author}{T.~Stathopoulos}, \bibinfo{author}{B.~A. Baskaran},
\newblock \bibinfo{title}{Computer simulation of wind environmental conditions
  around buildings},
\newblock \bibinfo{journal}{Eng. Struct.} \bibinfo{volume}{18}
  (\bibinfo{year}{1996}) \bibinfo{pages}{876--885}.
\bibitem[{Grant et~al.(2002)Grant, Heisler, and Gao}]{Grant2002-vp}
\bibinfo{author}{R.~H. Grant}, \bibinfo{author}{G.~M. Heisler},
  \bibinfo{author}{W.~Gao},
\newblock \bibinfo{title}{Estimation of pedestrian level {UV} exposure under
  trees},
\newblock \bibinfo{journal}{Photochem. Photobiol.} \bibinfo{volume}{75}
  (\bibinfo{year}{2002}) \bibinfo{pages}{369--376}.
\bibitem[{Galal et~al.(2020)Galal, Mahmoud, and Sailor}]{Galal2020-uc}
\bibinfo{author}{O.~M. Galal}, \bibinfo{author}{H.~Mahmoud},
  \bibinfo{author}{D.~Sailor},
\newblock \bibinfo{title}{Impact of evolving building morphology on
  microclimate in a hot arid climate},
\newblock \bibinfo{journal}{Sustain. Cities Soc.} \bibinfo{volume}{54}
  (\bibinfo{year}{2020}) \bibinfo{pages}{102011}.
\bibitem[{Zhang et~al.(2022)Zhang, Li, Wei, and Hu}]{Zhang2022-av}
\bibinfo{author}{J.~Zhang}, \bibinfo{author}{Z.~Li}, \bibinfo{author}{Y.~Wei},
  \bibinfo{author}{D.~Hu},
\newblock \bibinfo{title}{The impact of the building morphology on microclimate
  and thermal comfort-a case study in beijing},
\newblock \bibinfo{journal}{Build. Environ.} \bibinfo{volume}{223}
  (\bibinfo{year}{2022}) \bibinfo{pages}{109469}.
\bibitem[{Hong and Lin(2015)}]{Hong2015-ik}
\bibinfo{author}{B.~Hong}, \bibinfo{author}{B.~Lin},
\newblock \bibinfo{title}{Numerical studies of the outdoor wind environment and
  thermal comfort at pedestrian level in housing blocks with different building
  layout patterns and trees arrangement},
\newblock \bibinfo{journal}{Renew. Energy} \bibinfo{volume}{73}
  (\bibinfo{year}{2015}) \bibinfo{pages}{18--27}.
\bibitem[{Wu et~al.(2019)Wu, Dou, and Chen}]{Wu2019-hp}
\bibinfo{author}{Z.~Wu}, \bibinfo{author}{P.~Dou}, \bibinfo{author}{L.~Chen},
\newblock \bibinfo{title}{Comparative and combinative cooling effects of
  different spatial arrangements of buildings and trees on microclimate},
\newblock \bibinfo{journal}{Sustain. Cities Soc.} \bibinfo{volume}{51}
  (\bibinfo{year}{2019}) \bibinfo{pages}{101711}.
\bibitem[{Morakinyo and Lam(2016)}]{Morakinyo2016-pm}
\bibinfo{author}{T.~E. Morakinyo}, \bibinfo{author}{Y.~F. Lam},
\newblock \bibinfo{title}{Simulation study on the impact of tree-configuration,
  planting pattern and wind condition on street-canyon's micro-climate and
  thermal comfort},
\newblock \bibinfo{journal}{Build. Environ.} \bibinfo{volume}{103}
  (\bibinfo{year}{2016}) \bibinfo{pages}{262--275}.
\bibitem[{Li et~al.(2023)Li, Zeng, Zhao, Wu, Niu, Wang, Gao, and
  Shi}]{Li2023-uq}
\bibinfo{author}{R.~Li}, \bibinfo{author}{F.~Zeng}, \bibinfo{author}{Y.~Zhao},
  \bibinfo{author}{Y.~Wu}, \bibinfo{author}{J.~Niu}, \bibinfo{author}{L.~l.
  Wang}, \bibinfo{author}{N.~Gao}, \bibinfo{author}{X.~Shi},
\newblock \bibinfo{title}{{CFD} simulations of the tree effect on the outdoor
  microclimate by coupling the canopy energy balance model},
\newblock \bibinfo{journal}{Build. Environ.} \bibinfo{volume}{230}
  (\bibinfo{year}{2023}) \bibinfo{pages}{109995}.
\bibitem[{Wang et~al.(2015)Wang, Bakker, de~Groot, Wortche, and
  Leemans}]{Wang2015-nb}
\bibinfo{author}{Y.~Wang}, \bibinfo{author}{F.~Bakker},
  \bibinfo{author}{R.~de~Groot}, \bibinfo{author}{H.~Wortche},
  \bibinfo{author}{R.~Leemans},
\newblock \bibinfo{title}{Effects of urban trees on local outdoor microclimate:
  synthesizing field measurements by numerical modelling},
\newblock \bibinfo{journal}{Urban Ecosyst.} \bibinfo{volume}{18}
  (\bibinfo{year}{2015}) \bibinfo{pages}{1305--1331}.
\bibitem[{Oshio et~al.(2021)Oshio, Kiyono, and Asawa}]{Oshio2021-vr}
\bibinfo{author}{H.~Oshio}, \bibinfo{author}{T.~Kiyono},
  \bibinfo{author}{T.~Asawa},
\newblock \bibinfo{title}{Numerical simulation of the nocturnal cooling effect
  of urban trees considering the leaf area density distribution},
\newblock \bibinfo{journal}{Urban For. Urban Greening} \bibinfo{volume}{66}
  (\bibinfo{year}{2021}) \bibinfo{pages}{127391}.
\bibitem[{Li et~al.(2024)Li, Zhao, Chang, Zeng, Wu, Wang, Niu, Shi, and
  Gao}]{Li2024-zi}
\bibinfo{author}{R.~Li}, \bibinfo{author}{Y.~Zhao}, \bibinfo{author}{M.~Chang},
  \bibinfo{author}{F.~Zeng}, \bibinfo{author}{Y.~Wu}, \bibinfo{author}{L.~l.
  Wang}, \bibinfo{author}{J.~Niu}, \bibinfo{author}{X.~Shi},
  \bibinfo{author}{N.~Gao},
\newblock \bibinfo{title}{Numerical simulation methods of tree effects on
  microclimate: A review},
\newblock \bibinfo{journal}{Renew. Sustain. Energy Rev.} \bibinfo{volume}{205}
  (\bibinfo{year}{2024}) \bibinfo{pages}{114852}.
\bibitem[{He et~al.(2022)He, Lin, Zhang, Tan, Tan, and Wong}]{He2022-fs}
\bibinfo{author}{Y.~He}, \bibinfo{author}{E.~S. Lin},
  \bibinfo{author}{W.~Zhang}, \bibinfo{author}{C.~L. Tan},
  \bibinfo{author}{P.~Y. Tan}, \bibinfo{author}{N.~H. Wong},
\newblock \bibinfo{title}{Local microclimate above shrub and grass in tropical
  city: A case study in singapore},
\newblock \bibinfo{journal}{Urban Clim.} \bibinfo{volume}{43}
  (\bibinfo{year}{2022}) \bibinfo{pages}{101142}.
\bibitem[{Yang et~al.(2013)Yang, Zhao, Bruse, and Meng}]{Yang2013}
\bibinfo{author}{X.~Yang}, \bibinfo{author}{L.~Zhao},
  \bibinfo{author}{M.~Bruse}, \bibinfo{author}{Q.~Meng},
\newblock \bibinfo{title}{{Evaluation of a microclimate model for predicting
  the thermal behavior of different ground surfaces}},
\newblock \bibinfo{journal}{Building and Environment} \bibinfo{volume}{60}
  (\bibinfo{year}{2013}) \bibinfo{pages}{93--104}.
  \DOIprefix\doi{10.1016/j.buildenv.2012.11.008}.
\bibitem[{Wang et~al.(2021)Wang, Liu, and Xu}]{Wang2021-ua}
\bibinfo{author}{X.~Wang}, \bibinfo{author}{P.~Liu}, \bibinfo{author}{G.~Xu},
\newblock \bibinfo{title}{Influence of grass lawns on the summer thermal
  environment and microclimate of heritage sites: a case study of fuling
  mausoleum, china},
\newblock \bibinfo{journal}{Herit. Sci.} \bibinfo{volume}{9}
  (\bibinfo{year}{2021}).
\bibitem[{Robitu et~al.(2006)Robitu, Musy, Inard, and Groleau}]{Robitu2006-lp}
\bibinfo{author}{M.~Robitu}, \bibinfo{author}{M.~Musy},
  \bibinfo{author}{C.~Inard}, \bibinfo{author}{D.~Groleau},
\newblock \bibinfo{title}{Modeling the influence of vegetation and water pond
  on urban microclimate},
\newblock \bibinfo{journal}{Sol. Energy} \bibinfo{volume}{80}
  (\bibinfo{year}{2006}) \bibinfo{pages}{435--447}.
\bibitem[{Du et~al.(2019)Du, Cai, Zhou, Jiang, Jiang, and Xu}]{Du2019-ab}
\bibinfo{author}{H.~Du}, \bibinfo{author}{Y.~Cai}, \bibinfo{author}{F.~Zhou},
  \bibinfo{author}{H.~Jiang}, \bibinfo{author}{W.~Jiang},
  \bibinfo{author}{Y.~Xu},
\newblock \bibinfo{title}{Urban blue-green space planning based on thermal
  environment simulation: A case study of shanghai, china},
\newblock \bibinfo{journal}{Ecol. Indic.} \bibinfo{volume}{106}
  (\bibinfo{year}{2019}) \bibinfo{pages}{105501}.
\bibitem[{Imam~Syafii et~al.(2017)Imam~Syafii, Ichinose, Kumakura, Jusuf,
  Chigusa, and Wong}]{Imam-Syafii2017-to}
\bibinfo{author}{N.~Imam~Syafii}, \bibinfo{author}{M.~Ichinose},
  \bibinfo{author}{E.~Kumakura}, \bibinfo{author}{S.~K. Jusuf},
  \bibinfo{author}{K.~Chigusa}, \bibinfo{author}{N.~H. Wong},
\newblock \bibinfo{title}{Thermal environment assessment around bodies of water
  in urban canyons: A scale model study},
\newblock \bibinfo{journal}{Sustain. Cities Soc.} \bibinfo{volume}{34}
  (\bibinfo{year}{2017}) \bibinfo{pages}{79--89}.
\bibitem[{Sun et~al.(2012)Sun, Nottrott, and Kleissl}]{Sun2012-fw}
\bibinfo{author}{L.~Sun}, \bibinfo{author}{A.~Nottrott},
  \bibinfo{author}{J.~Kleissl},
\newblock \bibinfo{title}{Effect of hilly urban morphology on dispersion in the
  urban boundary layer},
\newblock \bibinfo{journal}{Build. Environ.} \bibinfo{volume}{48}
  (\bibinfo{year}{2012}) \bibinfo{pages}{195--205}.
\bibitem[{Abd-Alhamid et~al.(2023)Abd-Alhamid, Kent, and
  Wu}]{Abd-Alhamid2023-on}
\bibinfo{author}{F.~Abd-Alhamid}, \bibinfo{author}{M.~Kent},
  \bibinfo{author}{Y.~Wu},
\newblock \bibinfo{title}{Quantifying window view quality: A review on view
  perception assessment and representation methods},
\newblock \bibinfo{journal}{Build. Environ.} \bibinfo{volume}{227}
  (\bibinfo{year}{2023}) \bibinfo{pages}{109742}.
\bibitem[{Yu et~al.(2016)Yu, Yu, Song, Wu, Zhou, Huang, Wu, Zhao, and
  Mao}]{Yu2016-gb}
\bibinfo{author}{S.~Yu}, \bibinfo{author}{B.~Yu}, \bibinfo{author}{W.~Song},
  \bibinfo{author}{B.~Wu}, \bibinfo{author}{J.~Zhou},
  \bibinfo{author}{Y.~Huang}, \bibinfo{author}{J.~Wu},
  \bibinfo{author}{F.~Zhao}, \bibinfo{author}{W.~Mao},
\newblock \bibinfo{title}{View-based greenery: A three-dimensional assessment
  of city buildings’ green visibility using floor green view index},
\newblock \bibinfo{journal}{Landsc. Urban Plan.} \bibinfo{volume}{152}
  (\bibinfo{year}{2016}) \bibinfo{pages}{13--26}.
\bibitem[{Yi and Kim(2017)}]{Yi2017-vf}
\bibinfo{author}{Y.~K. Yi}, \bibinfo{author}{H.~Kim},
\newblock \bibinfo{title}{Universal visible sky factor: A method for
  calculating the three-dimensional visible sky ratio},
\newblock \bibinfo{journal}{Build. Environ.} \bibinfo{volume}{123}
  (\bibinfo{year}{2017}) \bibinfo{pages}{390--403}.
\bibitem[{Fujiwara et~al.(2022)Fujiwara, Asawa, and Kiyono}]{Fujiwara2022-yb}
\bibinfo{author}{K.~Fujiwara}, \bibinfo{author}{T.~Asawa},
  \bibinfo{author}{T.~Kiyono},
\newblock \bibinfo{title}{Multi-objective optimization for tree arrangement in
  urban open space considering thermal radiant environment and scenery},
\newblock \bibinfo{journal}{AIJ J. Technol. Des.} \bibinfo{volume}{28}
  (\bibinfo{year}{2022}) \bibinfo{pages}{320--325}.
\bibitem[{Labib et~al.(2021)Labib, Huck, and Lindley}]{Labib2021-xj}
\bibinfo{author}{S.~M. Labib}, \bibinfo{author}{J.~J. Huck},
  \bibinfo{author}{S.~Lindley},
\newblock \bibinfo{title}{Modelling and mapping eye-level greenness visibility
  exposure using multi-source data at high spatial resolutions},
\newblock \bibinfo{journal}{Sci. Total Environ.} \bibinfo{volume}{755}
  (\bibinfo{year}{2021}) \bibinfo{pages}{143050}.
\bibitem[{Biljecki et~al.(2023)Biljecki, Zhao, Liang, and
  Hou}]{Biljecki2023-rx}
\bibinfo{author}{F.~Biljecki}, \bibinfo{author}{T.~Zhao},
  \bibinfo{author}{X.~Liang}, \bibinfo{author}{Y.~Hou},
\newblock \bibinfo{title}{Sensitivity of measuring the urban form and greenery
  using street-level imagery: A comparative study of approaches and visual
  perspectives},
\newblock \bibinfo{journal}{Int. J. Appl. Earth Obs. Geoinf.}
  \bibinfo{volume}{122} (\bibinfo{year}{2023}) \bibinfo{pages}{103385}.
\bibitem[{Ki and Lee(2021)}]{Ki2021-qs}
\bibinfo{author}{D.~Ki}, \bibinfo{author}{S.~Lee},
\newblock \bibinfo{title}{Analyzing the effects of green view index of
  neighborhood streets on walking time using google street view and deep
  learning},
\newblock \bibinfo{journal}{Landsc. Urban Plan.} \bibinfo{volume}{205}
  (\bibinfo{year}{2021}) \bibinfo{pages}{103920}.
\bibitem[{Zhang and Zeng(2024)}]{Zhang2024-fj}
\bibinfo{author}{W.~Zhang}, \bibinfo{author}{H.~Zeng},
\newblock \bibinfo{title}{Spatial differentiation characteristics and
  influencing factors of the green view index in urban areas based on street
  view images: A case study of futian district, shenzhen, china},
\newblock \bibinfo{journal}{Urban For. Urban Greening} \bibinfo{volume}{93}
  (\bibinfo{year}{2024}) \bibinfo{pages}{128219}.
\bibitem[{Liang et~al.(2023)Liang, Zhao, and Biljecki}]{Liang2023-cy}
\bibinfo{author}{X.~Liang}, \bibinfo{author}{T.~Zhao},
  \bibinfo{author}{F.~Biljecki},
\newblock \bibinfo{title}{Revealing spatio-temporal evolution of urban visual
  environments with street view imagery},
\newblock \bibinfo{journal}{Landsc. Urban Plan.} \bibinfo{volume}{237}
  (\bibinfo{year}{2023}) \bibinfo{pages}{104802}.
\bibitem[{Ito et~al.(2025)Ito, Zhu, Abdelrahman, Liang, Fan, Hou, Zhao, Ma,
  Fujiwara, Ouyang, Quintana, and Biljecki}]{Ito2025-mf}
\bibinfo{author}{K.~Ito}, \bibinfo{author}{Y.~Zhu},
  \bibinfo{author}{M.~Abdelrahman}, \bibinfo{author}{X.~Liang},
  \bibinfo{author}{Z.~Fan}, \bibinfo{author}{Y.~Hou},
  \bibinfo{author}{T.~Zhao}, \bibinfo{author}{R.~Ma},
  \bibinfo{author}{K.~Fujiwara}, \bibinfo{author}{J.~Ouyang},
  \bibinfo{author}{M.~Quintana}, \bibinfo{author}{F.~Biljecki},
\newblock \bibinfo{title}{{ZenSVI}: An open-source software for the integrated
  acquisition, processing and analysis of street view imagery towards scalable
  urban science},
\newblock \bibinfo{journal}{Comput. Environ. Urban Syst.} \bibinfo{volume}{119}
  (\bibinfo{year}{2025}) \bibinfo{pages}{102283}.
\bibitem[{Lu et~al.(2018)Lu, Sarkar, and Xiao}]{Lu2018-bu}
\bibinfo{author}{Y.~Lu}, \bibinfo{author}{C.~Sarkar},
  \bibinfo{author}{Y.~Xiao},
\newblock \bibinfo{title}{The effect of street-level greenery on walking
  behavior: Evidence from hong kong},
\newblock \bibinfo{journal}{Soc. Sci. Med.} \bibinfo{volume}{208}
  (\bibinfo{year}{2018}) \bibinfo{pages}{41--49}.
\bibitem[{Ito and Biljecki(2021)}]{Ito2021-mk}
\bibinfo{author}{K.~Ito}, \bibinfo{author}{F.~Biljecki},
\newblock \bibinfo{title}{Assessing bikeability with street view imagery and
  computer vision},
\newblock \bibinfo{journal}{Transp. Res. Part C: Emerg. Technol.}
  \bibinfo{volume}{132} (\bibinfo{year}{2021}) \bibinfo{pages}{103371}.
\bibitem[{Wang et~al.(2020)Wang, Yang, Yao, Bloom, Feng, Yuan, Zhang, Liu, Wu,
  Lu, Baranyi, Wu, Liu, and Dong}]{Wang2020-kx}
\bibinfo{author}{R.~Wang}, \bibinfo{author}{B.~Yang}, \bibinfo{author}{Y.~Yao},
  \bibinfo{author}{M.~S. Bloom}, \bibinfo{author}{Z.~Feng},
  \bibinfo{author}{Y.~Yuan}, \bibinfo{author}{J.~Zhang},
  \bibinfo{author}{P.~Liu}, \bibinfo{author}{W.~Wu}, \bibinfo{author}{Y.~Lu},
  \bibinfo{author}{G.~Baranyi}, \bibinfo{author}{R.~Wu},
  \bibinfo{author}{Y.~Liu}, \bibinfo{author}{G.~Dong},
\newblock \bibinfo{title}{Residential greenness, air pollution and
  psychological well-being among urban residents in guangzhou, china},
\newblock \bibinfo{journal}{Sci. Total Environ.} \bibinfo{volume}{711}
  (\bibinfo{year}{2020}) \bibinfo{pages}{134843}.
\bibitem[{Wu et~al.(2022)Wu, Du, Li, Liu, and Ye}]{Wu2022-ve}
\bibinfo{author}{C.~Wu}, \bibinfo{author}{Y.~Du}, \bibinfo{author}{S.~Li},
  \bibinfo{author}{P.~Liu}, \bibinfo{author}{X.~Ye},
\newblock \bibinfo{title}{Does visual contact with green space impact housing
  prices? an integrated approach of machine learning and hedonic modeling based
  on the perception of green space},
\newblock \bibinfo{journal}{Land Use Policy} \bibinfo{volume}{115}
  (\bibinfo{year}{2022}) \bibinfo{pages}{106048}.
\bibitem[{Yang et~al.(2021)Yang, Rong, Kang, Zhang, and Chegut}]{Yang2021-xu}
\bibinfo{author}{J.~Yang}, \bibinfo{author}{H.~Rong},
  \bibinfo{author}{Y.~Kang}, \bibinfo{author}{F.~Zhang},
  \bibinfo{author}{A.~Chegut},
\newblock \bibinfo{title}{The financial impact of street-level greenery on new
  york commercial buildings},
\newblock \bibinfo{journal}{Landsc. Urban Plan.} \bibinfo{volume}{214}
  (\bibinfo{year}{2021}) \bibinfo{pages}{104162}.
\bibitem[{Luo et~al.(2022)Luo, Zhao, Cao, and Biljecki}]{Luo2022-la}
\bibinfo{author}{J.~Luo}, \bibinfo{author}{T.~Zhao}, \bibinfo{author}{L.~Cao},
  \bibinfo{author}{F.~Biljecki},
\newblock \bibinfo{title}{Water view imagery: Perception and evaluation of
  urban waterscapes worldwide},
\newblock \bibinfo{journal}{Ecol. Indic.} \bibinfo{volume}{145}
  (\bibinfo{year}{2022}) \bibinfo{pages}{109615}.
\bibitem[{Wang et~al.(2024)Wang, He, Zhai, Wang, and Zhao}]{Wang2024-gg}
\bibinfo{author}{Y.~Wang}, \bibinfo{author}{Z.~He}, \bibinfo{author}{W.~Zhai},
  \bibinfo{author}{S.~Wang}, \bibinfo{author}{C.~Zhao},
\newblock \bibinfo{title}{How do the {3D} urban morphological characteristics
  spatiotemporally affect the urban thermal environment? a case study of san
  antonio},
\newblock \bibinfo{journal}{Build. Environ.} \bibinfo{volume}{261}
  (\bibinfo{year}{2024}) \bibinfo{pages}{111738}.
\bibitem[{Peters et~al.(2022)Peters, Dukai, Vitalis, van Liempt, and
  Stoter}]{Peters2022-qq}
\bibinfo{author}{R.~Peters}, \bibinfo{author}{B.~Dukai},
  \bibinfo{author}{S.~Vitalis}, \bibinfo{author}{J.~van Liempt},
  \bibinfo{author}{J.~Stoter},
\newblock \bibinfo{title}{Automated {3D} reconstruction of {LoD2} and {LoD1}
  models for all 10 million buildings of the netherlands},
\newblock \bibinfo{journal}{Photogramm. Eng. Remote Sensing}
  \bibinfo{volume}{88} (\bibinfo{year}{2022}) \bibinfo{pages}{165--170}.
\bibitem[{Rong et~al.(2020)Rong, Zhang, Zheng, Hu, Peng, and
  Feng}]{Rong2020-wk}
\bibinfo{author}{Y.~Rong}, \bibinfo{author}{T.~Zhang},
  \bibinfo{author}{Y.~Zheng}, \bibinfo{author}{C.~Hu},
  \bibinfo{author}{L.~Peng}, \bibinfo{author}{P.~Feng},
\newblock \bibinfo{title}{Three-dimensional urban flood inundation simulation
  based on digital aerial photogrammetry},
\newblock \bibinfo{journal}{J. Hydrol. (Amst.)} \bibinfo{volume}{584}
  (\bibinfo{year}{2020}) \bibinfo{pages}{124308}.
\bibitem[{Matsuoka et~al.(2024)Matsuoka, Takemoto, Takahashi, Inazawa, and
  Sogo}]{Matsuoka2024-ij}
\bibinfo{author}{R.~Matsuoka}, \bibinfo{author}{T.~Takemoto},
  \bibinfo{author}{G.~Takahashi}, \bibinfo{author}{T.~Inazawa},
  \bibinfo{author}{S.~Sogo},
\newblock \bibinfo{title}{Estimation of photovoltaic potential of urban
  buildings considering a solar panel arrangement using a {3D} city model},
\newblock \bibinfo{journal}{ISPRS Ann. Photogramm. Remote Sens. Spat. Inf.
  Sci.} \bibinfo{volume}{X-4/W4-2024} (\bibinfo{year}{2024})
  \bibinfo{pages}{123--130}.
\bibitem[{Zhang and Ludwig(2025)}]{Zhang2025-bz}
\bibinfo{author}{X.~Zhang}, \bibinfo{author}{F.~Ludwig},
\newblock \bibinfo{title}{Shade for pedestrians: A novel approach to calculate
  the spatio-temporal shade benefits of street trees considering pedestrian
  flow},
\newblock \bibinfo{journal}{Build. Environ.} \bibinfo{volume}{272}
  (\bibinfo{year}{2025}) \bibinfo{pages}{112662}.
\bibitem[{Nouvel et~al.(2015)Nouvel, Mastrucci, Leopold, Baume, Coors, and
  Eicker}]{Nouvel2015-ph}
\bibinfo{author}{R.~Nouvel}, \bibinfo{author}{A.~Mastrucci},
  \bibinfo{author}{U.~Leopold}, \bibinfo{author}{O.~Baume},
  \bibinfo{author}{V.~Coors}, \bibinfo{author}{U.~Eicker},
\newblock \bibinfo{title}{Combining {GIS}-based statistical and engineering
  urban heat consumption models: Towards a new framework for multi-scale policy
  support},
\newblock \bibinfo{journal}{Energy Build.} \bibinfo{volume}{107}
  (\bibinfo{year}{2015}) \bibinfo{pages}{204--212}.
\bibitem[{Harter et~al.(2023)Harter, Willenborg, Lang, and
  Kolbe}]{Harter2023-ar}
\bibinfo{author}{H.~Harter}, \bibinfo{author}{B.~Willenborg},
  \bibinfo{author}{W.~Lang}, \bibinfo{author}{T.~H. Kolbe},
\newblock \bibinfo{title}{Climate-neutral municipal building stock - life cycle
  assessment of large residential building stocks based on semantic {3D} city
  models},
\newblock \bibinfo{journal}{Energy Build.} \bibinfo{volume}{292}
  (\bibinfo{year}{2023}) \bibinfo{pages}{113141}.
\bibitem[{Malhotra et~al.(2022)Malhotra, Shamovich, Frisch, and van
  Treeck}]{Malhotra2022-yv}
\bibinfo{author}{A.~Malhotra}, \bibinfo{author}{M.~Shamovich},
  \bibinfo{author}{J.~Frisch}, \bibinfo{author}{C.~van Treeck},
\newblock \bibinfo{title}{Urban energy simulations using open {CityGML} models:
  A comparative analysis},
\newblock \bibinfo{journal}{Energy Build.} \bibinfo{volume}{255}
  (\bibinfo{year}{2022}) \bibinfo{pages}{111658}.
\bibitem[{Farr and Kobrick(2000)}]{Farr2000-dl}
\bibinfo{author}{T.~G. Farr}, \bibinfo{author}{M.~Kobrick},
\newblock \bibinfo{title}{Shuttle radar topography mission produces a wealth of
  data},
\newblock \bibinfo{journal}{Eos, Transactions American Geophysical Union}
  \bibinfo{volume}{81} (\bibinfo{year}{2000}) \bibinfo{pages}{583--585}.
\bibitem[{Ding et~al.(2024)Ding, Zhao, Strebel, Fan, Ge, and
  Carmeliet}]{Ding2024-kj}
\bibinfo{author}{X.~Ding}, \bibinfo{author}{Y.~Zhao},
  \bibinfo{author}{D.~Strebel}, \bibinfo{author}{Y.~Fan},
  \bibinfo{author}{J.~Ge}, \bibinfo{author}{J.~Carmeliet},
\newblock \bibinfo{title}{A {WRF}-{UCM}-{SOLWEIG} framework for mapping thermal
  comfort and quantifying urban climate drivers: Advancing spatial and temporal
  resolutions at city scale},
\newblock \bibinfo{journal}{Sustain. Cities Soc.} \bibinfo{volume}{112}
  (\bibinfo{year}{2024}) \bibinfo{pages}{105628}.
\bibitem[{Yap et~al.(2023)Yap, Stouffs, and Biljecki}]{Yap2023-my}
\bibinfo{author}{W.~Yap}, \bibinfo{author}{R.~Stouffs},
  \bibinfo{author}{F.~Biljecki},
\newblock \bibinfo{title}{Urbanity: automated modelling and analysis of
  multidimensional networks in cities},
\newblock \bibinfo{journal}{npj Urban Sustainability} \bibinfo{volume}{3}
  (\bibinfo{year}{2023}) \bibinfo{pages}{1--11}.
\bibitem[{Gholami et~al.(2024)Gholami, Middel, Torreggiani, Tassinari, and
  Barbaresi}]{Gholami2024-xs}
\bibinfo{author}{M.~Gholami}, \bibinfo{author}{A.~Middel},
  \bibinfo{author}{D.~Torreggiani}, \bibinfo{author}{P.~Tassinari},
  \bibinfo{author}{A.~Barbaresi},
\newblock \bibinfo{title}{A hybrid python approach to assess microscale human
  thermal stress in urban environments},
\newblock \bibinfo{journal}{Build. Environ.} \bibinfo{volume}{248}
  (\bibinfo{year}{2024}) \bibinfo{pages}{111054}.
\bibitem[{Balakrishnan and Cassottana(2022)}]{Balakrishnan2022-pz}
\bibinfo{author}{S.~Balakrishnan}, \bibinfo{author}{B.~Cassottana},
\newblock \bibinfo{title}{{InfraRisk}: An open-source simulation platform for
  resilience analysis in interconnected power–water–transport networks},
\newblock \bibinfo{journal}{Sustain. Cities Soc.} \bibinfo{volume}{83}
  (\bibinfo{year}{2022}) \bibinfo{pages}{103963}.
\bibitem[{Boeing(2017)}]{Boeing2017-es}
\bibinfo{author}{G.~Boeing},
\newblock \bibinfo{title}{{OSMnx}: New methods for acquiring, constructing,
  analyzing, and visualizing complex street networks},
\newblock \bibinfo{journal}{Comput. Environ. Urban Syst.} \bibinfo{volume}{65}
  (\bibinfo{year}{2017}) \bibinfo{pages}{126--139}.
\bibitem[{Martin et~al.(2023)Martin, Hong, Wiedemann, Bucher, and
  Raubal}]{Martin2023-mi}
\bibinfo{author}{H.~Martin}, \bibinfo{author}{Y.~Hong},
  \bibinfo{author}{N.~Wiedemann}, \bibinfo{author}{D.~Bucher},
  \bibinfo{author}{M.~Raubal},
\newblock \bibinfo{title}{Trackintel: An open-source python library for human
  mobility analysis},
\newblock \bibinfo{journal}{Comput. Environ. Urban Syst.} \bibinfo{volume}{101}
  (\bibinfo{year}{2023}) \bibinfo{pages}{101938}.
\bibitem[{Huttner(2012)}]{huttner2012further}
\bibinfo{author}{S.~Huttner}, \bibinfo{title}{Further development and
  application of the 3D microclimate simulation ENVI-met}, Ph.D. thesis, Mainz,
  Univ., Diss., 2012, \bibinfo{year}{2012}.
\bibitem[{Liu et~al.(2023)Liu, Zhang, Shao, and Wu}]{Liu2023-vx}
\bibinfo{author}{Z.~Liu}, \bibinfo{author}{S.~Zhang},
  \bibinfo{author}{X.~Shao}, \bibinfo{author}{Z.~Wu},
\newblock \bibinfo{title}{Accurate and efficient urban wind prediction at
  city-scale with memory-scalable graph neural network},
\newblock \bibinfo{journal}{Sustain. Cities Soc.} \bibinfo{volume}{99}
  (\bibinfo{year}{2023}) \bibinfo{pages}{104935}.
\bibitem[{Maronga et~al.(2020)Maronga, Banzhaf, Burmeister, Esch, Forkel,
  Fröhlich, Fuka, Gehrke, Geletič, Giersch, Gronemeier, Groß, Heldens,
  Hellsten, Hoffmann, Inagaki, Kadasch, Kanani-Sühring, Ketelsen, Khan,
  Knigge, Knoop, Krč, Kurppa, Maamari, Matzarakis, Mauder, Pallasch, Pavlik,
  Pfafferott, Resler, Rissmann, Russo, Salim, Schrempf, Schwenkel, Seckmeyer,
  Schubert, Sühring, von Tils, Vollmer, Ward, Witha, Wurps, Zeidler, and
  Raasch}]{Maronga2020-bn}
\bibinfo{author}{B.~Maronga}, \bibinfo{author}{S.~Banzhaf},
  \bibinfo{author}{C.~Burmeister}, \bibinfo{author}{T.~Esch},
  \bibinfo{author}{R.~Forkel}, \bibinfo{author}{D.~Fröhlich},
  \bibinfo{author}{V.~Fuka}, \bibinfo{author}{K.~F. Gehrke},
  \bibinfo{author}{J.~Geletič}, \bibinfo{author}{S.~Giersch},
  \bibinfo{author}{T.~Gronemeier}, \bibinfo{author}{G.~Groß},
  \bibinfo{author}{W.~Heldens}, \bibinfo{author}{A.~Hellsten},
  \bibinfo{author}{F.~Hoffmann}, \bibinfo{author}{A.~Inagaki},
  \bibinfo{author}{E.~Kadasch}, \bibinfo{author}{F.~Kanani-Sühring},
  \bibinfo{author}{K.~Ketelsen}, \bibinfo{author}{B.~A. Khan},
  \bibinfo{author}{C.~Knigge}, \bibinfo{author}{H.~Knoop},
  \bibinfo{author}{P.~Krč}, \bibinfo{author}{M.~Kurppa},
  \bibinfo{author}{H.~Maamari}, \bibinfo{author}{A.~Matzarakis},
  \bibinfo{author}{M.~Mauder}, \bibinfo{author}{M.~Pallasch},
  \bibinfo{author}{D.~Pavlik}, \bibinfo{author}{J.~Pfafferott},
  \bibinfo{author}{J.~Resler}, \bibinfo{author}{S.~Rissmann},
  \bibinfo{author}{E.~Russo}, \bibinfo{author}{M.~Salim},
  \bibinfo{author}{M.~Schrempf}, \bibinfo{author}{J.~Schwenkel},
  \bibinfo{author}{G.~Seckmeyer}, \bibinfo{author}{S.~Schubert},
  \bibinfo{author}{M.~Sühring}, \bibinfo{author}{R.~von Tils},
  \bibinfo{author}{L.~Vollmer}, \bibinfo{author}{S.~Ward},
  \bibinfo{author}{B.~Witha}, \bibinfo{author}{H.~Wurps},
  \bibinfo{author}{J.~Zeidler}, \bibinfo{author}{S.~Raasch},
\newblock \bibinfo{title}{Overview of the {PALM} model system 6.0},
\newblock \bibinfo{journal}{Geosci. Model Dev.} \bibinfo{volume}{13}
  (\bibinfo{year}{2020}) \bibinfo{pages}{1335--1372}.
\bibitem[{Milojevic-Dupont et~al.(2023)Milojevic-Dupont, Wagner, Nachtigall,
  Hu, Brüser, Zumwald, Biljecki, Heeren, Kaack, Pichler, and
  Creutzig}]{Milojevic-Dupont2023-bw}
\bibinfo{author}{N.~Milojevic-Dupont}, \bibinfo{author}{F.~Wagner},
  \bibinfo{author}{F.~Nachtigall}, \bibinfo{author}{J.~Hu},
  \bibinfo{author}{G.~B. Brüser}, \bibinfo{author}{M.~Zumwald},
  \bibinfo{author}{F.~Biljecki}, \bibinfo{author}{N.~Heeren},
  \bibinfo{author}{L.~H. Kaack}, \bibinfo{author}{P.-P. Pichler},
  \bibinfo{author}{F.~Creutzig},
\newblock \bibinfo{title}{{EUBUCCO} {v0}.1: European building stock
  characteristics in a common and open database for {200+} million individual
  buildings},
\newblock \bibinfo{journal}{Sci Data} \bibinfo{volume}{10}
  (\bibinfo{year}{2023}) \bibinfo{pages}{147}.
\bibitem[{{Microsoft}(2024)}]{microsoft2024}
\bibinfo{author}{{Microsoft}}, \bibinfo{title}{Microsoft building footprints},
  \bibinfo{year}{2024}. \URLprefix
  \url{https://github.com/microsoft/GlobalMLBuildingFootprints/},
  \bibinfo{note}{accessed: 2024-11-14}.
\bibitem[{Herfort et~al.(2023)Herfort, Lautenbach, Porto~de Albuquerque,
  Anderson, and Zipf}]{Herfort2023-gj}
\bibinfo{author}{B.~Herfort}, \bibinfo{author}{S.~Lautenbach},
  \bibinfo{author}{J.~Porto~de Albuquerque}, \bibinfo{author}{J.~Anderson},
  \bibinfo{author}{A.~Zipf},
\newblock \bibinfo{title}{A spatio-temporal analysis investigating completeness
  and inequalities of global urban building data in {OpenStreetMap}},
\newblock \bibinfo{journal}{Nat. Commun.} \bibinfo{volume}{14}
  (\bibinfo{year}{2023}) \bibinfo{pages}{3985}.
\bibitem[{Biljecki et~al.(2023)Biljecki, Chow, and Lee}]{2023_bae_osm_qa}
\bibinfo{author}{F.~Biljecki}, \bibinfo{author}{Y.~S. Chow},
  \bibinfo{author}{K.~Lee},
\newblock \bibinfo{title}{{Quality of crowdsourced geospatial building
  information: A global assessment of OpenStreetMap attributes}},
\newblock \bibinfo{journal}{Building and Environment} \bibinfo{volume}{237}
  (\bibinfo{year}{2023}) \bibinfo{pages}{110295}.
\bibitem[{Van~Tricht et~al.(2023)Van~Tricht, Degerickx, Gilliams, Zanaga,
  Battude, Grosu, Brombacher, Lesiv, Bayas, Karanam, Fritz, Becker-Reshef,
  Franch, Mollà-Bononad, Boogaard, Pratihast, Koetz, and
  Szantoi}]{Van-Tricht2023-jd}
\bibinfo{author}{K.~Van~Tricht}, \bibinfo{author}{J.~Degerickx},
  \bibinfo{author}{S.~Gilliams}, \bibinfo{author}{D.~Zanaga},
  \bibinfo{author}{M.~Battude}, \bibinfo{author}{A.~Grosu},
  \bibinfo{author}{J.~Brombacher}, \bibinfo{author}{M.~Lesiv},
  \bibinfo{author}{J.~C.~L. Bayas}, \bibinfo{author}{S.~Karanam},
  \bibinfo{author}{S.~Fritz}, \bibinfo{author}{I.~Becker-Reshef},
  \bibinfo{author}{B.~Franch}, \bibinfo{author}{B.~Mollà-Bononad},
  \bibinfo{author}{H.~Boogaard}, \bibinfo{author}{A.~K. Pratihast},
  \bibinfo{author}{B.~Koetz}, \bibinfo{author}{Z.~Szantoi},
\newblock \bibinfo{title}{{WorldCereal}: a dynamic open-source system for
  global-scale, seasonal, and reproducible crop and irrigation mapping},
\newblock \bibinfo{journal}{Earth Syst. Sci. Data} \bibinfo{volume}{15}
  (\bibinfo{year}{2023}) \bibinfo{pages}{5491--5515}.
\bibitem[{Shimada et~al.(2014)Shimada, Itoh, Motooka, Watanabe, Shiraishi,
  Thapa, and Lucas}]{Shimada2014-xw}
\bibinfo{author}{M.~Shimada}, \bibinfo{author}{T.~Itoh},
  \bibinfo{author}{T.~Motooka}, \bibinfo{author}{M.~Watanabe},
  \bibinfo{author}{T.~Shiraishi}, \bibinfo{author}{R.~Thapa},
  \bibinfo{author}{R.~Lucas},
\newblock \bibinfo{title}{New global forest/non-forest maps from {ALOS}
  {PALSAR} data ({2007–2010})},
\newblock \bibinfo{journal}{Remote Sens. Environ.} \bibinfo{volume}{155}
  (\bibinfo{year}{2014}) \bibinfo{pages}{13--31}.
\bibitem[{{GLIMS Consortium}(2005)}]{GLIMS2005}
\bibinfo{author}{{GLIMS Consortium}}, \bibinfo{title}{Glims glacier database},
  \bibinfo{year}{2005}. \URLprefix \url{https://doi.org/10.7265/N5V98602},
  \bibinfo{note}{nSIDC-0272, Version 1}.
\bibitem[{Raup et~al.(2007)Raup, Racoviteanu, Khalsa, Helm, Armstrong, and
  Arnaud}]{Raup2007}
\bibinfo{author}{B.~H. Raup}, \bibinfo{author}{A.~Racoviteanu},
  \bibinfo{author}{S.~J. Khalsa}, \bibinfo{author}{C.~Helm},
  \bibinfo{author}{R.~Armstrong}, \bibinfo{author}{Y.~Arnaud},
\newblock \bibinfo{title}{The glims geospatial glacier database: A new tool for
  studying glacier change},
\newblock \bibinfo{journal}{Global and Planetary Change} \bibinfo{volume}{56}
  (\bibinfo{year}{2007}).
\bibitem[{Karra et~al.(2021)Karra, Kontgis, Statman-Weil, Mazzariello, Mathis,
  and Brumby}]{Karra2021-tc}
\bibinfo{author}{K.~Karra}, \bibinfo{author}{C.~Kontgis},
  \bibinfo{author}{Z.~Statman-Weil}, \bibinfo{author}{J.~C. Mazzariello},
  \bibinfo{author}{M.~Mathis}, \bibinfo{author}{S.~P. Brumby},
\newblock \bibinfo{title}{Global land use / land cover with sentinel 2 and deep
  learning},
\newblock in: \bibinfo{booktitle}{2021 IEEE International Geoscience and Remote
  Sensing Symposium IGARSS}, \bibinfo{publisher}{IEEE}, \bibinfo{year}{2021},
  pp. \bibinfo{pages}{4704--4707}.
\bibitem[{Brown et~al.(2022)Brown, Brumby, Guzder-Williams, Birch, Hyde,
  Mazzariello, Czerwinski, Pasquarella, Haertel, Ilyushchenko, Schwehr, Weisse,
  Stolle, Hanson, Guinan, Moore, and Tait}]{Brown2022-ua}
\bibinfo{author}{C.~F. Brown}, \bibinfo{author}{S.~P. Brumby},
  \bibinfo{author}{B.~Guzder-Williams}, \bibinfo{author}{T.~Birch},
  \bibinfo{author}{S.~B. Hyde}, \bibinfo{author}{J.~Mazzariello},
  \bibinfo{author}{W.~Czerwinski}, \bibinfo{author}{V.~J. Pasquarella},
  \bibinfo{author}{R.~Haertel}, \bibinfo{author}{S.~Ilyushchenko},
  \bibinfo{author}{K.~Schwehr}, \bibinfo{author}{M.~Weisse},
  \bibinfo{author}{F.~Stolle}, \bibinfo{author}{C.~Hanson},
  \bibinfo{author}{O.~Guinan}, \bibinfo{author}{R.~Moore},
  \bibinfo{author}{A.~M. Tait},
\newblock \bibinfo{title}{Dynamic world, near real-time global 10 m land use
  land cover mapping},
\newblock \bibinfo{journal}{Sci. Data} \bibinfo{volume}{9}
  (\bibinfo{year}{2022}) \bibinfo{pages}{251}.
\bibitem[{Yokoya et~al.(2024)Yokoya, Xia, and Broni-Bediako}]{Yokoya2024-me}
\bibinfo{author}{N.~Yokoya}, \bibinfo{author}{J.~Xia},
  \bibinfo{author}{C.~Broni-Bediako},
\newblock \bibinfo{title}{Submeter-level land cover mapping of japan},
\newblock \bibinfo{journal}{Int. J. Appl. Earth Obs. Geoinf.}
  \bibinfo{volume}{127} (\bibinfo{year}{2024}) \bibinfo{pages}{103660}.
\bibitem[{Zhang et~al.(2022)Zhang, Chen, Myint, Zhou, Hay, Vukomanovic, and
  Meentemeyer}]{Zhang2022-oa}
\bibinfo{author}{Y.~Zhang}, \bibinfo{author}{G.~Chen}, \bibinfo{author}{S.~W.
  Myint}, \bibinfo{author}{Y.~Zhou}, \bibinfo{author}{G.~J. Hay},
  \bibinfo{author}{J.~Vukomanovic}, \bibinfo{author}{R.~K. Meentemeyer},
\newblock \bibinfo{title}{{UrbanWatch}: A 1-meter resolution land cover and
  land use database for 22 major cities in the united states},
\newblock \bibinfo{journal}{Remote Sens. Environ.} \bibinfo{volume}{278}
  (\bibinfo{year}{2022}) \bibinfo{pages}{113106}.
\bibitem[{Potapov et~al.(2021)Potapov, Li, Hernandez-Serna, Tyukavina, Hansen,
  Kommareddy, Pickens, Turubanova, Tang, Silva, Armston, Dubayah, Blair, and
  Hofton}]{Potapov2021-ff}
\bibinfo{author}{P.~Potapov}, \bibinfo{author}{X.~Li},
  \bibinfo{author}{A.~Hernandez-Serna}, \bibinfo{author}{A.~Tyukavina},
  \bibinfo{author}{M.~C. Hansen}, \bibinfo{author}{A.~Kommareddy},
  \bibinfo{author}{A.~Pickens}, \bibinfo{author}{S.~Turubanova},
  \bibinfo{author}{H.~Tang}, \bibinfo{author}{C.~E. Silva},
  \bibinfo{author}{J.~Armston}, \bibinfo{author}{R.~Dubayah},
  \bibinfo{author}{J.~B. Blair}, \bibinfo{author}{M.~Hofton},
\newblock \bibinfo{title}{Mapping global forest canopy height through
  integration of {GEDI} and landsat data},
\newblock \bibinfo{journal}{Remote Sens. Environ.} \bibinfo{volume}{253}
  (\bibinfo{year}{2021}) \bibinfo{pages}{112165}.
\bibitem[{Simard et~al.(2019)Simard, Fatoyinbo, Smetanka, Rivera-Monroy,
  Castañeda-Moya, Thomas, and Van~der Stocken}]{Simard2019-kw}
\bibinfo{author}{M.~Simard}, \bibinfo{author}{L.~Fatoyinbo},
  \bibinfo{author}{C.~Smetanka}, \bibinfo{author}{V.~H. Rivera-Monroy},
  \bibinfo{author}{E.~Castañeda-Moya}, \bibinfo{author}{N.~Thomas},
  \bibinfo{author}{T.~Van~der Stocken},
\newblock \bibinfo{title}{Mangrove canopy height globally related to
  precipitation, temperature and cyclone frequency},
\newblock \bibinfo{journal}{Nat. Geosci.} \bibinfo{volume}{12}
  (\bibinfo{year}{2019}) \bibinfo{pages}{40--45}.
\bibitem[{Ma et~al.(2021)Ma, Hauer, Östberg, Koeser, Wei, and Xu}]{Ma2021-ky}
\bibinfo{author}{B.~Ma}, \bibinfo{author}{R.~J. Hauer},
  \bibinfo{author}{J.~Östberg}, \bibinfo{author}{A.~K. Koeser},
  \bibinfo{author}{H.~Wei}, \bibinfo{author}{C.~Xu},
\newblock \bibinfo{title}{A global basis of urban tree inventories: What comes
  first the inventory or the program},
\newblock \bibinfo{journal}{Urban For. Urban Greening} \bibinfo{volume}{60}
  (\bibinfo{year}{2021}) \bibinfo{pages}{127087}.
\bibitem[{Nielsen et~al.(2014)Nielsen, Östberg, and
  Delshammar}]{Nielsen2014-iw}
\bibinfo{author}{A.~B. Nielsen}, \bibinfo{author}{J.~Östberg},
  \bibinfo{author}{T.~Delshammar},
\newblock \bibinfo{title}{Review of urban tree inventory methods used to
  collect data at single-tree level},
\newblock \bibinfo{journal}{Arboriculture \& Urban Forestry (AUF)}
  \bibinfo{volume}{40} (\bibinfo{year}{2014}) \bibinfo{pages}{96--111}.
\bibitem[{Ossola et~al.(2020)Ossola, Hoeppner, Burley, Gallagher, Beaumont, and
  Leishman}]{Ossola2020-vd}
\bibinfo{author}{A.~Ossola}, \bibinfo{author}{M.~J. Hoeppner},
  \bibinfo{author}{H.~M. Burley}, \bibinfo{author}{R.~V. Gallagher},
  \bibinfo{author}{L.~J. Beaumont}, \bibinfo{author}{M.~R. Leishman},
\newblock \bibinfo{title}{The global urban tree inventory: A database of the
  diverse tree flora that inhabits the world’s cities},
\newblock \bibinfo{journal}{Glob. Ecol. Biogeogr.} \bibinfo{volume}{29}
  (\bibinfo{year}{2020}) \bibinfo{pages}{1907--1914}.
\bibitem[{{Direction des Espaces Verts et de l'Environnement - Ville de
  Paris}(2024)}]{paris2024trees}
\bibinfo{author}{{Direction des Espaces Verts et de l'Environnement - Ville de
  Paris}}, \bibinfo{title}{Arbres}, \bibinfo{year}{2024}. \URLprefix
  \url{https://opendata.paris.fr/explore/dataset/les-arbres/information/},
  \bibinfo{note}{accessed: 2024-11-14}.
\bibitem[{{Gerència d'Àrea Agenda 2030, Transició Digital i
  Esports}(2024)}]{barcelona2015trees}
\bibinfo{author}{{Gerència d'Àrea Agenda 2030, Transició Digital i
  Esports}}, \bibinfo{title}{Local interest trees from barcelona},
  \bibinfo{year}{2024}. \URLprefix
  \url{https://opendata-ajuntament.barcelona.cat/data/es/dataset/arbres-interes-local},
  \bibinfo{note}{accessed: 2024-11-14}.
\bibitem[{{Department of Parks and Recreation (DPR)}(2024)}]{newyork2024trees}
\bibinfo{author}{{Department of Parks and Recreation (DPR)}},
  \bibinfo{title}{2015 street tree census - tree data}, \bibinfo{year}{2024}.
  \URLprefix
  \url{https://data.cityofnewyork.us/Environment/2015-Street-Tree-Census-Tree-Data/uvpi-gqnh/about_data},
  \bibinfo{note}{accessed: 2024-11-14}.
\bibitem[{Lang et~al.(2023)Lang, Jetz, Schindler, and Wegner}]{Lang2023-cy}
\bibinfo{author}{N.~Lang}, \bibinfo{author}{W.~Jetz},
  \bibinfo{author}{K.~Schindler}, \bibinfo{author}{J.~D. Wegner},
\newblock \bibinfo{title}{A high-resolution canopy height model of the earth},
\newblock \bibinfo{journal}{Nat Ecol Evol}  (\bibinfo{year}{2023}).
\bibitem[{{NASA JPL}(2020)}]{nasadem2020}
\bibinfo{author}{{NASA JPL}}, \bibinfo{title}{{NASADEM} {M}erged {DEM} {G}lobal
  1 arc second {V}001}, \bibinfo{year}{2020}.
  \DOIprefix\doi{10.5067/MEaSUREs/NASADEM/NASADEM_HGT.001},
  \bibinfo{note}{accessed 2024-10-28}.
\bibitem[{{European Space Agency}(2024)}]{copernicus2024dem}
\bibinfo{author}{{European Space Agency}}, \bibinfo{title}{Copernicus global
  digital elevation model}, \bibinfo{year}{2024}. \bibinfo{note}{Accessed:
  2024-11-13}.
\bibitem[{Tadono et~al.(2014)Tadono, Ishida, Oda, Naito, Minakawa, and
  Iwamoto}]{Tadono2014-xo}
\bibinfo{author}{T.~Tadono}, \bibinfo{author}{H.~Ishida},
  \bibinfo{author}{F.~Oda}, \bibinfo{author}{S.~Naito},
  \bibinfo{author}{K.~Minakawa}, \bibinfo{author}{H.~Iwamoto},
\newblock \bibinfo{title}{Precise global {DEM} generation by {ALOS} {PRISM}},
\newblock \bibinfo{journal}{ISPRS Ann. Photogramm. Remote Sens. Spat. Inf.
  Sci.} \bibinfo{volume}{II-4} (\bibinfo{year}{2014}) \bibinfo{pages}{71--76}.
\bibitem[{O'Loughlin et~al.(2016)O'Loughlin, Paiva, Durand, Alsdorf, and
  Bates}]{O-Loughlin2016-ju}
\bibinfo{author}{F.~E. O'Loughlin}, \bibinfo{author}{R.~C.~D. Paiva},
  \bibinfo{author}{M.~Durand}, \bibinfo{author}{D.~E. Alsdorf},
  \bibinfo{author}{P.~D. Bates},
\newblock \bibinfo{title}{A multi-sensor approach towards a global vegetation
  corrected {SRTM} {DEM} product},
\newblock \bibinfo{journal}{Remote Sens. Environ.} \bibinfo{volume}{182}
  (\bibinfo{year}{2016}) \bibinfo{pages}{49--59}.
\bibitem[{Pronk et~al.(2024)Pronk, Hooijer, Eilander, Haag, de~Jong,
  Vousdoukas, Vernimmen, Ledoux, and Eleveld}]{Pronk2024-sp}
\bibinfo{author}{M.~Pronk}, \bibinfo{author}{A.~Hooijer},
  \bibinfo{author}{D.~Eilander}, \bibinfo{author}{A.~Haag},
  \bibinfo{author}{T.~de~Jong}, \bibinfo{author}{M.~Vousdoukas},
  \bibinfo{author}{R.~Vernimmen}, \bibinfo{author}{H.~Ledoux},
  \bibinfo{author}{M.~Eleveld},
\newblock \bibinfo{title}{{DeltaDTM}: A global coastal digital terrain model},
\newblock \bibinfo{journal}{Sci. Data} \bibinfo{volume}{11}
  (\bibinfo{year}{2024}) \bibinfo{pages}{273}.
\bibitem[{{U.S. Geological Survey}(2023)}]{usgs2023}
\bibinfo{author}{{U.S. Geological Survey}}, \bibinfo{title}{1 meter digital
  elevation models (dems) - usgs national map 3dep downloadable data
  collection}, \bibinfo{year}{2023}. \URLprefix
  \url{https://data.usgs.gov/datacatalog/data/USGS:77ae0551-c61e-4979-aedd-d797abdcde0e},
  \bibinfo{note}{accessed: 2024-11-13}.
\bibitem[{{Environment Agency}(2024)}]{envagency2024}
\bibinfo{author}{{Environment Agency}}, \bibinfo{title}{Lidar composite digital
  terrain model (dtm) - 1m}, \bibinfo{year}{2024}. \URLprefix
  \url{https://environment.data.gov.uk/dataset/13787b9a-26a4-4775-8523-806d13af58fc},
  \bibinfo{note}{accessed: 2024-11-13}.
\bibitem[{{Geoscience Australia}(2015)}]{geoscienceaus2015dem}
\bibinfo{author}{{Geoscience Australia}}, \bibinfo{title}{Digital elevation
  model (dem) of australia derived from lidar 5 metre grid},
  \bibinfo{year}{2015}. \bibinfo{note}{Accessed: 2024-11-13}.
\bibitem[{{National Institute of Geographic and Forest
  Information}(2024)}]{rgealti2024}
\bibinfo{author}{{National Institute of Geographic and Forest Information}},
  \bibinfo{title}{Rge alti® 1m}, \bibinfo{year}{2024}. \URLprefix
  \url{https://geoservices.ign.fr/rgealti}, \bibinfo{note}{accessed:
  2024-11-13}.
\bibitem[{Liang and Gong(2017)}]{Liang2017-yu}
\bibinfo{author}{J.~Liang}, \bibinfo{author}{J.~Gong},
\newblock \bibinfo{title}{A sparse voxel octree-based framework for computing
  solar radiation using {3D} city models},
\newblock \bibinfo{journal}{ISPRS Int. J. Geoinf.} \bibinfo{volume}{6}
  (\bibinfo{year}{2017}) \bibinfo{pages}{106}.
\bibitem[{Park et~al.(2021)Park, Guldmann, and Liu}]{Park2021-dw}
\bibinfo{author}{Y.~Park}, \bibinfo{author}{J.-M. Guldmann},
  \bibinfo{author}{D.~Liu},
\newblock \bibinfo{title}{Impacts of tree and building shades on the urban heat
  island: Combining remote sensing, {3D} digital city and spatial regression
  approaches},
\newblock \bibinfo{journal}{Comput. Environ. Urban Syst.} \bibinfo{volume}{88}
  (\bibinfo{year}{2021}) \bibinfo{pages}{101655}.
\bibitem[{Huang et~al.(2018)Huang, Yang, Matzarakis, and Lin}]{Huang2018-ak}
\bibinfo{author}{K.-T. Huang}, \bibinfo{author}{S.-R. Yang},
  \bibinfo{author}{A.~Matzarakis}, \bibinfo{author}{T.-P. Lin},
\newblock \bibinfo{title}{Identifying outdoor thermal risk areas and evaluation
  of future thermal comfort concerning shading orientation in a traditional
  settlement},
\newblock \bibinfo{journal}{Sci. Total Environ.} \bibinfo{volume}{626}
  (\bibinfo{year}{2018}) \bibinfo{pages}{567--580}.
\bibitem[{Coccolo et~al.(2016)Coccolo, K{\"a}mpf, Scartezzini, and
  Pearlmutter}]{Coccolo2016-jd}
\bibinfo{author}{S.~Coccolo}, \bibinfo{author}{J.~K{\"a}mpf},
  \bibinfo{author}{J.-L. Scartezzini}, \bibinfo{author}{D.~Pearlmutter},
\newblock \bibinfo{title}{Outdoor human comfort and thermal stress: A
  comprehensive review on models and standards},
\newblock \bibinfo{journal}{Urban Climate} \bibinfo{volume}{18}
  (\bibinfo{year}{2016}) \bibinfo{pages}{33--57}.
\bibitem[{Palliwal et~al.(2021)Palliwal, Song, Tan, and
  Biljecki}]{Palliwal2021-ft}
\bibinfo{author}{A.~Palliwal}, \bibinfo{author}{S.~Song},
  \bibinfo{author}{H.~T.~W. Tan}, \bibinfo{author}{F.~Biljecki},
\newblock \bibinfo{title}{{3D} city models for urban farming site
  identification in buildings},
\newblock \bibinfo{journal}{Comput. Environ. Urban Syst.} \bibinfo{volume}{86}
  (\bibinfo{year}{2021}) \bibinfo{pages}{101584}.
\bibitem[{Hofierka and Ka{\v n}uk(2009)}]{Hofierka2009-eb}
\bibinfo{author}{J.~Hofierka}, \bibinfo{author}{J.~Ka{\v n}uk},
\newblock \bibinfo{title}{Assessment of photovoltaic potential in urban areas
  using open-source solar radiation tools},
\newblock \bibinfo{journal}{Renewable Energy} \bibinfo{volume}{34}
  (\bibinfo{year}{2009}) \bibinfo{pages}{2206--2214}.
\bibitem[{Mondol et~al.(2008)Mondol, Yohanis, and Norton}]{Mondol2008-hh}
\bibinfo{author}{J.~D. Mondol}, \bibinfo{author}{Y.~G. Yohanis},
  \bibinfo{author}{B.~Norton},
\newblock \bibinfo{title}{Solar radiation modelling for the simulation of
  photovoltaic systems},
\newblock \bibinfo{journal}{Renewable Energy} \bibinfo{volume}{33}
  (\bibinfo{year}{2008}) \bibinfo{pages}{1109--1120}.
\bibitem[{Lam and Li(1999)}]{Lam1999-im}
\bibinfo{author}{J.~C. Lam}, \bibinfo{author}{D.~H.~W. Li},
\newblock \bibinfo{title}{An analysis of daylighting and solar heat for
  cooling-dominated office buildings},
\newblock \bibinfo{journal}{Sol. Energy} \bibinfo{volume}{65}
  (\bibinfo{year}{1999}) \bibinfo{pages}{251--262}.
\bibitem[{Causone et~al.(2010)Causone, Corgnati, Filippi, and
  Olesen}]{Causone2010-ug}
\bibinfo{author}{F.~Causone}, \bibinfo{author}{S.~P. Corgnati},
  \bibinfo{author}{M.~Filippi}, \bibinfo{author}{B.~W. Olesen},
\newblock \bibinfo{title}{Solar radiation and cooling load calculation for
  radiant systems: Definition and evaluation of the direct solar load},
\newblock \bibinfo{journal}{Energy Build.} \bibinfo{volume}{42}
  (\bibinfo{year}{2010}) \bibinfo{pages}{305--314}.
\bibitem[{Pružinec and Ďuračiová(2022)}]{Pruzinec2022-uo}
\bibinfo{author}{F.~Pružinec}, \bibinfo{author}{R.~Ďuračiová},
\newblock \bibinfo{title}{A point-cloud solar radiation tool},
\newblock \bibinfo{journal}{Energies} \bibinfo{volume}{15}
  (\bibinfo{year}{2022}) \bibinfo{pages}{7018}.
\bibitem[{Monsi and Saeki(2004)}]{Monsi2004-eb}
\bibinfo{author}{M.~Monsi}, \bibinfo{author}{T.~Saeki},
\newblock \bibinfo{title}{On the factor light in plant communities and its
  importance for matter production. 1953},
\newblock \bibinfo{journal}{Ann. Bot.} \bibinfo{volume}{95}
  (\bibinfo{year}{2004}) \bibinfo{pages}{549--567}.
\bibitem[{Mor et~al.(2021)Mor, Fisher-Gewirtzman, Yosifof, and
  Dalyot}]{Mor2021-di}
\bibinfo{author}{M.~Mor}, \bibinfo{author}{D.~Fisher-Gewirtzman},
  \bibinfo{author}{R.~Yosifof}, \bibinfo{author}{S.~Dalyot},
\newblock \bibinfo{title}{{3D} visibility analysis for evaluating the
  attractiveness of tourism routes computed from social media photos},
\newblock \bibinfo{journal}{ISPRS Int. J. Geoinf.} \bibinfo{volume}{10}
  (\bibinfo{year}{2021}) \bibinfo{pages}{275}.
\bibitem[{Wróżyński et~al.(2024)Wróżyński, Pyszny, and
  Wróżyńska}]{Wrozynski2024-cd}
\bibinfo{author}{R.~Wróżyński}, \bibinfo{author}{K.~Pyszny},
  \bibinfo{author}{M.~Wróżyńska},
\newblock \bibinfo{title}{Reaching beyond {GIS} for comprehensive {3D}
  visibility analysis},
\newblock \bibinfo{journal}{Landsc. Urban Plan.} \bibinfo{volume}{247}
  (\bibinfo{year}{2024}) \bibinfo{pages}{105074}.
\bibitem[{Filomena and Verstegen(2021)}]{filomena_modelling_2021}
\bibinfo{author}{G.~Filomena}, \bibinfo{author}{J.~A. Verstegen},
\newblock \bibinfo{title}{Modelling the effect of landmarks on pedestrian
  dynamics in urban environments},
\newblock \bibinfo{journal}{Computers, Environment and Urban Systems}
  \bibinfo{volume}{86} (\bibinfo{year}{2021}) \bibinfo{pages}{101573}.
  \DOIprefix\doi{10.1016/j.compenvurbsys.2020.101573}.
\bibitem[{Yesiltepe et~al.(2021)Yesiltepe, Conroy~Dalton, and
  Ozbil~Torun}]{yesiltepe_landmarks_2021}
\bibinfo{author}{D.~Yesiltepe}, \bibinfo{author}{R.~Conroy~Dalton},
  \bibinfo{author}{A.~Ozbil~Torun},
\newblock \bibinfo{title}{Landmarks in wayfinding: a review of the existing
  literature},
\newblock \bibinfo{journal}{Cognitive Processing} \bibinfo{volume}{22}
  (\bibinfo{year}{2021}) \bibinfo{pages}{369--410}.
  \DOIprefix\doi{10.1007/s10339-021-01012-x}.
\bibitem[{Sabesan et~al.(2024)Sabesan, Meetiyagoda, and
  Rathnasekara}]{Sabesan2024-vy}
\bibinfo{author}{L.~Sabesan}, \bibinfo{author}{L.~Meetiyagoda},
  \bibinfo{author}{S.~Rathnasekara},
\newblock \bibinfo{title}{Landmarks and walkability: wayfinding during
  nighttime in a tourism-based city. case study of jaffna, sri lanka},
\newblock \bibinfo{journal}{GeoJournal} \bibinfo{volume}{89}
  (\bibinfo{year}{2024}) \bibinfo{pages}{1--15}.
\bibitem[{Yuan et~al.(2024)Yuan, Yuyao, and Miao}]{Yuan2024-ls}
\bibinfo{author}{G.~Yuan}, \bibinfo{author}{W.~Yuyao},
  \bibinfo{author}{Y.~Miao},
\newblock \bibinfo{title}{Selecting building height control indicators of
  landmark skylines: A visual perception experiment},
\newblock \bibinfo{journal}{Environ. Plan. B Urban Anal. City Sci.}
  (\bibinfo{year}{2024}).
\bibitem[{Turan et~al.(2021)Turan, Chegut, Fink, and
  Reinhart}]{turan_development_2021}
\bibinfo{author}{I.~Turan}, \bibinfo{author}{A.~Chegut},
  \bibinfo{author}{D.~Fink}, \bibinfo{author}{C.~Reinhart},
\newblock \bibinfo{title}{Development of view potential metrics and the
  financial impact of views on office rents},
\newblock \bibinfo{journal}{Landscape and Urban Planning} \bibinfo{volume}{215}
  (\bibinfo{year}{2021}) \bibinfo{pages}{104193}.
  \DOIprefix\doi{10.1016/j.landurbplan.2021.104193}.
\bibitem[{Liu et~al.(2021)Liu, Cheng, Jim, Morakinyo, Shi, and Ng}]{Liu2021}
\bibinfo{author}{Z.~Liu}, \bibinfo{author}{W.~Cheng}, \bibinfo{author}{C.~Jim},
  \bibinfo{author}{T.~E. Morakinyo}, \bibinfo{author}{Y.~Shi},
  \bibinfo{author}{E.~Ng},
\newblock \bibinfo{title}{{Heat mitigation benefits of urban green and blue
  infrastructures: A systematic review of modeling techniques, validation and
  scenario simulation in ENVI-met V4}},
\newblock \bibinfo{journal}{Building and Environment} \bibinfo{volume}{200}
  (\bibinfo{year}{2021}) \bibinfo{pages}{107939}.
  \DOIprefix\doi{10.1016/j.buildenv.2021.107939}.
\bibitem[{Ouyang et~al.(2022)Ouyang, Sinsel, Simon, Morakinyo, Liu, and
  Ng}]{Ouyang2022-tv}
\bibinfo{author}{W.~Ouyang}, \bibinfo{author}{T.~Sinsel},
  \bibinfo{author}{H.~Simon}, \bibinfo{author}{T.~E. Morakinyo},
  \bibinfo{author}{H.~Liu}, \bibinfo{author}{E.~Ng},
\newblock \bibinfo{title}{Evaluating the thermal-radiative performance of
  {ENVI}-met model for green infrastructure typologies: Experience from a
  subtropical climate},
\newblock \bibinfo{journal}{Build. Environ.} \bibinfo{volume}{207}
  (\bibinfo{year}{2022}) \bibinfo{pages}{108427}.
\bibitem[{Harris et~al.(2020)Harris, Millman, van~der Walt, Gommers, Virtanen,
  Cournapeau, Wieser, Taylor, Berg, Smith, Kern, Picus, Hoyer, van Kerkwijk,
  Brett, Haldane, del R{\'{i}}o, Wiebe, Peterson, G{\'{e}}rard-Marchant,
  Sheppard, Reddy, Weckesser, Abbasi, Gohlke, and Oliphant}]{harris2020array}
\bibinfo{author}{C.~R. Harris}, \bibinfo{author}{K.~J. Millman},
  \bibinfo{author}{S.~J. van~der Walt}, \bibinfo{author}{R.~Gommers},
  \bibinfo{author}{P.~Virtanen}, \bibinfo{author}{D.~Cournapeau},
  \bibinfo{author}{E.~Wieser}, \bibinfo{author}{J.~Taylor},
  \bibinfo{author}{S.~Berg}, \bibinfo{author}{N.~J. Smith},
  \bibinfo{author}{R.~Kern}, \bibinfo{author}{M.~Picus},
  \bibinfo{author}{S.~Hoyer}, \bibinfo{author}{M.~H. van Kerkwijk},
  \bibinfo{author}{M.~Brett}, \bibinfo{author}{A.~Haldane},
  \bibinfo{author}{J.~F. del R{\'{i}}o}, \bibinfo{author}{M.~Wiebe},
  \bibinfo{author}{P.~Peterson}, \bibinfo{author}{P.~G{\'{e}}rard-Marchant},
  \bibinfo{author}{K.~Sheppard}, \bibinfo{author}{T.~Reddy},
  \bibinfo{author}{W.~Weckesser}, \bibinfo{author}{H.~Abbasi},
  \bibinfo{author}{C.~Gohlke}, \bibinfo{author}{T.~E. Oliphant},
\newblock \bibinfo{title}{Array programming with {NumPy}},
\newblock \bibinfo{journal}{Nature} \bibinfo{volume}{585}
  (\bibinfo{year}{2020}) \bibinfo{pages}{357--362}.
\bibitem[{Bröde et~al.(2012)Bröde, Fiala, Błażejczyk, Holmér, Jendritzky,
  Kampmann, Tinz, and Havenith}]{Brode2012-yj}
\bibinfo{author}{P.~Bröde}, \bibinfo{author}{D.~Fiala},
  \bibinfo{author}{K.~Błażejczyk}, \bibinfo{author}{I.~Holmér},
  \bibinfo{author}{G.~Jendritzky}, \bibinfo{author}{B.~Kampmann},
  \bibinfo{author}{B.~Tinz}, \bibinfo{author}{G.~Havenith},
\newblock \bibinfo{title}{Deriving the operational procedure for the universal
  thermal climate index ({UTCI})},
\newblock \bibinfo{journal}{Int. J. Biometeorol.} \bibinfo{volume}{56}
  (\bibinfo{year}{2012}) \bibinfo{pages}{481--494}.
\bibitem[{Tsoka et~al.(2018)Tsoka, Tsikaloudaki, and Theodosiou}]{Tsoka2018}
\bibinfo{author}{S.~Tsoka}, \bibinfo{author}{A.~Tsikaloudaki},
  \bibinfo{author}{T.~Theodosiou},
\newblock \bibinfo{title}{{Analyzing the ENVI-met microclimate model's
  performance and assessing cool materials and urban vegetation
  applications–A review}},
\newblock \bibinfo{journal}{Sustainable Cities and Society}
  \bibinfo{volume}{43} (\bibinfo{year}{2018}) \bibinfo{pages}{55--76}.
  \DOIprefix\doi{10.1016/j.scs.2018.08.009}.
\bibitem[{Biljecki and Arroyo~Ohori(2015)}]{2015_udmv_citygml_obj}
\bibinfo{author}{F.~Biljecki}, \bibinfo{author}{K.~Arroyo~Ohori},
\newblock \bibinfo{title}{{Automatic Semantic-preserving Conversion Between OBJ
  and CityGML}},
\newblock in: \bibinfo{booktitle}{Eurographics Workshop on Urban Data Modelling
  and Visualisation 2015}, \bibinfo{address}{Delft, Netherlands},
  \bibinfo{year}{2015}, pp. \bibinfo{pages}{25--30}.
\bibitem[{Ledoux et~al.(2019)Ledoux, Arroyo~Ohori, Kumar, Dukai, Labetski, and
  Vitalis}]{ledoux2019cityjson}
\bibinfo{author}{H.~Ledoux}, \bibinfo{author}{K.~Arroyo~Ohori},
  \bibinfo{author}{K.~Kumar}, \bibinfo{author}{B.~Dukai},
  \bibinfo{author}{A.~Labetski}, \bibinfo{author}{S.~Vitalis},
\newblock \bibinfo{title}{Cityjson: A compact and easy-to-use encoding of the
  citygml data model},
\newblock \bibinfo{journal}{Open Geospatial Data, Software and Standards}
  \bibinfo{volume}{4} (\bibinfo{year}{2019}) \bibinfo{pages}{1--12}.
\bibitem[{Vitalis et~al.(2020)Vitalis, Labetski, Boersma, Dahle, Li,
  Arroyo~Ohori, Ledoux, and Stoter}]{vitalis2020cityjson+}
\bibinfo{author}{S.~Vitalis}, \bibinfo{author}{A.~Labetski},
  \bibinfo{author}{F.~Boersma}, \bibinfo{author}{F.~Dahle},
  \bibinfo{author}{X.~Li}, \bibinfo{author}{K.~Arroyo~Ohori},
  \bibinfo{author}{H.~Ledoux}, \bibinfo{author}{J.~Stoter},
\newblock \bibinfo{title}{Cityjson+ web= ninja},
\newblock \bibinfo{journal}{ISPRS Annals of the Photogrammetry, Remote Sensing
  and Spatial Information Sciences} \bibinfo{volume}{6} (\bibinfo{year}{2020})
  \bibinfo{pages}{167--173}.
\bibitem[{Petrova-Antonova et~al.(2024)Petrova-Antonova, Malinov, Mrosla, and
  Petrov}]{petrova2024towards}
\bibinfo{author}{D.~Petrova-Antonova}, \bibinfo{author}{S.~Malinov},
  \bibinfo{author}{L.~Mrosla}, \bibinfo{author}{A.~Petrov},
\newblock \bibinfo{title}{Towards a conceptual model of citygml 3.0 vegetation
  ade},
\newblock \bibinfo{journal}{The International Archives of the Photogrammetry,
  Remote Sensing and Spatial Information Sciences} \bibinfo{volume}{48}
  (\bibinfo{year}{2024}) \bibinfo{pages}{155--161}.
\bibitem[{Lei et~al.(2024)Lei, Liang, and Biljecki}]{lei2024integrating}
\bibinfo{author}{B.~Lei}, \bibinfo{author}{X.~Liang},
  \bibinfo{author}{F.~Biljecki},
\newblock \bibinfo{title}{Integrating human perception in 3d city models and
  urban digital twins},
\newblock \bibinfo{journal}{ISPRS Annals of the Photogrammetry, Remote Sensing
  and Spatial Information Sciences} \bibinfo{volume}{10} (\bibinfo{year}{2024})
  \bibinfo{pages}{211--218}.
\bibitem[{Marino et~al.(2017)Marino, Nucara, and Pietrafesa}]{marino_does_2017}
\bibinfo{author}{C.~Marino}, \bibinfo{author}{A.~Nucara},
  \bibinfo{author}{M.~Pietrafesa},
\newblock \bibinfo{title}{Does window-to-wall ratio have a significant effect
  on the energy consumption of buildings? a parametric analysis in italian
  climate conditions},
\newblock \bibinfo{journal}{Journal of Building Engineering}
  \bibinfo{volume}{13} (\bibinfo{year}{2017}) \bibinfo{pages}{169--183}.
\bibitem[{Troup et~al.(2019)Troup, Phillips, Eckelman, and
  Fannon}]{troup_effect_2019}
\bibinfo{author}{L.~Troup}, \bibinfo{author}{R.~Phillips},
  \bibinfo{author}{M.~J. Eckelman}, \bibinfo{author}{D.~Fannon},
\newblock \bibinfo{title}{Effect of window-to-wall ratio on measured energy
  consumption in {US} office buildings},
\newblock \bibinfo{journal}{Energy and Buildings} \bibinfo{volume}{203}
  (\bibinfo{year}{2019}) \bibinfo{pages}{109434}.
\bibitem[{Chi et~al.(2020)Chi, Wang, Wang, Li, and
  Peng}]{chi_investigation_2020}
\bibinfo{author}{F.~Chi}, \bibinfo{author}{Y.~Wang}, \bibinfo{author}{R.~Wang},
  \bibinfo{author}{G.~Li}, \bibinfo{author}{C.~Peng},
\newblock \bibinfo{title}{An investigation of optimal window-to-wall ratio
  based on changes in building orientations for traditional dwellings},
\newblock \bibinfo{journal}{Solar Energy} \bibinfo{volume}{195}
  (\bibinfo{year}{2020}) \bibinfo{pages}{64--81}.
\bibitem[{Liang et~al.(2025)Liang, Xie, Zhao, Stouffs, and
  Biljecki}]{liang2025openfacades}
\bibinfo{author}{X.~Liang}, \bibinfo{author}{J.~Xie},
  \bibinfo{author}{T.~Zhao}, \bibinfo{author}{R.~Stouffs},
  \bibinfo{author}{F.~Biljecki}, \bibinfo{title}{Openfacades: An open framework
  for architectural caption and attribute data enrichment via street view
  imagery}, \bibinfo{year}{2025}. \URLprefix
  \url{https://arxiv.org/abs/2504.02866}.
  \href{http://arxiv.org/abs/2504.02866}{{\tt arXiv:2504.02866}}.
\bibitem[{Kang et~al.(2018)Kang, Körner, Wang, Taubenböck, and
  Zhu}]{kang_building_2018}
\bibinfo{author}{J.~Kang}, \bibinfo{author}{M.~Körner},
  \bibinfo{author}{Y.~Wang}, \bibinfo{author}{H.~Taubenböck},
  \bibinfo{author}{X.~X. Zhu},
\newblock \bibinfo{title}{Building instance classification using street view
  images},
\newblock \bibinfo{journal}{ISPRS Journal of Photogrammetry and Remote Sensing}
  \bibinfo{volume}{145} (\bibinfo{year}{2018}) \bibinfo{pages}{44--59}.
\bibitem[{Sun et~al.(2022)Sun, Zhang, Duarte, and
  Ratti}]{sun_understanding_2022}
\bibinfo{author}{M.~Sun}, \bibinfo{author}{F.~Zhang},
  \bibinfo{author}{F.~Duarte}, \bibinfo{author}{C.~Ratti},
\newblock \bibinfo{title}{Understanding architecture age and style through deep
  learning},
\newblock \bibinfo{journal}{Cities} \bibinfo{volume}{128}
  (\bibinfo{year}{2022}) \bibinfo{pages}{103787}.
\bibitem[{Raghu et~al.(2023)Raghu, Bucher, and De~Wolf}]{raghu_towards_2023}
\bibinfo{author}{D.~Raghu}, \bibinfo{author}{M.~J.~J. Bucher},
  \bibinfo{author}{C.~De~Wolf},
\newblock \bibinfo{title}{Towards a ‘resource cadastre’ for a circular
  economy – {Urban}-scale building material detection using street view
  imagery and computer vision},
\newblock \bibinfo{journal}{Resources, Conservation and Recycling}
  \bibinfo{volume}{198} (\bibinfo{year}{2023}) \bibinfo{pages}{107140}.
\bibitem[{Fabbri et~al.(2020)Fabbri, Gaspari, Bartoletti, and
  Antonini}]{fabbri2020effect}
\bibinfo{author}{K.~Fabbri}, \bibinfo{author}{J.~Gaspari},
  \bibinfo{author}{S.~Bartoletti}, \bibinfo{author}{E.~Antonini},
\newblock \bibinfo{title}{Effect of facade reflectance on outdoor microclimate:
  An italian case study},
\newblock \bibinfo{journal}{Sustainable cities and society}
  \bibinfo{volume}{54} (\bibinfo{year}{2020}) \bibinfo{pages}{101984}.
\bibitem[{Harish and Kumar(2016)}]{harish2016review}
\bibinfo{author}{V.~Harish}, \bibinfo{author}{A.~Kumar},
\newblock \bibinfo{title}{A review on modeling and simulation of building
  energy systems},
\newblock \bibinfo{journal}{Renewable and sustainable energy reviews}
  \bibinfo{volume}{56} (\bibinfo{year}{2016}) \bibinfo{pages}{1272--1292}.
\bibitem[{Ren et~al.(2024)Ren, Liu, Zeng, Lin, Li, Cao, Chen, Huang, Chen, Yan,
  Zeng, Zhang, Li, Yang, Li, Jiang, and Zhang}]{ren_grounded_2024}
\bibinfo{author}{T.~Ren}, \bibinfo{author}{S.~Liu}, \bibinfo{author}{A.~Zeng},
  \bibinfo{author}{J.~Lin}, \bibinfo{author}{K.~Li}, \bibinfo{author}{H.~Cao},
  \bibinfo{author}{J.~Chen}, \bibinfo{author}{X.~Huang},
  \bibinfo{author}{Y.~Chen}, \bibinfo{author}{F.~Yan},
  \bibinfo{author}{Z.~Zeng}, \bibinfo{author}{H.~Zhang},
  \bibinfo{author}{F.~Li}, \bibinfo{author}{J.~Yang}, \bibinfo{author}{H.~Li},
  \bibinfo{author}{Q.~Jiang}, \bibinfo{author}{L.~Zhang},
  \bibinfo{title}{Grounded {SAM}: Assembling open-world models for diverse
  visual tasks}, \bibinfo{year}{2024}. \URLprefix
  \url{http://arxiv.org/abs/2401.14159}. \href{http://arxiv.org/abs/2401.14159
  [cs]}{{\tt arXiv:2401.14159 [cs]}}.
\bibitem[{Fath et~al.(2015)Fath, Stengel, Sprenger, Wilson, Schultmann, and
  Kuhn}]{Fath2015-jw}
\bibinfo{author}{K.~Fath}, \bibinfo{author}{J.~Stengel},
  \bibinfo{author}{W.~Sprenger}, \bibinfo{author}{H.~R. Wilson},
  \bibinfo{author}{F.~Schultmann}, \bibinfo{author}{T.~E. Kuhn},
\newblock \bibinfo{title}{A method for predicting the economic potential of
  (building-integrated) photovoltaics in urban areas based on hourly radiance
  simulations},
\newblock \bibinfo{journal}{Sol. Energy} \bibinfo{volume}{116}
  (\bibinfo{year}{2015}) \bibinfo{pages}{357--370}.

\end{thebibliography}
\end{document}